\documentclass[aps,pra,superscriptaddress,twocolumn,longbibliography]{revtex4}
\usepackage{mathrsfs}
\usepackage{amsfonts}
\usepackage{amssymb}
\usepackage{amsmath}
\usepackage{graphicx}
\usepackage{lipsum} 
\usepackage{ulem}
\graphicspath{{figures/}{./}} 
\usepackage{sidecap}
\usepackage{color}
\usepackage[colorlinks=true, urlcolor=blue, citecolor=blue,linkcolor=blue,citebordercolor={1 0 0},linkbordercolor={0 0 1}]{hyperref}
\usepackage{subeqnarray}
\usepackage{verbatim}
\usepackage{euscript} 

\usepackage{shortcuts}

\usepackage[textsize=tiny]{todonotes}

\usepackage{array}
\usepackage{multirow}
\usepackage{multibib}
\usepackage{xcolor}

\newcommand{\ket}[1]{\vert #1 \rangle}

\newcommand{\brakets}[1]{{\langle #1 \rangle}}
\newcommand{\mean}[1]{\langle #1 \rangle}

\renewcommand{\eqref}[1]{Eq.~(\ref{#1})} 
\newcommand{\figref}[1]{Fig.~\ref{#1}} 

\newcommand\commentout[1]{}
\newboolean{hidecomments}
\setboolean{hidecomments}{false} 
\ifthenelse{\boolean{hidecomments}}
{\newcommand{\cb}[2]{}}
{\newcommand{\cb}[2]{
    \fbox{\bfseries\sffamily\scriptsize#1}
    {\sf\small$\blacktriangleright$ 
      {#2} $\blacktriangleleft$}}} 

\definecolor{mydarkblue}{rgb}{0,0.08,0.45}
\definecolor{myfavblue}{rgb}{0.1176, 0.392, 1.0}
\hypersetup{
    colorlinks=true,
    linkcolor=mydarkblue,
    citecolor=mydarkblue,
    filecolor=mydarkblue,
    urlcolor=mydarkblue}

\renewcommand{\epsilon}{\varepsilon}

\newcommand{\vh}{\mathbf{h}}

\newcommand{\hidden}{\vh}

\newcommand{\solvefunc}{\textnormal{ODESolve}}

\newcommand{\utchem}{Department  of  Chemistry,  University  of  Toronto,  Toronto,  Ontario  M5G 1Z8,  Canada}
\newcommand{\utcomp}{Department  of  Computer Science,  University  of  Toronto,  Toronto,  Ontario  M5S 2E4,  Canada}
\newcommand{\vectorinst}{Vector  Institute  for  Artificial  Intelligence,  Toronto,  Ontario  M5S  1M1,  Canada}
\newcommand{\cifar}{Canadian  Institute  for  Advanced  Research,  Toronto,  Ontario  M5G  1Z8,  Canada}

\begin{document}

\title{Learning quantum dynamics with latent neural ODEs}

\author{Matthew Choi}
\thanks{These authors contributed equally. \urlstyle{same} }
\affiliation{\utcomp}

\author{Daniel Flam-Shepherd}
\thanks{These authors contributed equally. \urlstyle{same} }
\affiliation{\utcomp}
\affiliation{\vectorinst}

\author{Thi Ha Kyaw}
\email{thihakyaw@cs.toronto.edu}
\affiliation{\utcomp}
\affiliation{\utchem}

\author{Al\'an Aspuru-Guzik}
\email{alan@aspuru.com}
\affiliation{\utcomp}
\affiliation{\vectorinst}
\affiliation{\utchem}
\affiliation{\cifar}

\date{\today}

\begin{abstract}
The core objective of machine-assisted scientific discovery is to learn physical laws from experimental data without prior knowledge of the systems in question.
In the area of quantum physics, making progress towards these goals is significantly more challenging due to the curse of dimensionality as well as the counter-intuitive nature of quantum mechanics.
Here, we present the QNODE, a latent neural ODE trained on expectation values of closed and open quantum systems dynamics. 
It can learn to generate such measurement data and extrapolate outside of its training region that satisfies the von Neumann and time-local Lindblad master equations for closed and open quantum systems respectively in an unsupervised means. 
Furthermore, the QNODE rediscovers quantum mechanical laws such as the Heisenberg's uncertainty principle in a data-driven way, without any constraint or guidance. 
Additionally, we show that trajectories that are generated from the QNODE that are close in its latent space have similar quantum dynamics while preserving the physics of the training system. 
\end{abstract}

\maketitle

\section{Introduction}
Deep learning and neural networks have recently become the powerhouse in machine learning (ML) and they have successfully been used to tackle complex problems in classical \cite{greydanus2019hamiltonian,iten2020discovering,tranter2018multiparameter} and quantum mechanics \cite{torlai2018neural,biamonte2017quantum,carrasquilla2017machine,rem2019identifying} (see Refs. \cite{carleo2019machine,sarma2019machine,alhousseini2019physicist,bharti2020machine,roscher2020explainable} for reviews). 
Machine-assisted scientific discovery is still in its infancy but progress has been made, mostly by building the correct inductive bias-- or structure into the model or loss function.  
For example physical conservation laws can be learned \cite{greydanus2019hamiltonian,iten2020discovering}. Other work has made progress, in a purely data-driven approach learning relationships between quantum experiments and entanglement 
using generative models \cite{flamshepherd2021learning}. 
Recently, neural ordinary differential equations (ODEs) were introduced \cite{chen2018neural,wiewel2019latent}, a neural network layer defined by differential equations. 
Neural ODEs provide the perfect model for physics, since many physical laws are governed by ODEs, and thus every neural ODE has the correct inductive bias built into the model itself.

Quantum computing is another prominent area of research currently, with the potential to outperform the capabilities of the best classical computers 
\cite{arute2019quantum,zhong2020quantum,wu2021strong}. 
To make advances in this so-called noisy intermediate-scale quantum (NISQ) era \cite{preskill2018quantum},
where a quantum computer can possess hundreds of qubits, several NISQ algorithms \cite{bharti2021noisy,cerezo2021variational,tilly2021variational,zhang2021geometric} have been proposed. 
One of the fundamental challenges for NISQ devices to scale-up is to understand noises involved in devices. 
However, accounting for these undesired environmental effects would require exponential classical compute resources \cite{iyer2018small}-- 
one solution attempt is to employ quantum computer-aided design  \cite{kyaw2020quantum,kottmann2021quantum} 
which still can't fully account for the environmental effects due to limited NISQ hardware availability.

In general, the study of open quantum systems are important for quantum computing as well as many other areas of physics from many-body phenomenon \cite{kyaw2020dynamical,ashida2020quantum}, light-matter interaction \cite{kyaw2017parity,kyaw2015creation,kyaw2019towards} to non-equilibrium physics \cite{bastidas2018floquet,heyl2018dynamical}.

Here, we demonstrate that latent ODEs can be trained to generate and extrapolate measurement data from dynamical quantum evolution in both closed and open quantum systems 
using only physical observations without specifying the physics a priori.
This is in line with treating the quantum system as a black box and the ``shut up and calculate" philosophy \cite{mermin2004could} all the while ignoring ontological interpretation \cite{bohm1994undivided} of quantum physics. 
The QNODE can predict and extrapolate quantum trajectories much longer than trained on without needing to solve the underlying Schrodinger equation of motion or  the time-local Lindblad master equation \cite{breuer2002theory,kyaw2019towards} (setting $\hbar=1$ onwards. See supplemental material (SM) \ref{append:master_eq} for detailed derivation): 
$
    {d\hat{\rho}_S (t)}/{dt}=-i [\hat{H},\hat{\rho}_S (t)] + \sum_{\nu} \Gamma_\nu \left({\textbf{A}}_\nu \hat{\rho}_S (t) {\textbf{A}}_\nu ^\dagger -\frac{1}{2} \{{\textbf{A}}_\nu^\dagger {\textbf{A}}_\nu ,\hat{\rho}_S (t)  \} \right).
$
Here, $\Gamma_\nu \geq 0$ are decay rates, $\textbf{A}_\nu$ are superoperators depending on physical noise model considered and $\{a ,b\}=ab+ba$ is an anti-commutator, while $\hat{H}$ is the quantum system Hamiltonian and $\hat{\rho}_S$ is the system density matrix.
By setting $\Gamma_\nu =0$, we arrive at the von Neumann equation for a closed quantum system.
There are some attempts to learn open quantum systems  \cite{flurin2020using,krastanov2020unboxing} using recurrent neural networks (RNNs) but these models
have inaccurate long-term predictions and poor extrapolation \cite{chen2020learning}.

\begin{figure*}[t]
\includegraphics[width=0.99\textwidth]{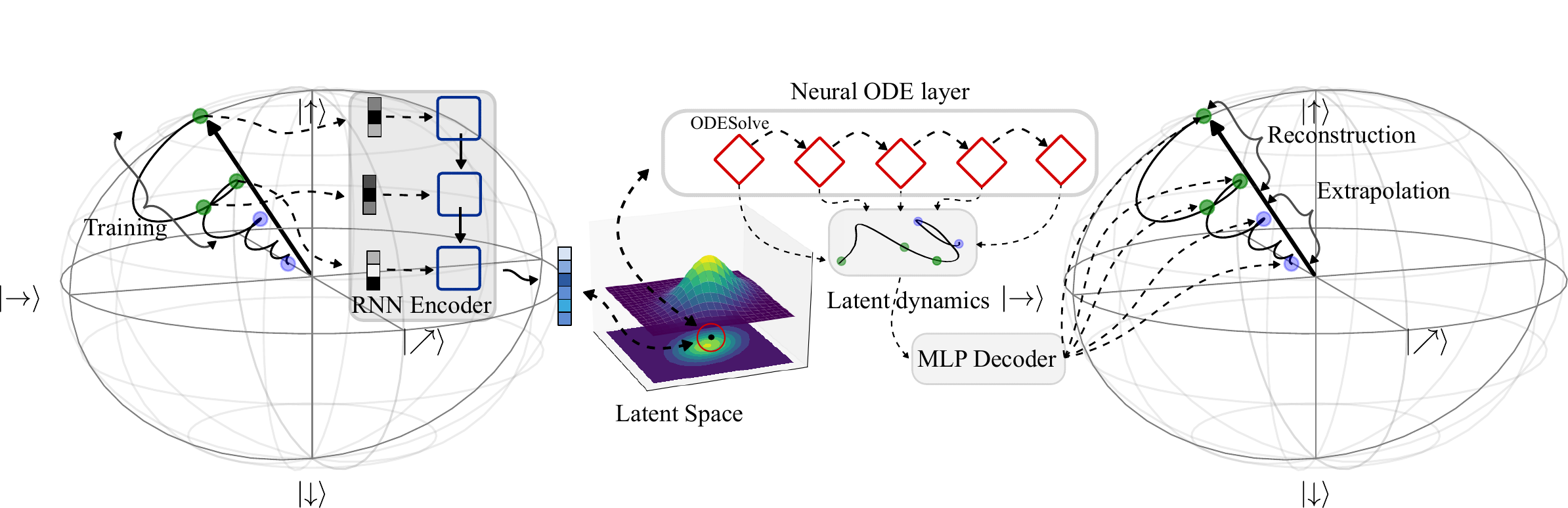}
(a) A visualization of the components of the QNODE 
\begin{tabular}{cc}
\rotatebox{90}{\hspace{0.35cm} \text{\footnotesize DATA}}  &
     \includegraphics[width=0.1\textwidth]{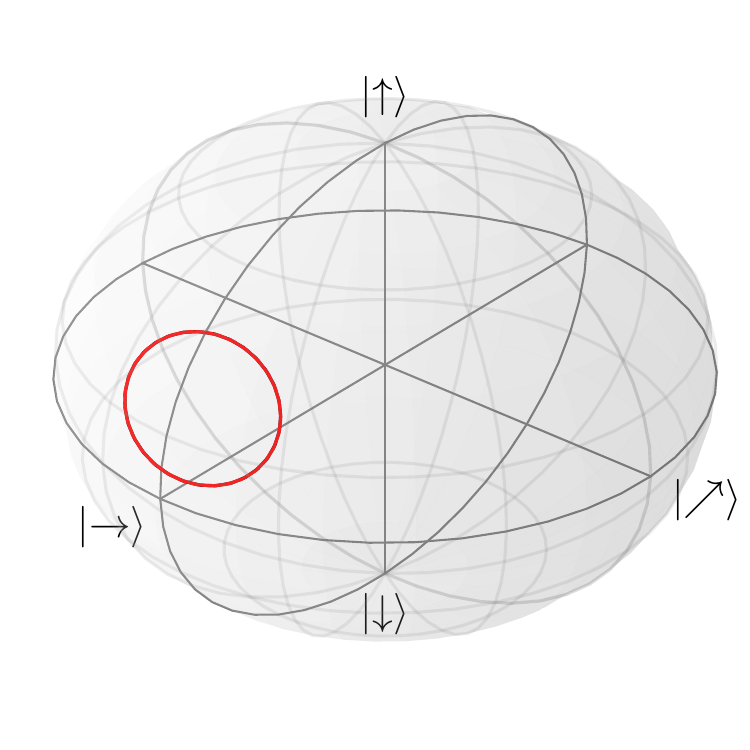}
     \includegraphics[width=0.1\textwidth]{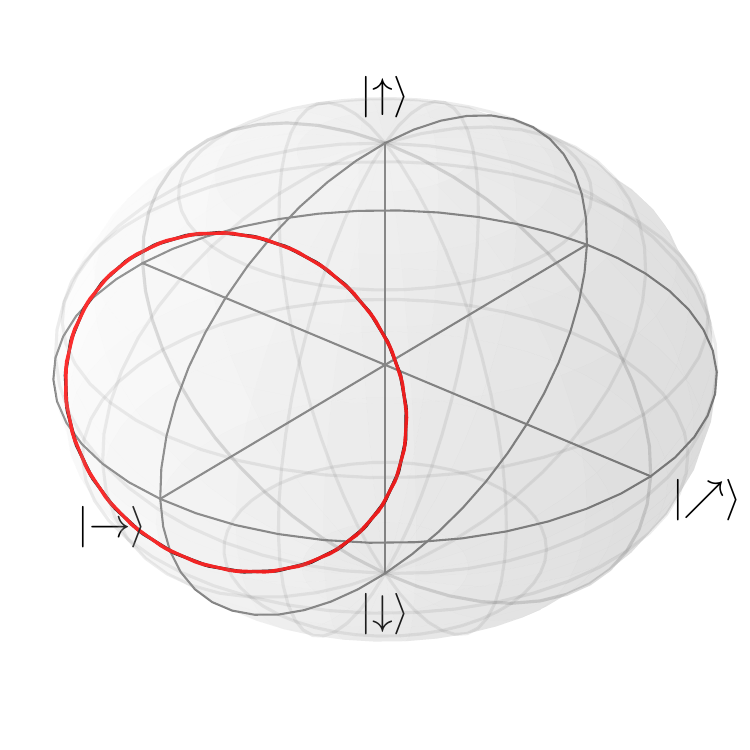}
     \includegraphics[width=0.1\textwidth]{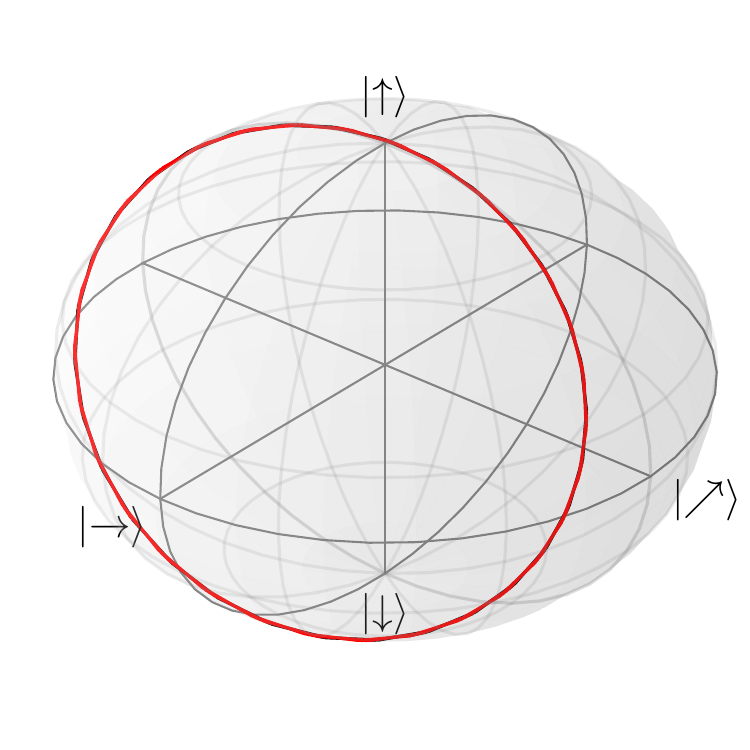}
     \includegraphics[width=0.1\textwidth]{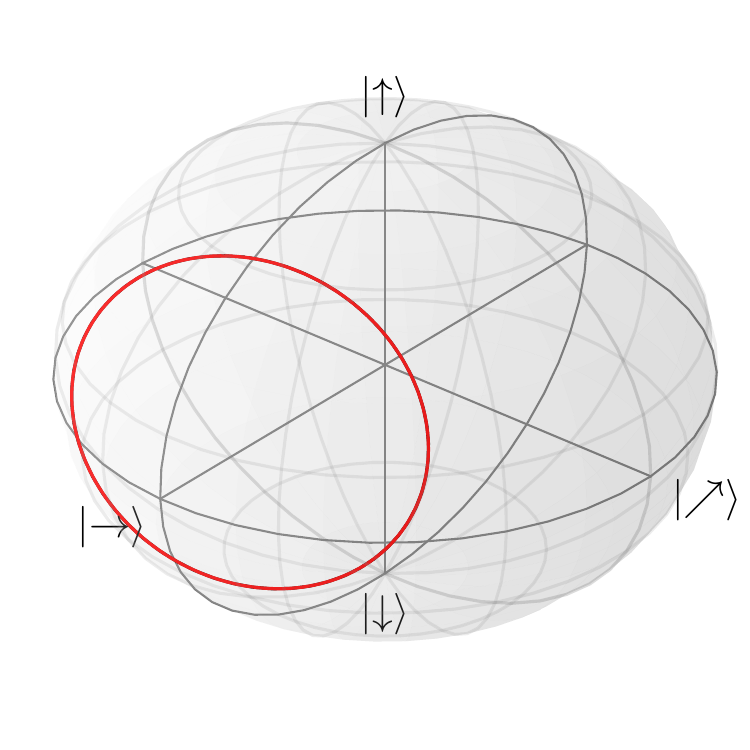}
     \includegraphics[width=0.1\textwidth]{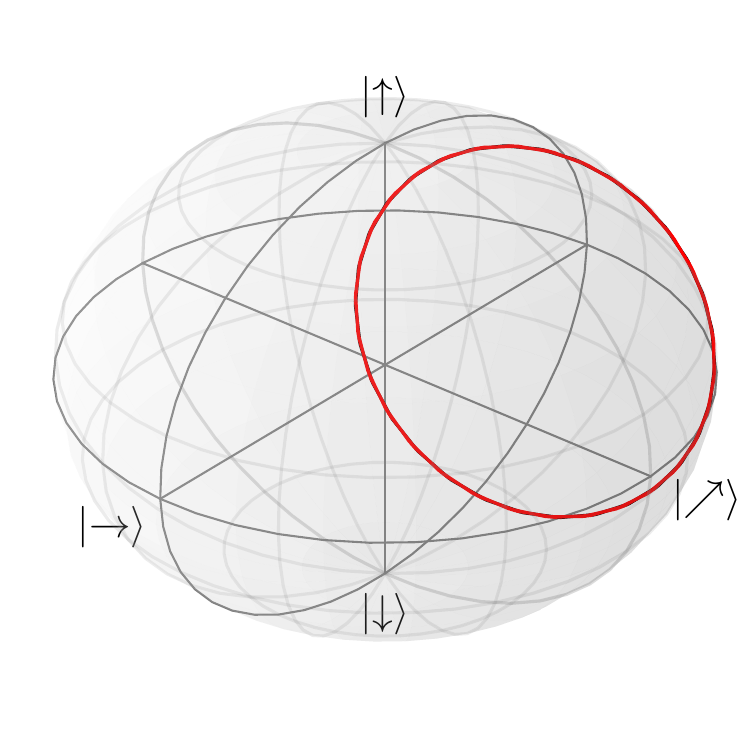}
     \includegraphics[width=0.1\textwidth]{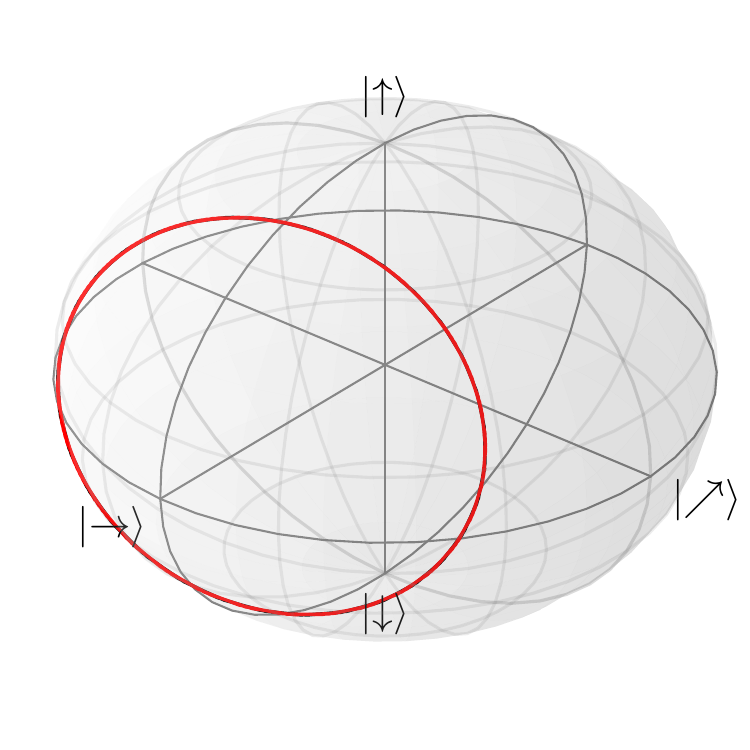}
     \includegraphics[width=0.1\textwidth]{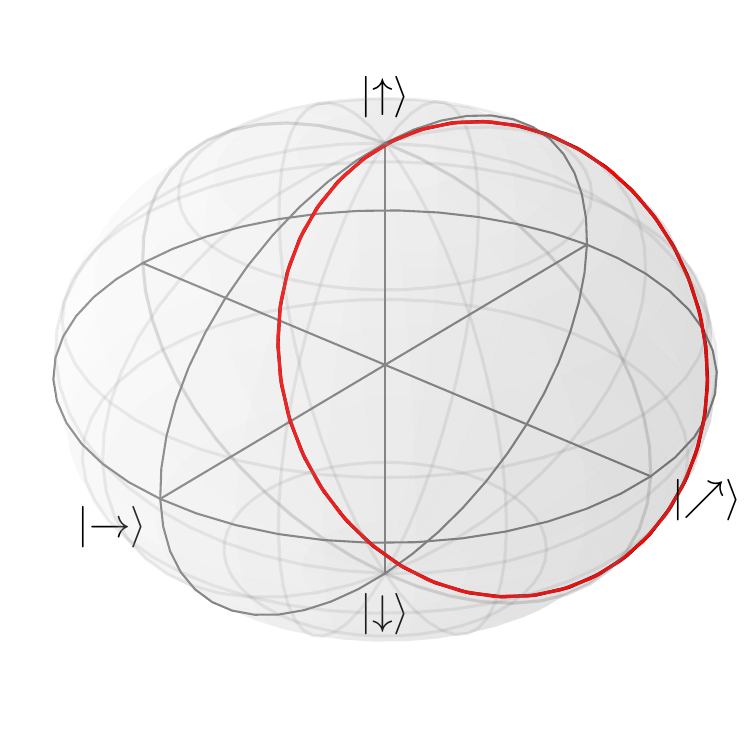}
     \includegraphics[width=0.1\textwidth]{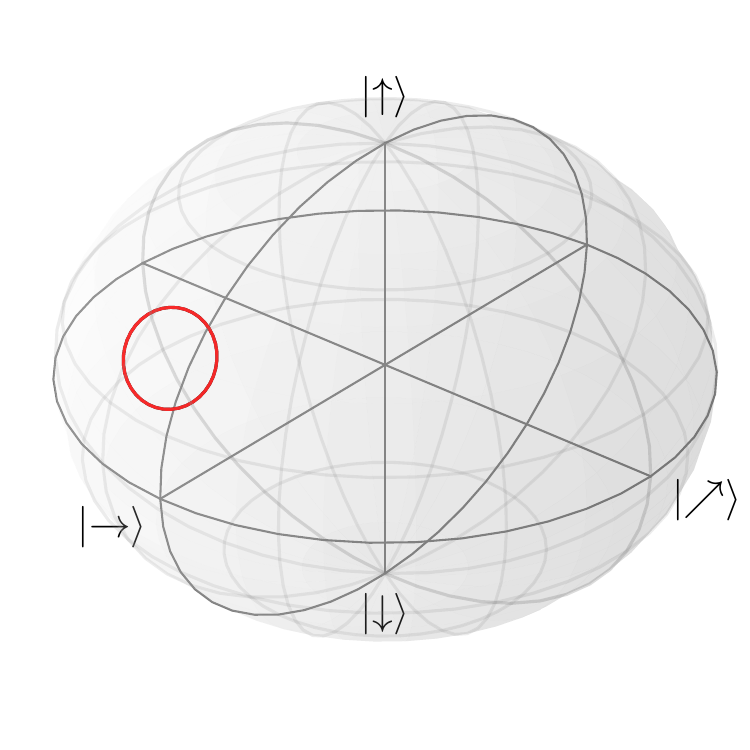}
     \includegraphics[width=0.1\textwidth]{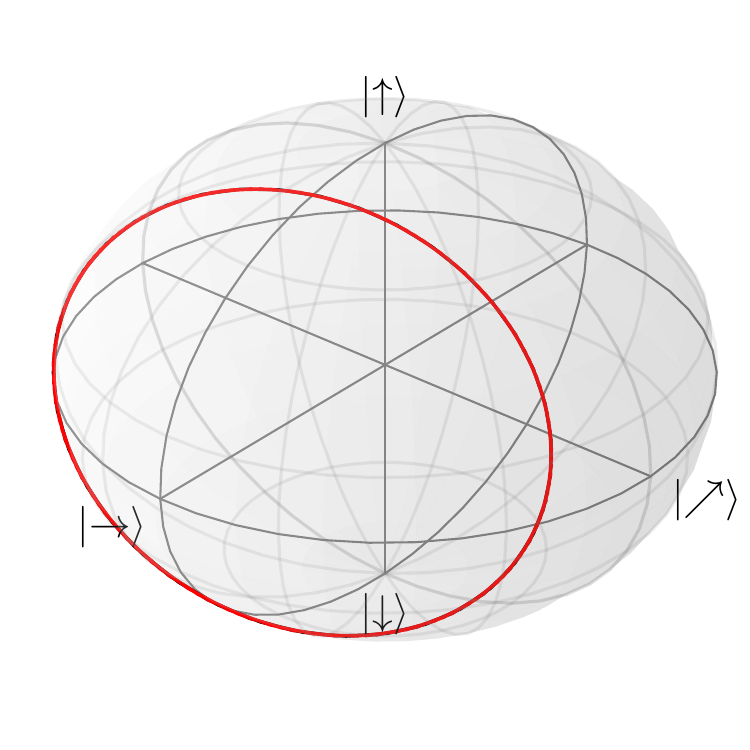}
\vspace{-0.25cm} \
\\ 
\rotatebox{90}{\hspace{-.1cm} $||\Psi||_{2}$}  &
     \includegraphics[width=0.1\textwidth]{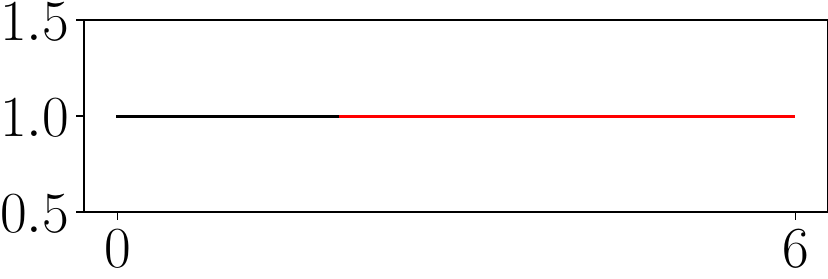}
     \includegraphics[width=0.1\textwidth]{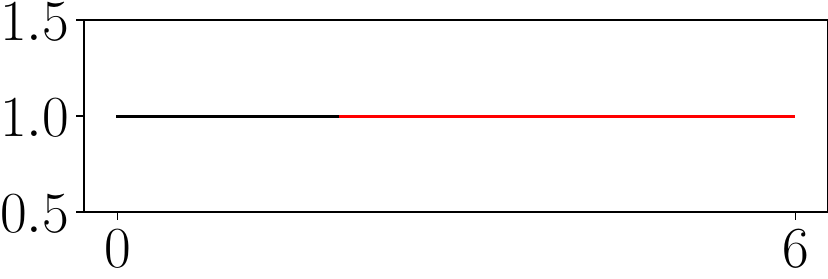}
     \includegraphics[width=0.1\textwidth]{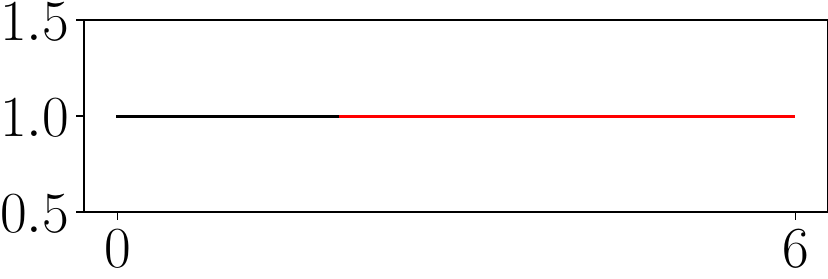}
     \includegraphics[width=0.1\textwidth]{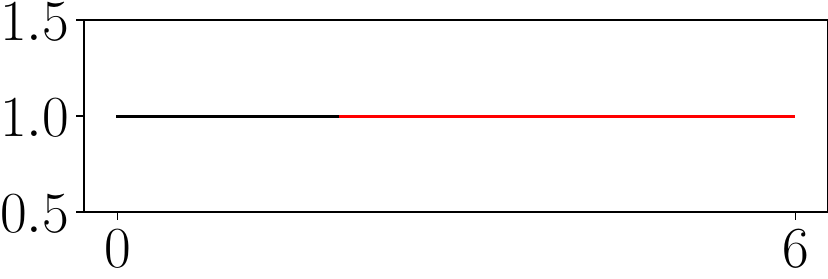}
     \includegraphics[width=0.1\textwidth]{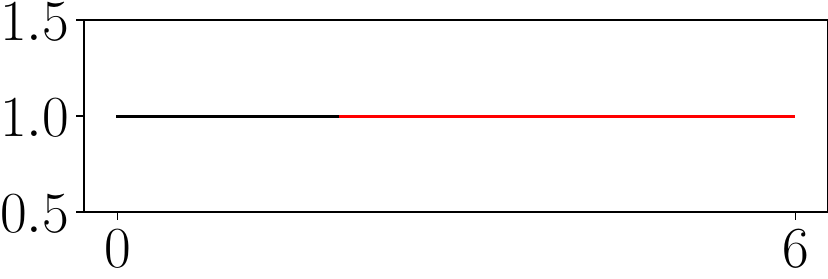}
     \includegraphics[width=0.1\textwidth]{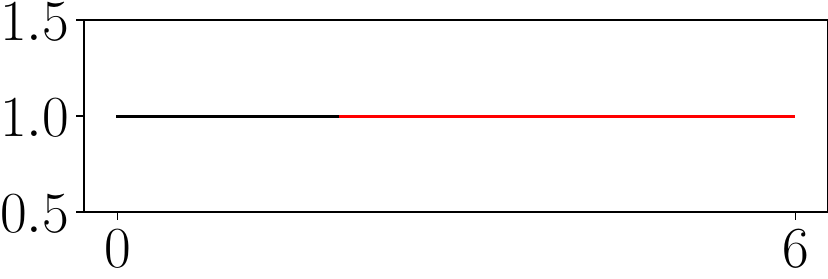}
     \includegraphics[width=0.1\textwidth]{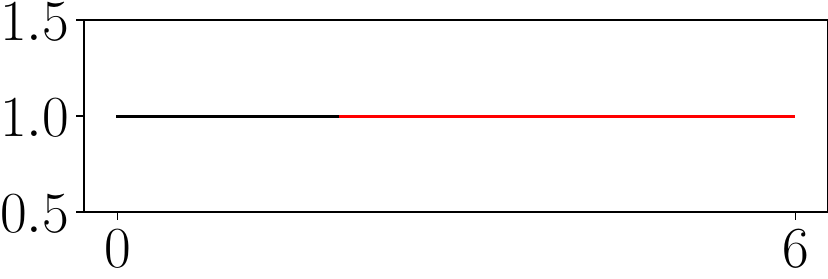}
     \includegraphics[width=0.1\textwidth]{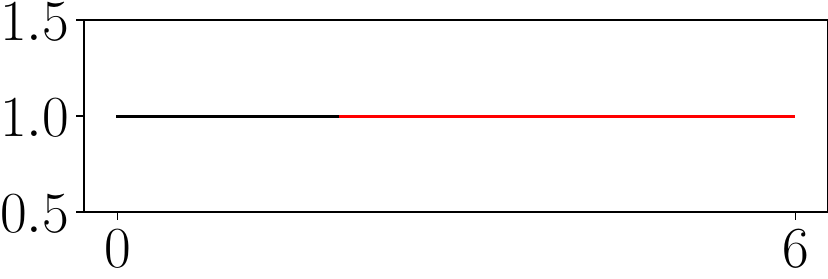}
     \includegraphics[width=0.1\textwidth]{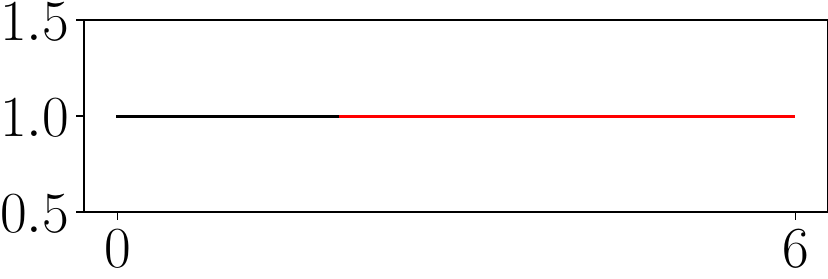}
 \\
\noalign{\smallskip} \hline \hline  \noalign{\smallskip}  \vspace{-0.25cm} \\ 
\rotatebox{90}{\hspace{0.25cm} \text{\footnotesize QNODE}}  &
     \includegraphics[width=0.1\textwidth]{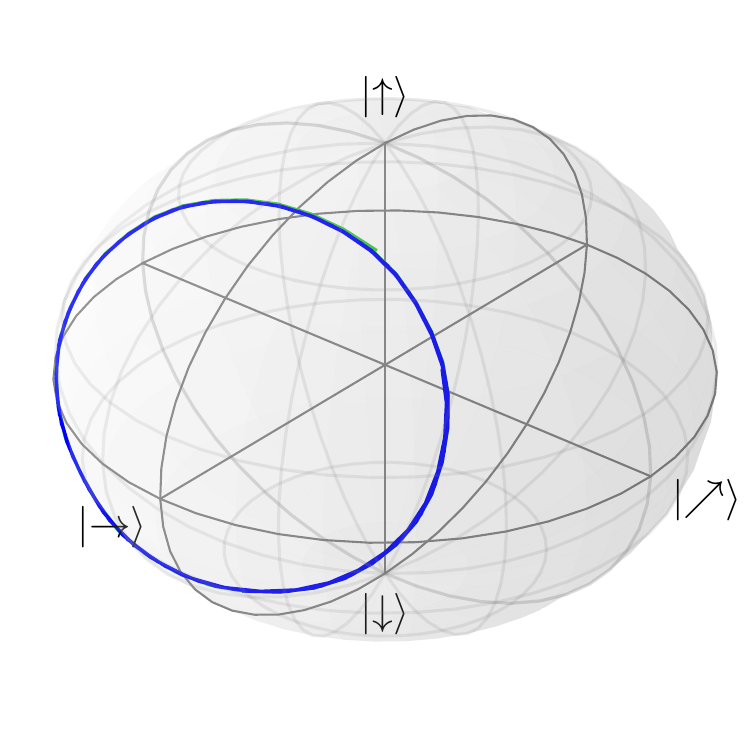}
     \includegraphics[width=0.1\textwidth]{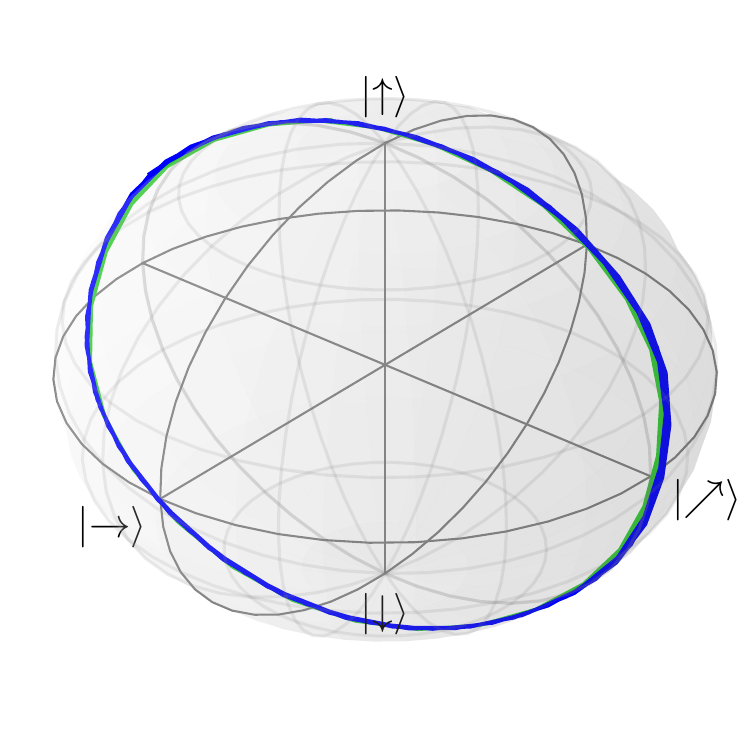}
     \includegraphics[width=0.1\textwidth]{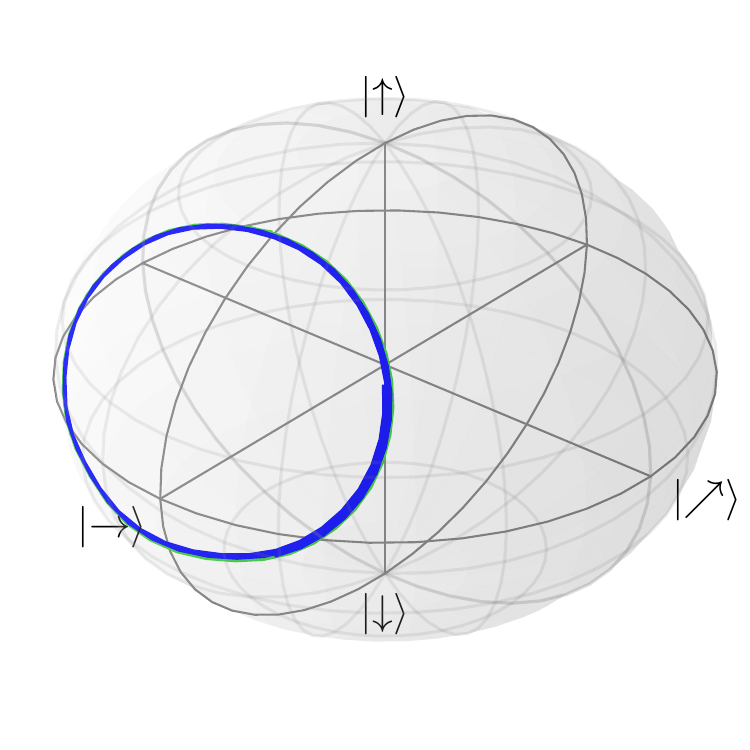}
     \includegraphics[width=0.1\textwidth]{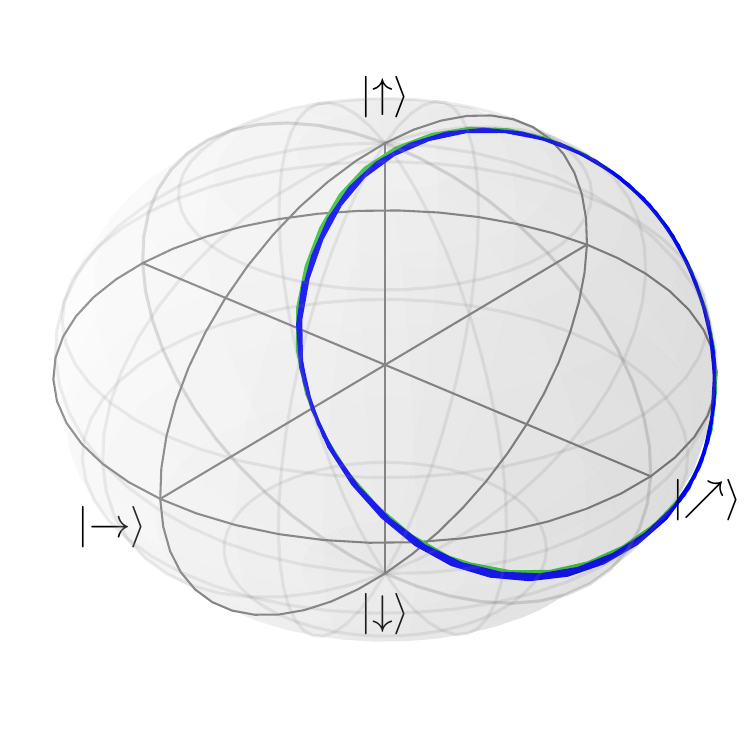}
     \includegraphics[width=0.1\textwidth]{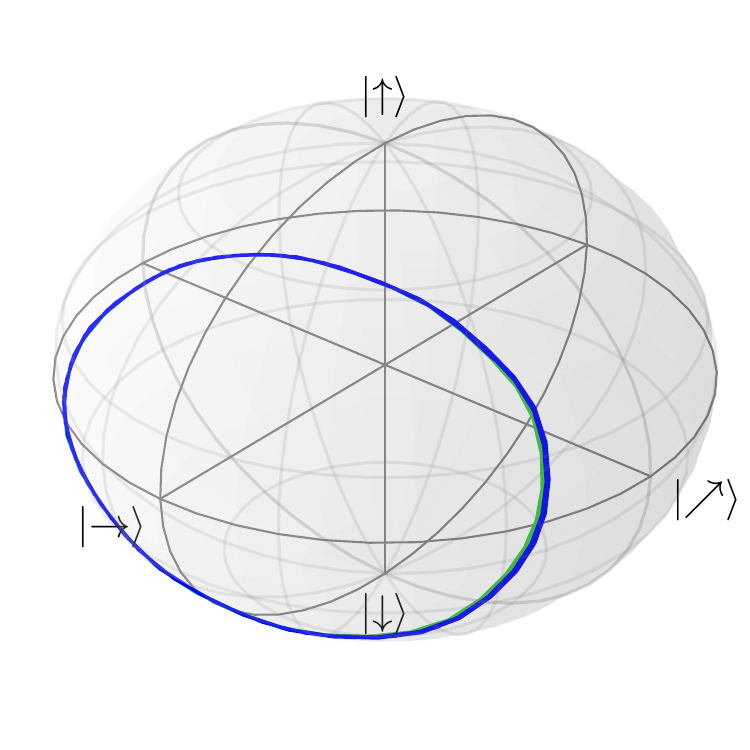}
     \includegraphics[width=0.1\textwidth]{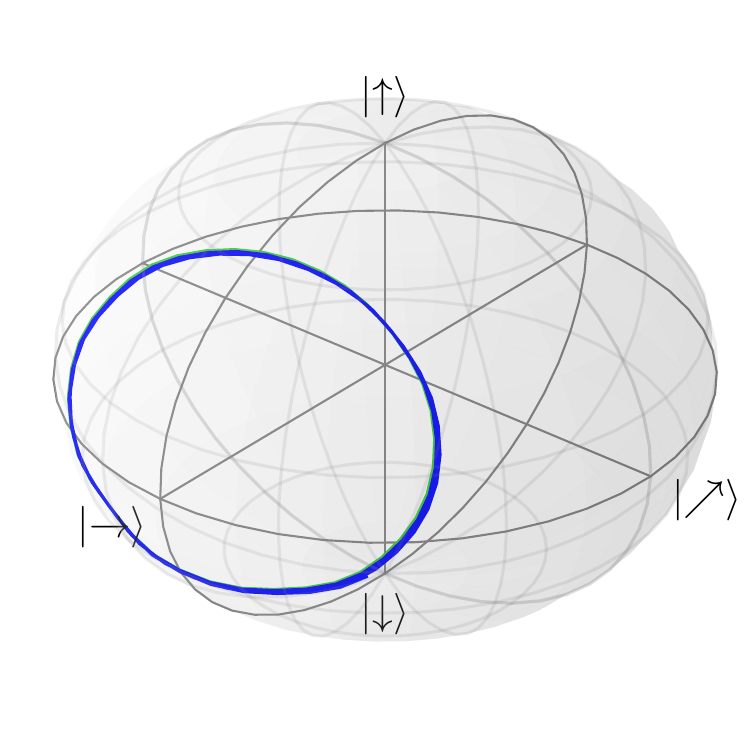}
     \includegraphics[width=0.1\textwidth]{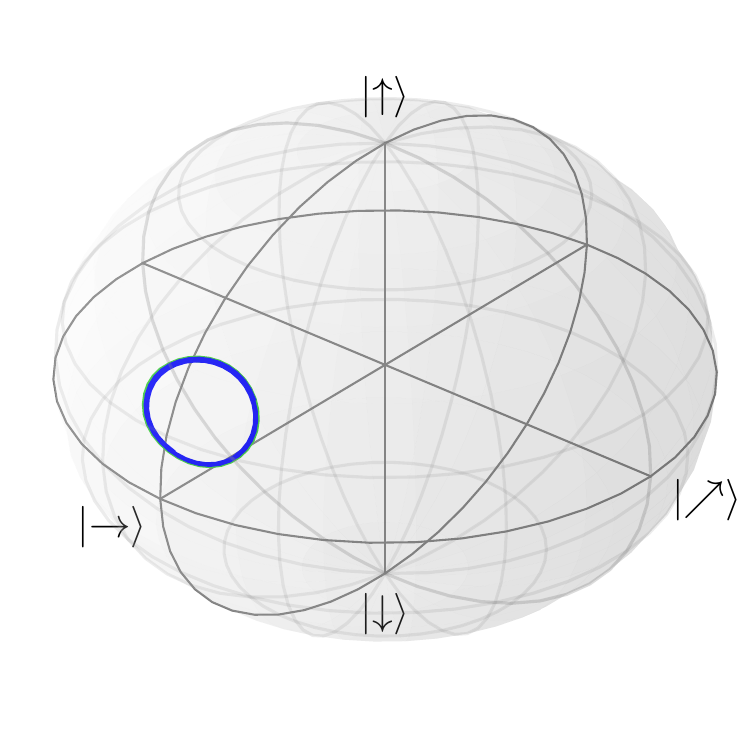}
     \includegraphics[width=0.1\textwidth]{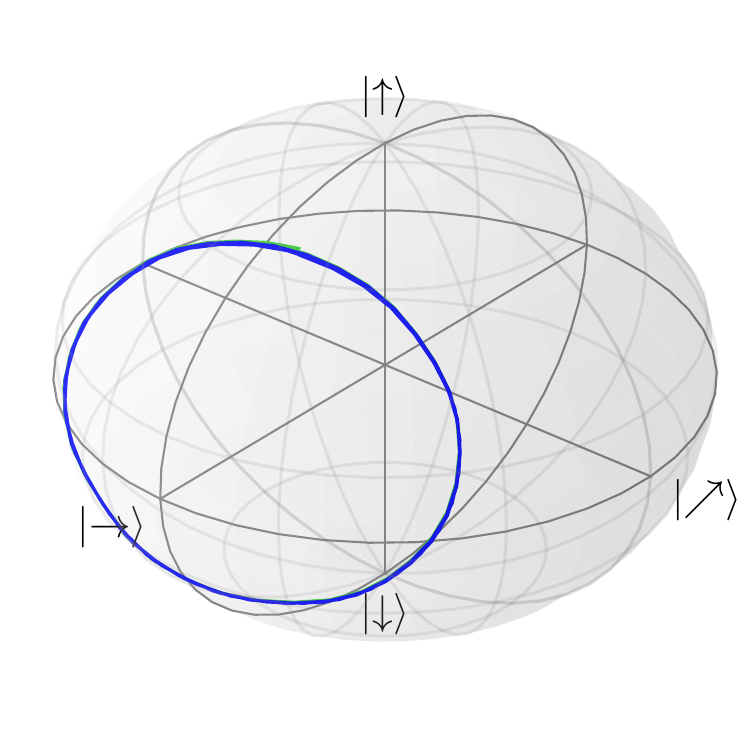}
     \includegraphics[width=0.1\textwidth]{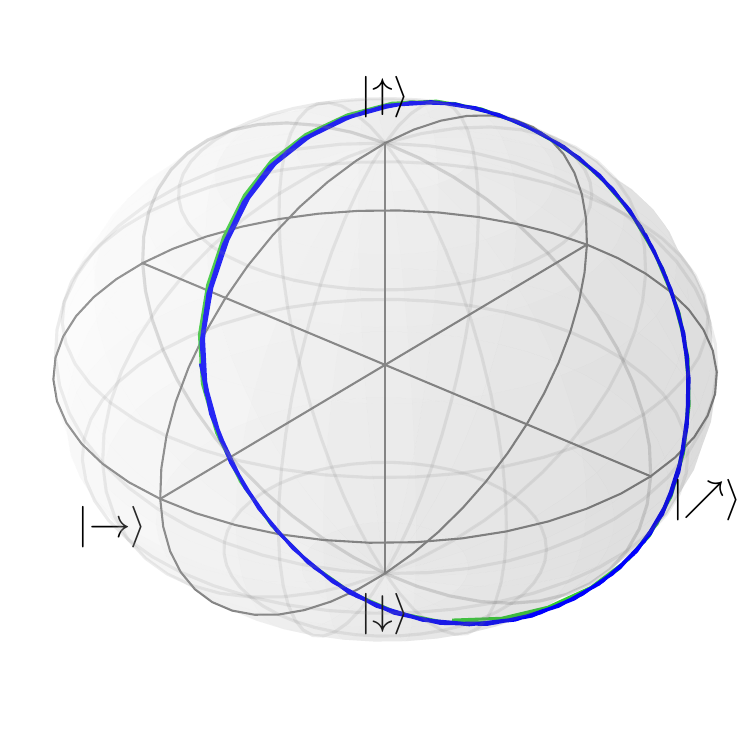}
\vspace{-0.25cm}
\\ 
\rotatebox{90}{\hspace{-.1cm} $||\Psi||_{2}$}  &
     \includegraphics[width=0.1\textwidth]{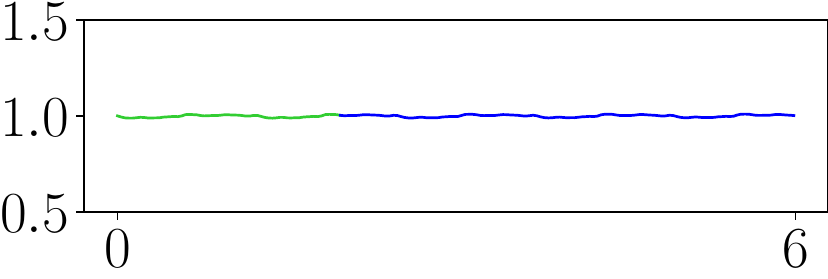}
     \includegraphics[width=0.1\textwidth]{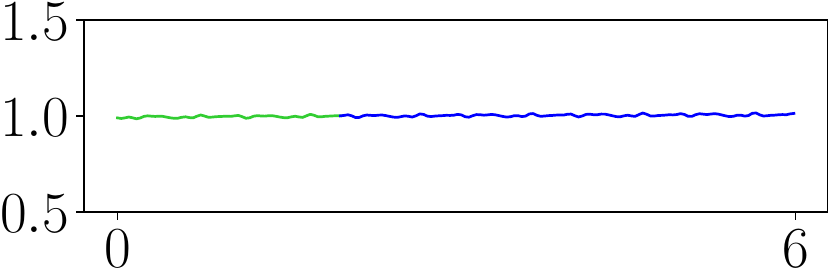}
     \includegraphics[width=0.1\textwidth]{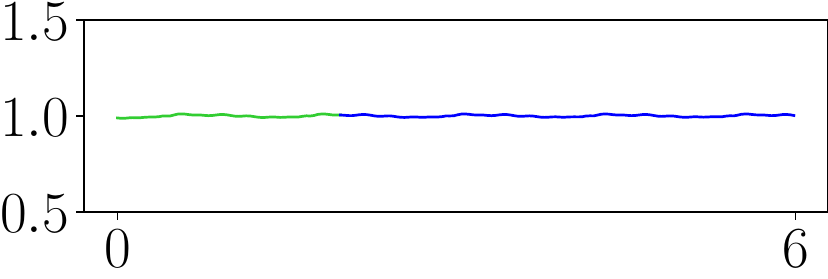}
     \includegraphics[width=0.1\textwidth]{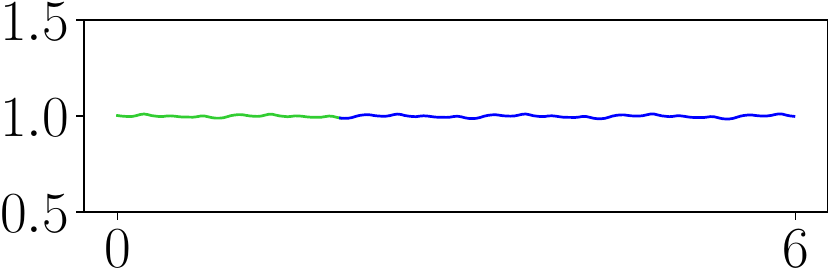}
     \includegraphics[width=0.1\textwidth]{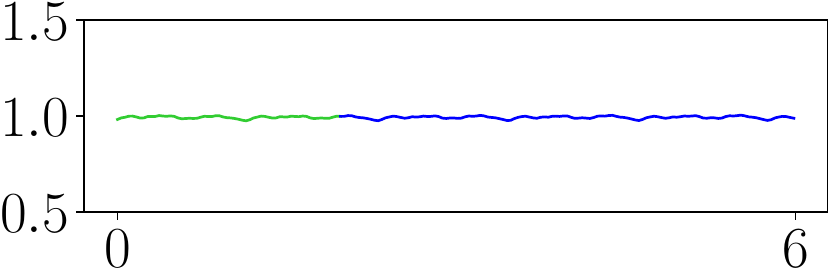}
     \includegraphics[width=0.1\textwidth]{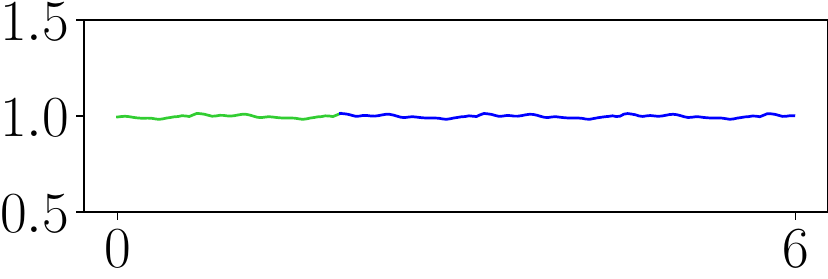}
     \includegraphics[width=0.1\textwidth]{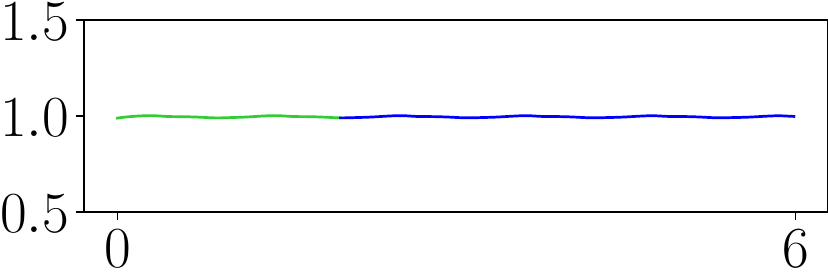}
     \includegraphics[width=0.1\textwidth]{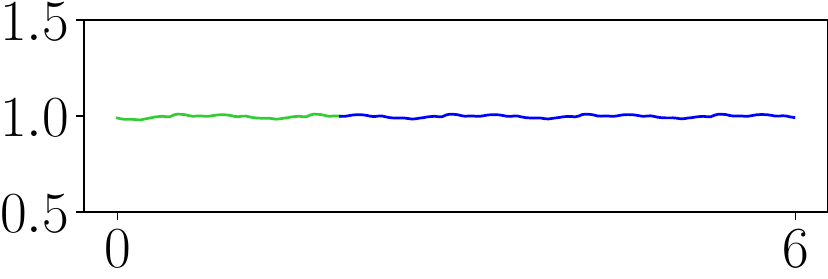}
     \includegraphics[width=0.1\textwidth]{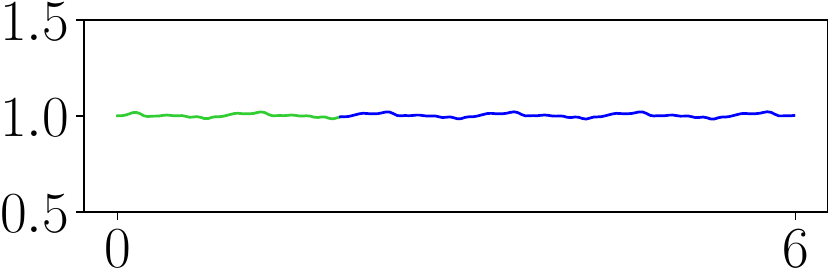} \\
\end{tabular}

(b) Closed quantum system dynamics generated from the QNODE and the training DATA

\caption{(a) \textbf{The Model}: visualizing QNODE's components. 
We first encode the dynamics of a qubit in the form of a time series of its observables.
In the case of a simple two-level quantum system, we are looking at the bloch vector evolution, 
which is mapped via a recurrent neural network to a latent representation of the dynamics. 
Then, a neural ODE layer outputs a latent representation of subsequent dynamics, which are mapped to quantum dynamics by an MLP decoder. 
QNODE can reconstruct the training dynamics and extrapolate the dynamics forward in time.
(b) \textbf{Closed quantum system dynamics}. A table comparing examples of closed system quantum dynamics from the training DATA (top row) with examples of closed system quantum dynamics generated by the QNODE (bottom row). Beneath each Bloch sphere is the time series plot of the norm $||\Psi||_2$ of the dynamics.
The green solid line is two arbitrary time units (\textbf{as}, i.e., when $\hbar=\omega=1$) of trained dynamics and blue is 4 \textbf{as} of extrapolated dynamics. Black and Red are the real quantum dynamics with black being the actual training region or two \textbf{as} and red being four \textbf{as} of dynamics unseen to the QNODE.
}
\label{fig:the_model}
\end{figure*}

\begin{figure*}[t]
\begin{tabular}{cc}
\rotatebox{90}{\hspace{0.35cm} \text{\footnotesize DATA}}  &
     \includegraphics[width=0.1\textwidth]{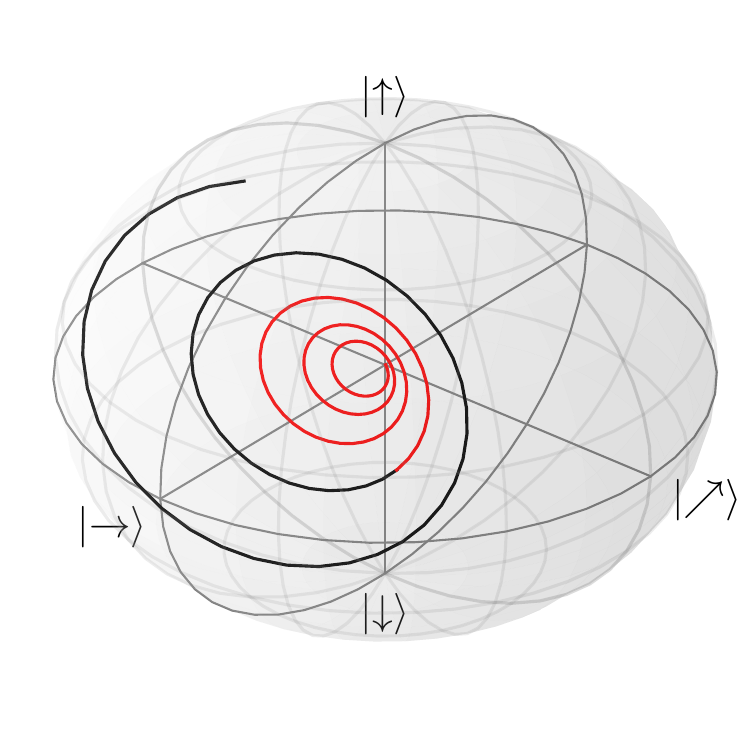}
    \includegraphics[width=0.1\textwidth]{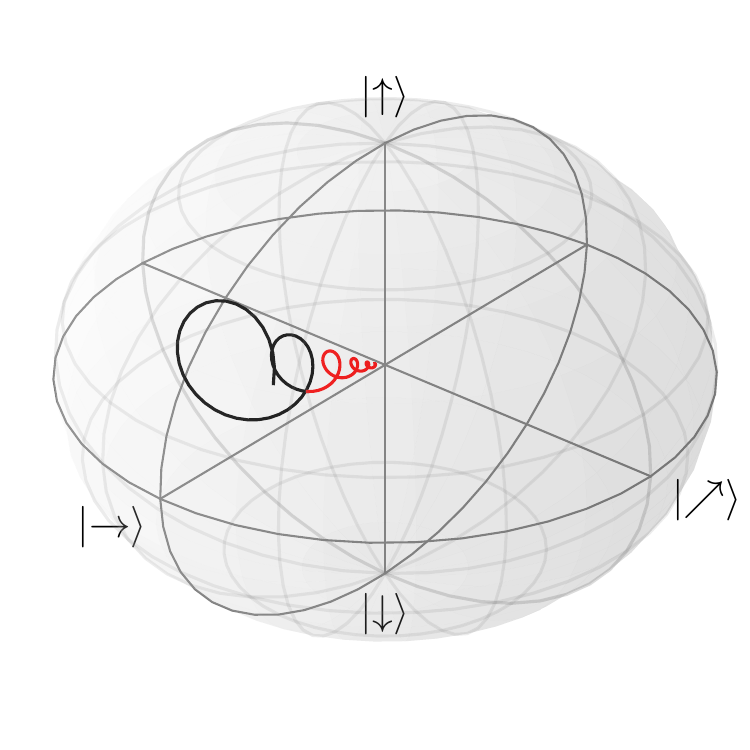}
    \includegraphics[width=0.1\textwidth]{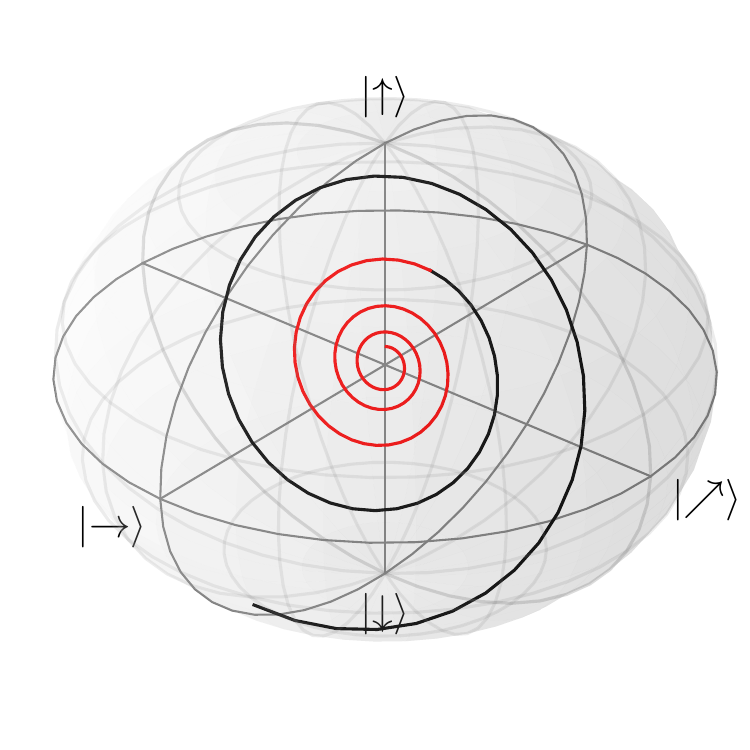}
    \includegraphics[width=0.1\textwidth]{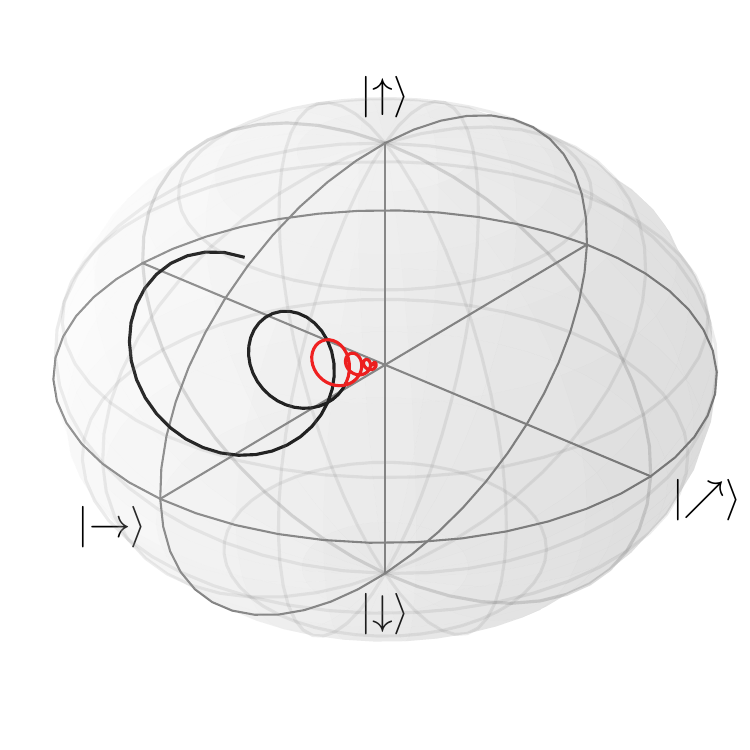}
    \includegraphics[width=0.1\textwidth]{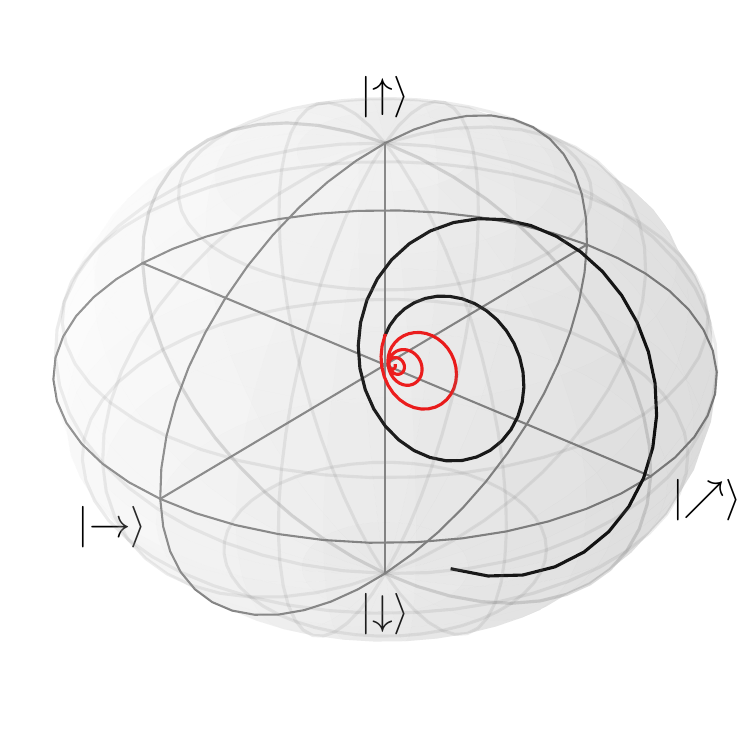}
    \includegraphics[width=0.1\textwidth]{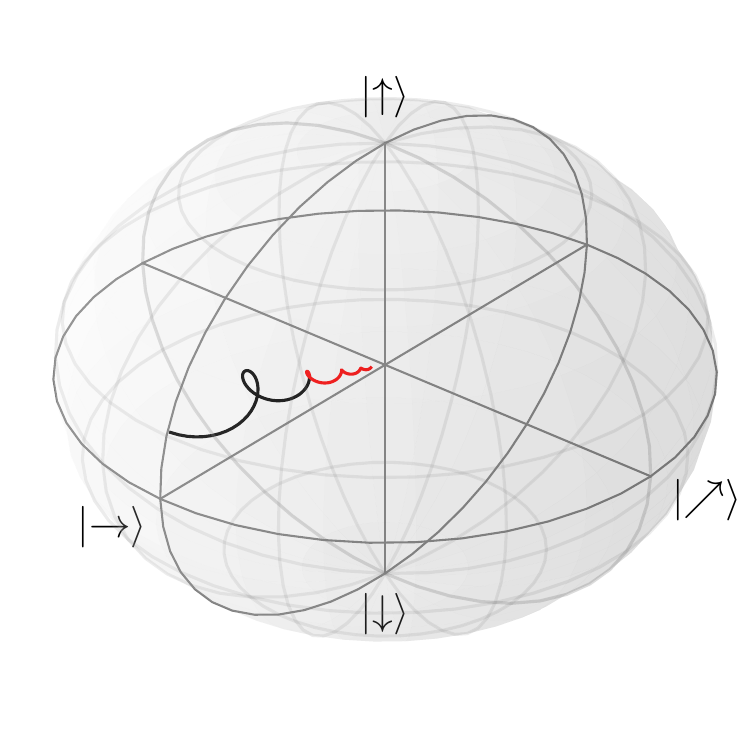}
    \includegraphics[width=0.1\textwidth]{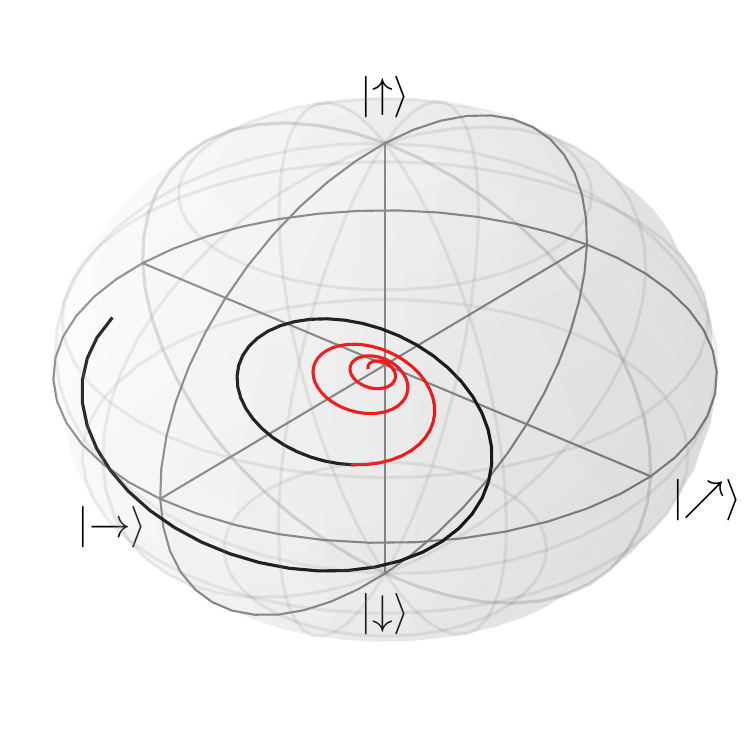}
    \includegraphics[width=0.1\textwidth]{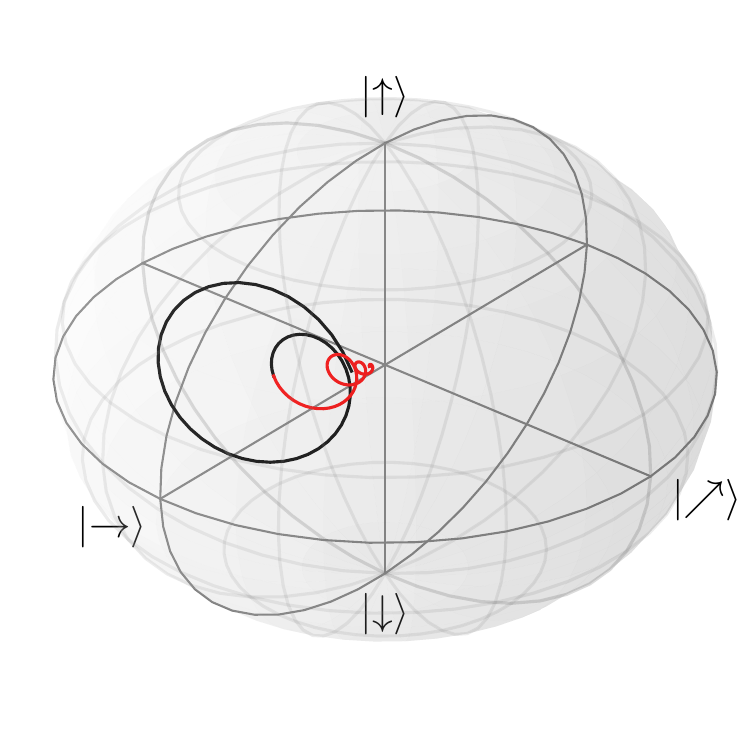}
    \includegraphics[width=0.1\textwidth]{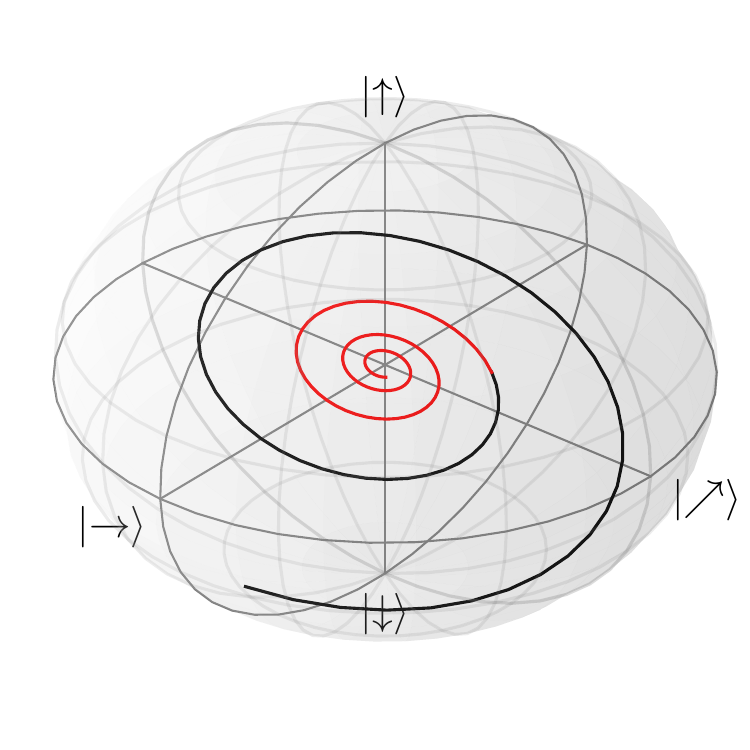}
\vspace{-0.25cm} \
\\ 
\rotatebox{90}{\hspace{-.1cm} $||\Psi||_{2}$}  &
     \includegraphics[width=0.1\textwidth]{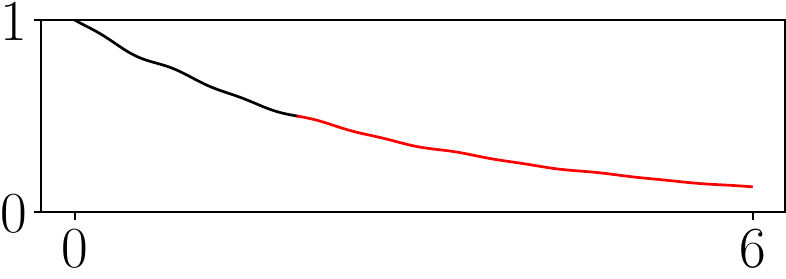}
    \includegraphics[width=0.1\textwidth]{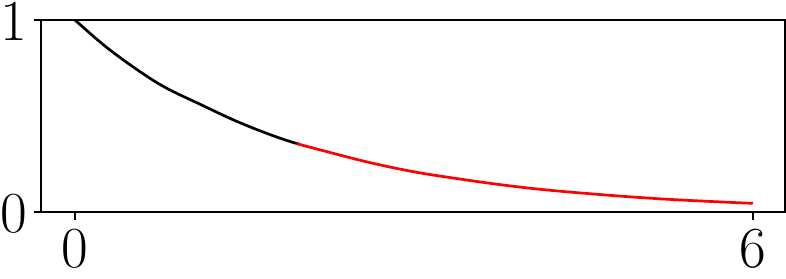}
    \includegraphics[width=0.1\textwidth]{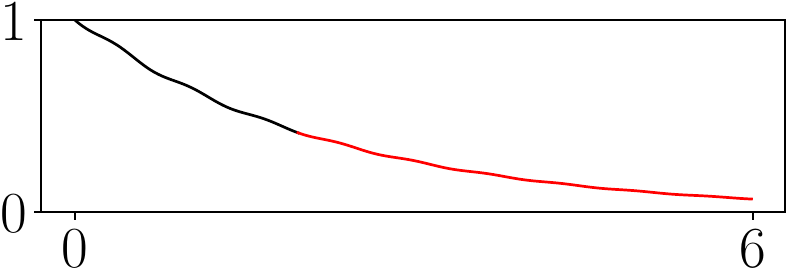}
    \includegraphics[width=0.1\textwidth]{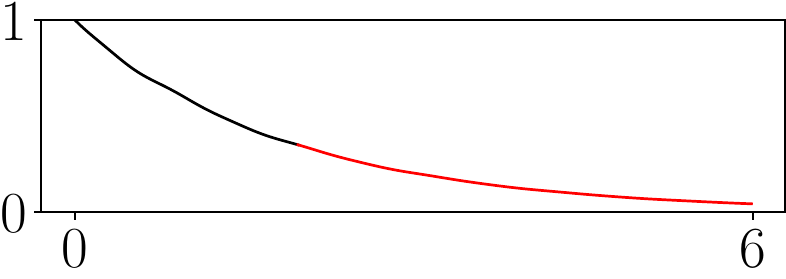}
    \includegraphics[width=0.1\textwidth]{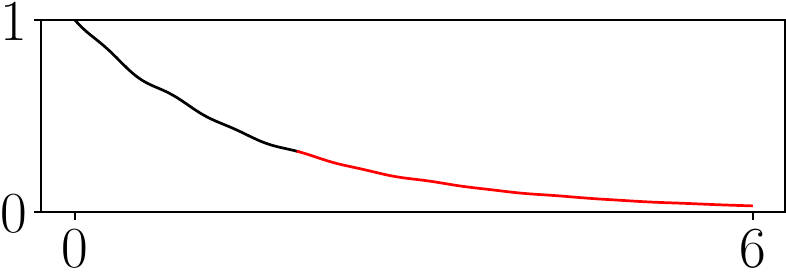}
    \includegraphics[width=0.1\textwidth]{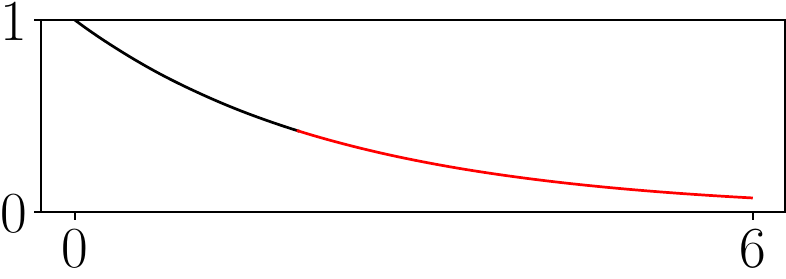}
    \includegraphics[width=0.1\textwidth]{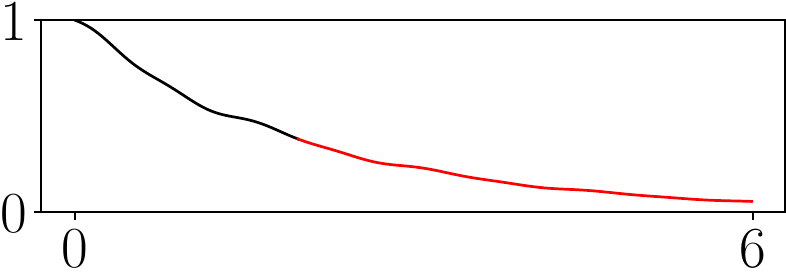}
    \includegraphics[width=0.1\textwidth]{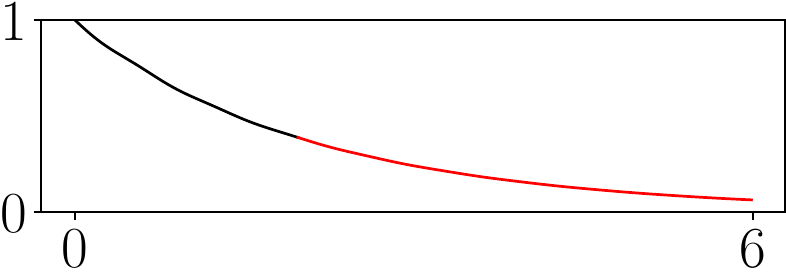}
    \includegraphics[width=0.1\textwidth]{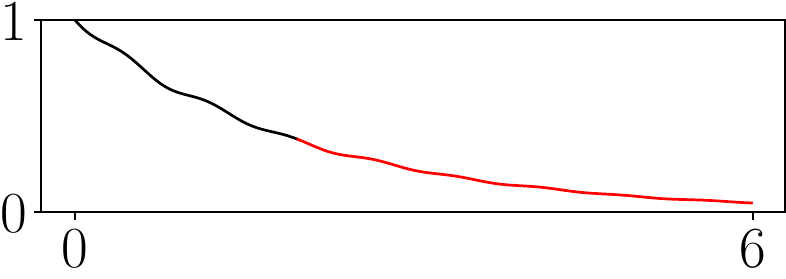}
 \\
\noalign{\smallskip} \hline  \noalign{\smallskip}  \vspace{-0.35cm} \\ 
\rotatebox{90}{\hspace{0.25cm} \text{\footnotesize QNODE}}  &
     \includegraphics[width=0.1\textwidth]{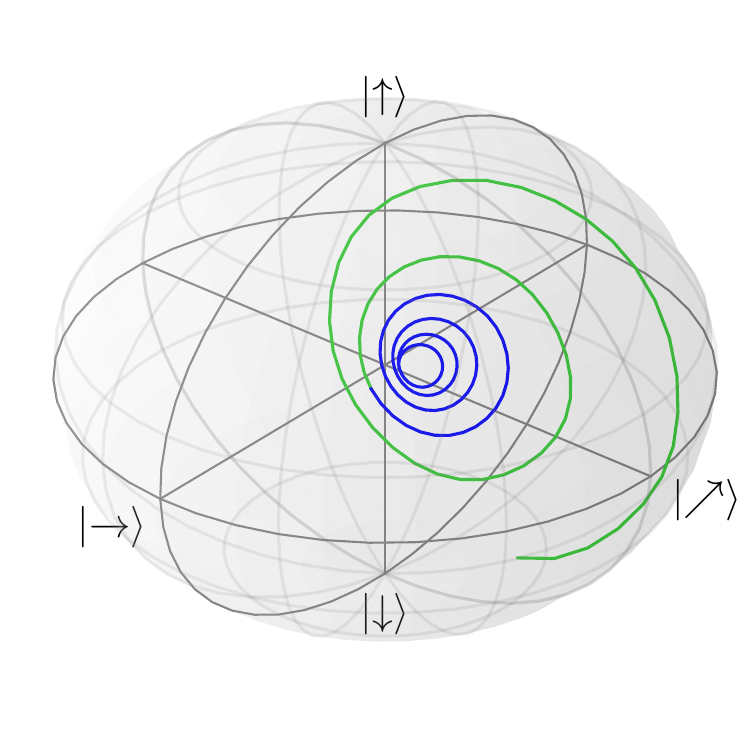}
     \includegraphics[width=0.1\textwidth]{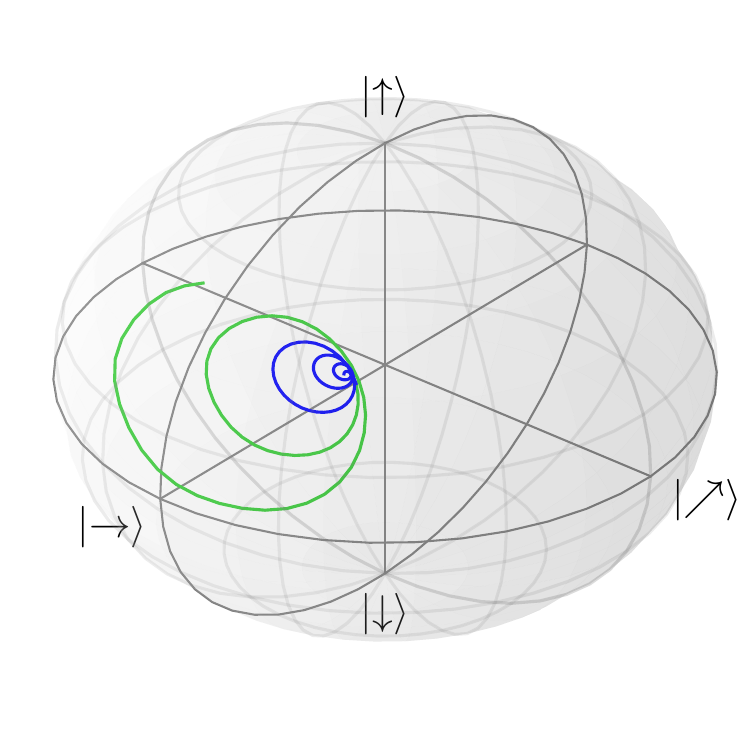}
     \includegraphics[width=0.1\textwidth]{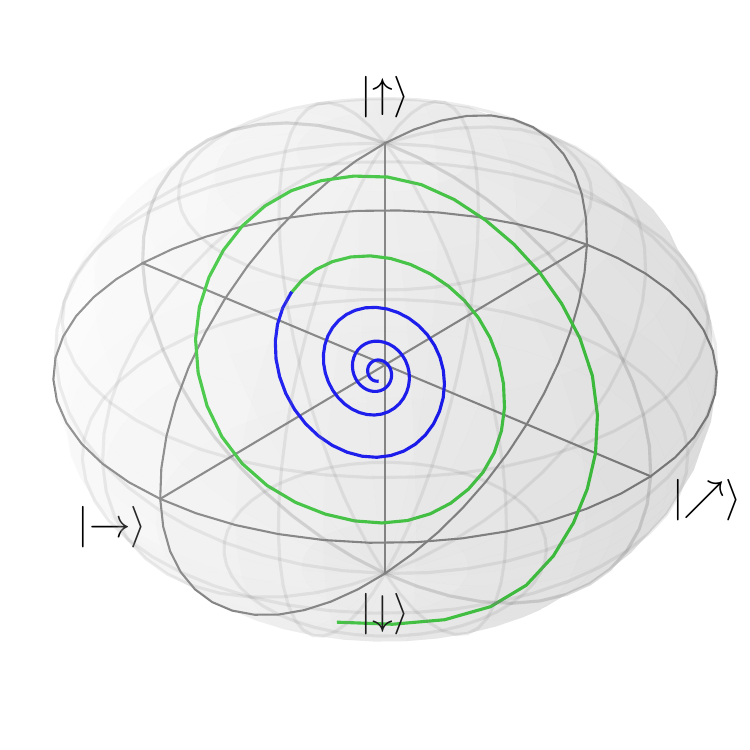}
     \includegraphics[width=0.1\textwidth]{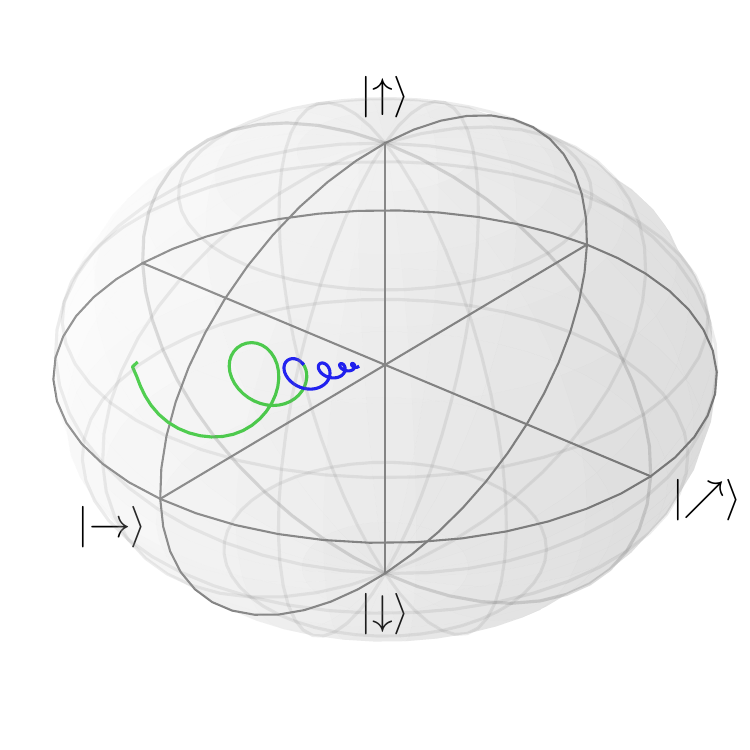}
     \includegraphics[width=0.1\textwidth]{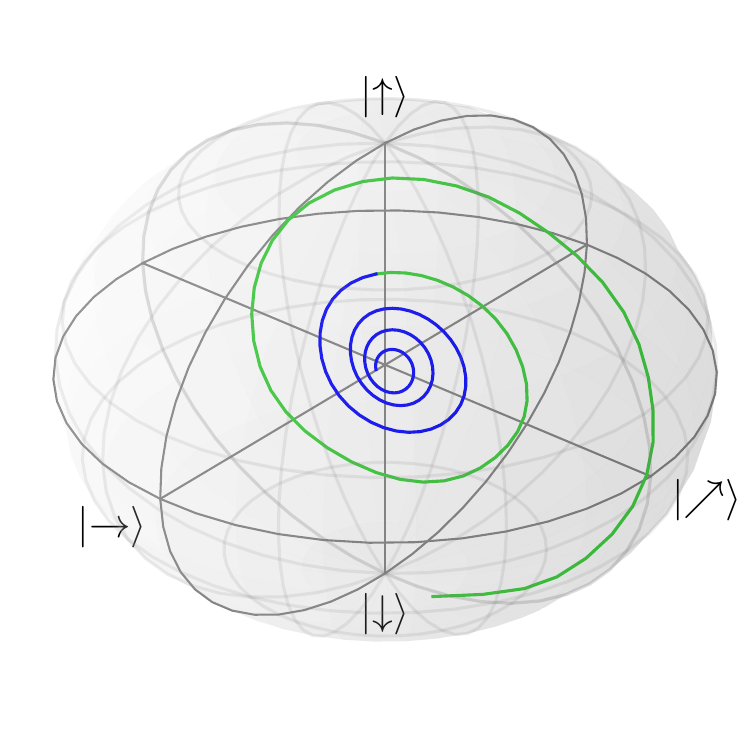}
     \includegraphics[width=0.1\textwidth]{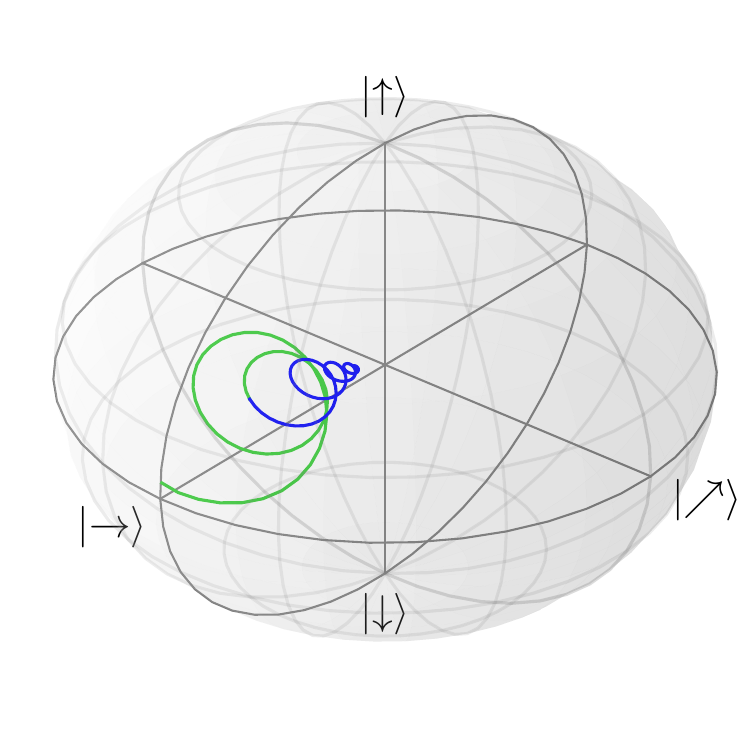}
     \includegraphics[width=0.1\textwidth]{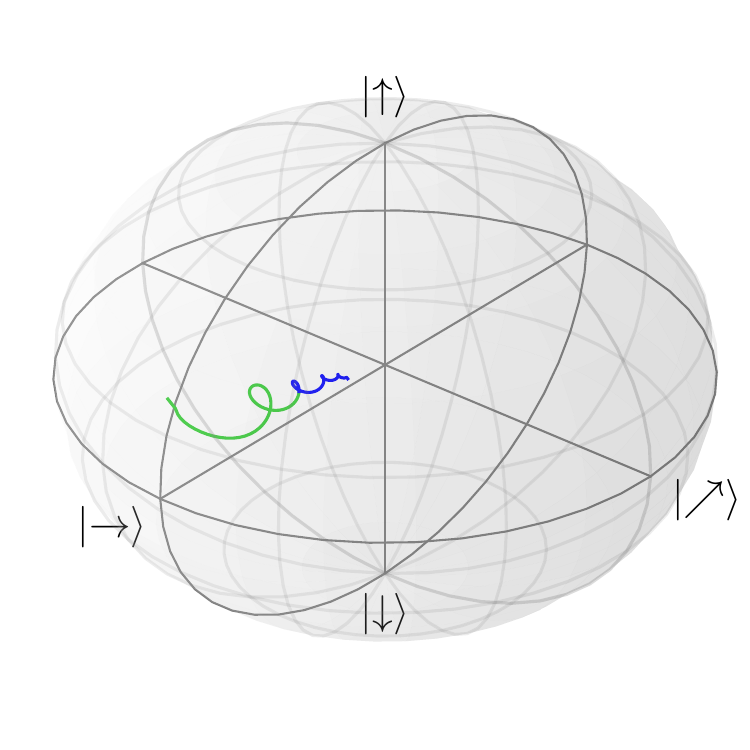}
     \includegraphics[width=0.1\textwidth]{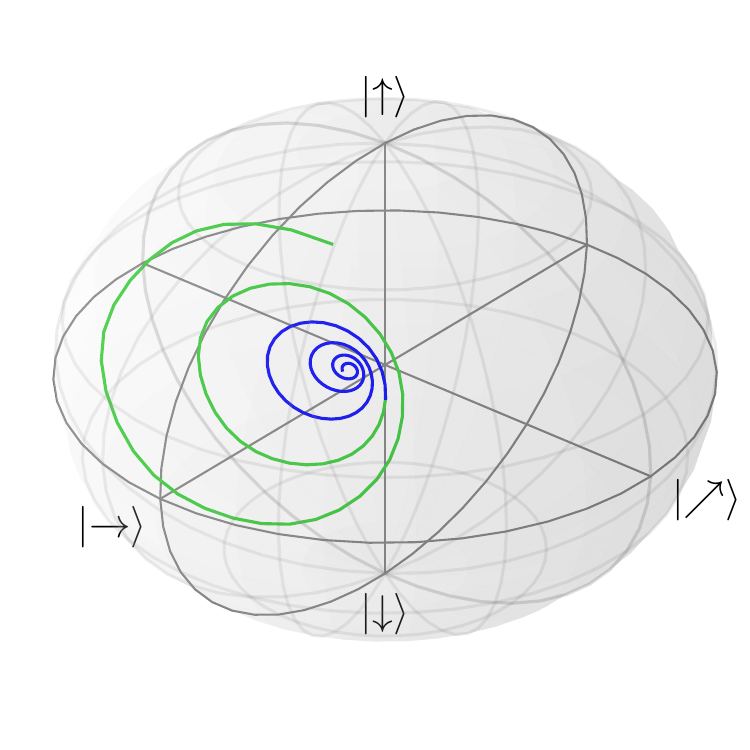}
     \includegraphics[width=0.1\textwidth]{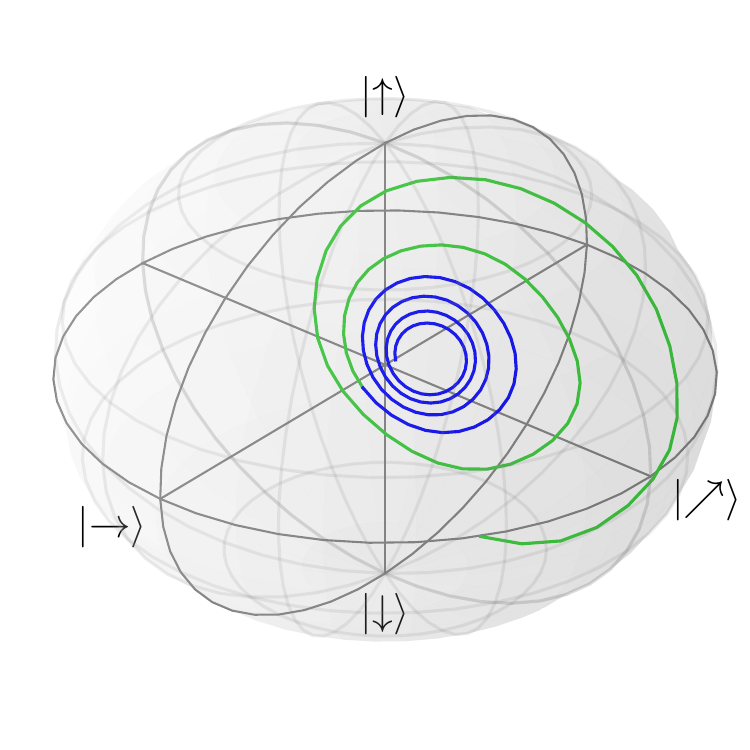}
\vspace{-0.25cm}
\\ 
\rotatebox{90}{\hspace{-.1cm} $||\Psi||_{2}$}  &
     \includegraphics[width=0.1\textwidth]{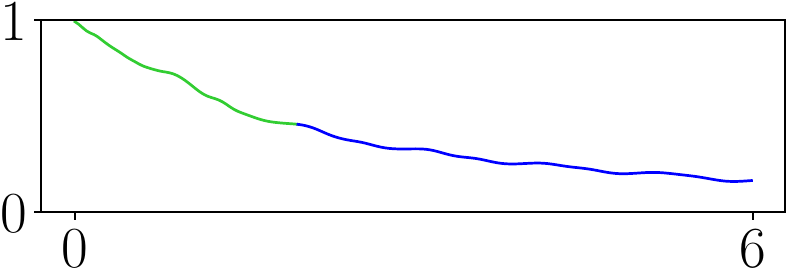}
     \includegraphics[width=0.1\textwidth]{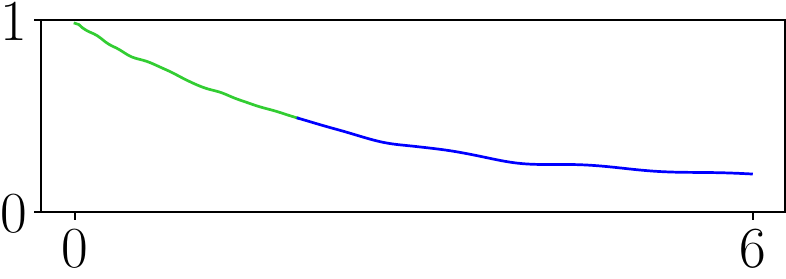}
     \includegraphics[width=0.1\textwidth]{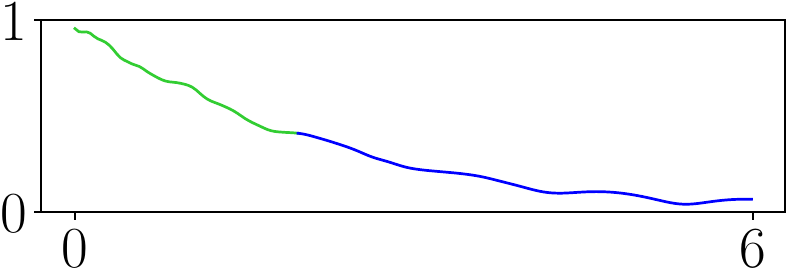}
     \includegraphics[width=0.1\textwidth]{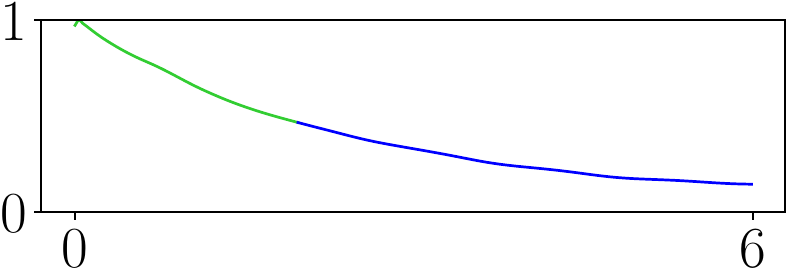}
     \includegraphics[width=0.1\textwidth]{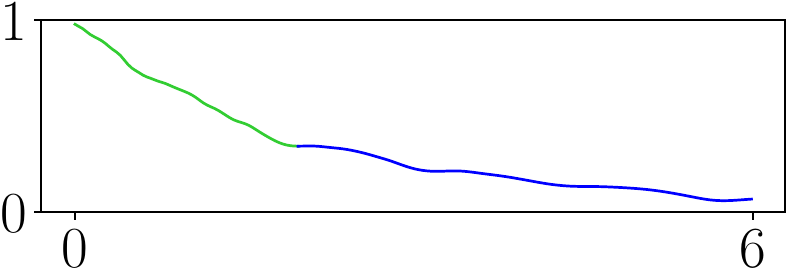}
     \includegraphics[width=0.1\textwidth]{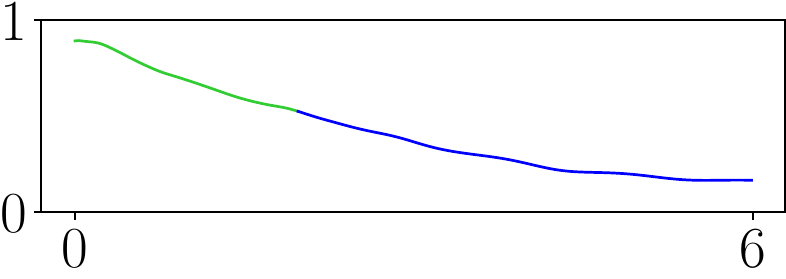}
     \includegraphics[width=0.1\textwidth]{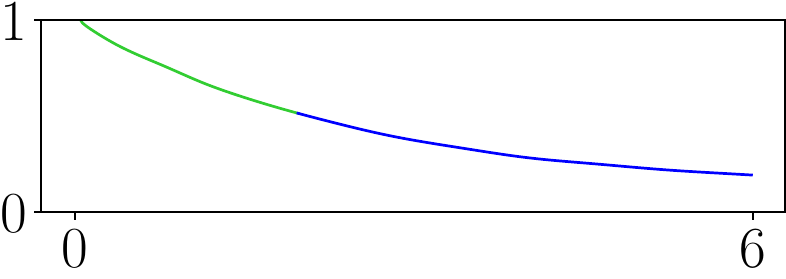}
     \includegraphics[width=0.1\textwidth]{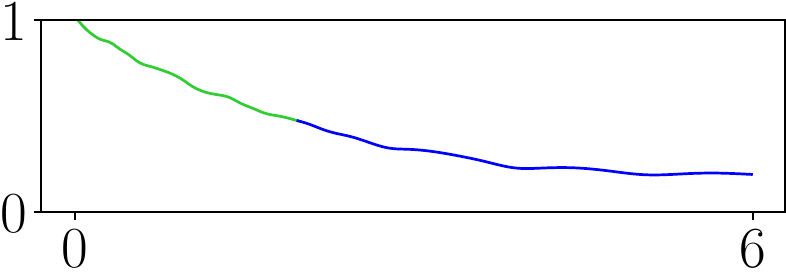}
     \includegraphics[width=0.1\textwidth]{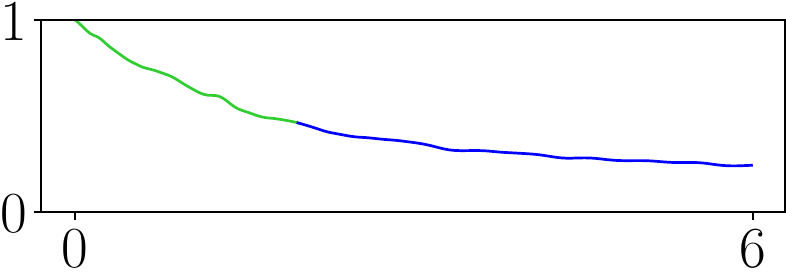}
\end{tabular}

(a) Generated quantum dynamics from QNODE compared with training dynamics on the open quantum system

\includegraphics[width=0.4\textwidth]{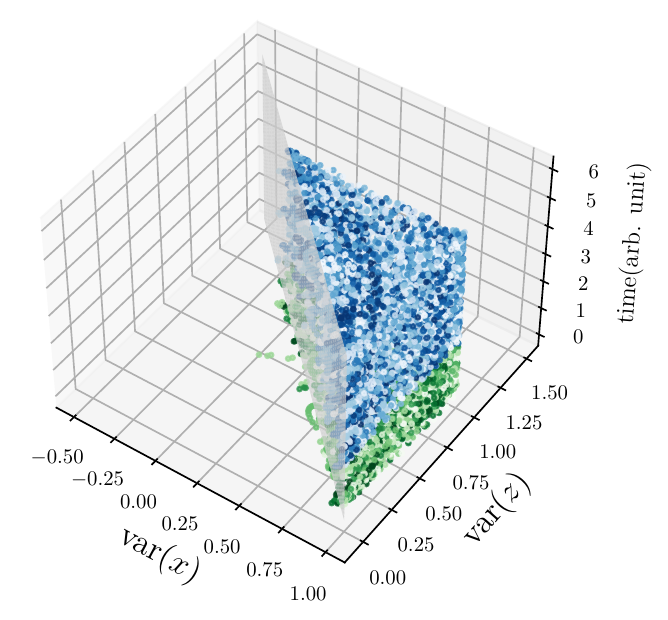} \hspace{1cm}
\includegraphics[width=0.4\textwidth]{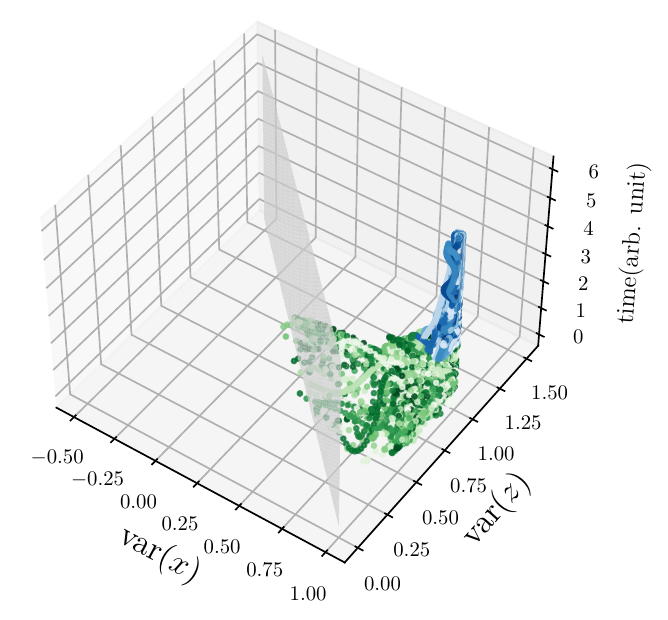}

(b) HUP Experiment for closed system \hspace{28mm}
(c) HUP Experiment for open system

\caption{\textbf{Open system dynamics}. (a) A table comparing examples of open system quantum dynamics that the QNODE is trained on (top row) with examples of closed system quantum dynamics generated by the QNODE (bottom row). Beneath each Bloch sphere is the time series plot of the norm $||\Psi||_2$ of the dynamics.
The green solid line is two \textbf{as} of trained dynamics and blue is 4 \textbf{as} of extrapolated dynamics. Black and red lines are the real quantum dynamics with black being the actual training region or two \textbf{as} and red being four \textbf{as} of dynamics unseen to the QNODE.
\textbf{Discovering the uncertainty principle}. (b) and (c) the variance plotted through time of generated quantum dynamics produced by the QNODE as well as plane separating dynamics that satisfy the uncertainty principle from those that do not.}
\label{fig:gen_hup}
\end{figure*}

\section{Preliminaries}
\subsection{QNODE model}
We re-purpose the latent variable neural ordinary differential equation (ODE) model \cite{chen2018neural} 
which is a generative latent function time-series model. 
The model is trained as a variational autoencoder \cite{kingma2014autoencoding, rezende2014stochastic}, 
on the quantum dynamics time series  
and in this context refer to the model as the Quantum dynamics latent Neural ODE (QNODE). We visualize each component of the Model in \figref{fig:the_model}.  
The encoder is a RNN, which takes in the training quantum dynamics $\B x ^{\mathcal O} (t_1) , \dots, \B x ^{\mathcal O} (t_n) $ 
sequentially backwards in time and outputs the parameters of the distribution on the latent representation $\B h ^{\mathcal O} (t_0) $ for some initial $t_0$.
Using a standard Gaussian prior on the latents, the dynamics $\B x ^{\mathcal O} (t) $ are modeled with a Gaussian likelihood.

The subsequent learned latent representation   $\B h^{\mathcal O} (t)$  of the quantum dynamics comes from a neural ODE layer \cite{chen2018neural} which parameterizes the continuous dynamics of the system using an ordinary differential equation specified by a multilayer perceptron (MLP) with parameters $\theta$:
\begin{align}\label{eq:ivp}
\frac{d\hidden ^{\mathcal O}(t)}{dt} = \text{MLP}_{\theta}(\hidden^{\mathcal O}(t),t). %
\end{align}
Starting from the latent space with $\hidden^{\mathcal O}(t_0)$, we can obtain the subsequent latent dynamics  $\hidden^{\mathcal O}(t)$ from the output of this layer which is the solution to this ODE initial value problem at any time $t$. A black-box differential equation solver computes this : 
\begin{align}
\hidden ^{\mathcal O}(t_{1:N}) = \solvefunc(\hidden ^{\mathcal O}(t_0), \text{MLP}_{\theta}, t_{0:N}), 
\end{align}
where $ \hidden ^{\mathcal O}(t_{1:N}) = \hidden ^{\mathcal O}(t_{1}),\dots, \hidden ^{\mathcal O}(t_N)$ is the latent representation of the quantum dynamics  $\B x ^{\mathcal O} (t_1) , \dots, \B x ^{\mathcal O} (t_N) $. A single MLP is used to decode the latent dynamics into quantum dynamics.    
For any latent state, the entire latent trajectory is uniquely defined and we can extrapolate this latent trajectory to make predictions arbitrarily far forwards in time. The continuously-defined dynamics provided by the neural ODE allow us to work over arbitrary times and avoid any discretization of the time intervals. 
This is akin to treating $\hidden ^{\mathcal O} (t)$ as $\hat{\rho}_S (t)$ in the quantum dynamics.

\begin{figure*}[t]

\begin{tabular}{cc} 
\vspace{-0.2cm} 
\rotatebox{90}{\hspace{0.5cm} $\hidden ^{\mathcal O}(t)$ }  &
     \includegraphics[width=0.11\textwidth]{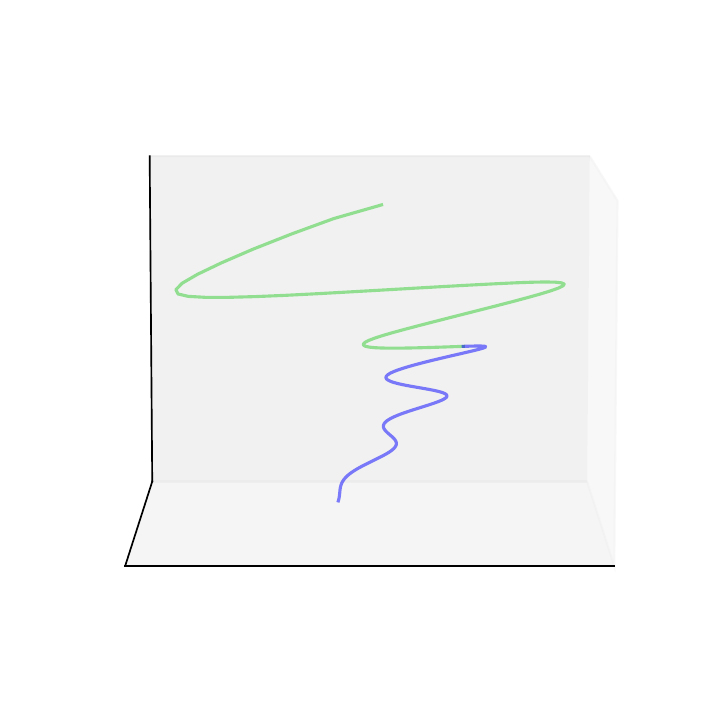}
     \includegraphics[width=0.11\textwidth]{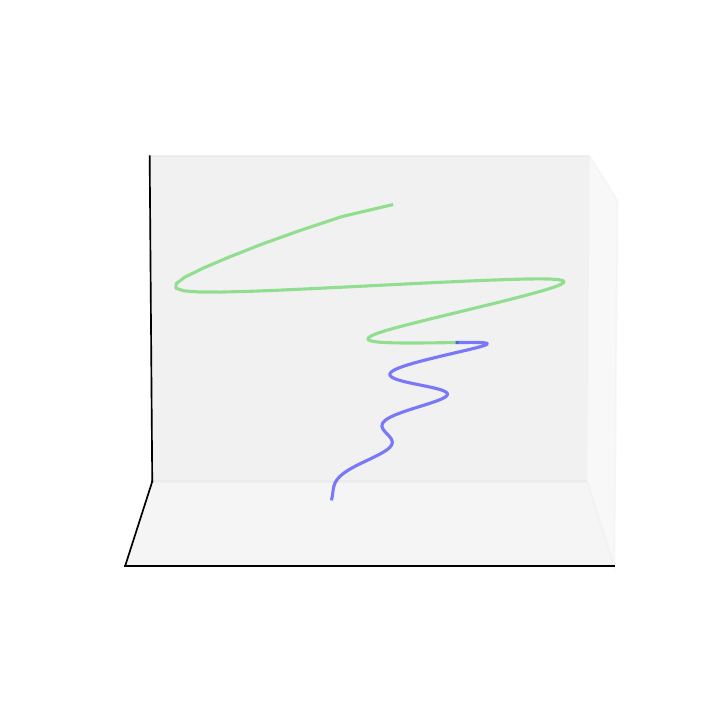}
     \includegraphics[width=0.11\textwidth]{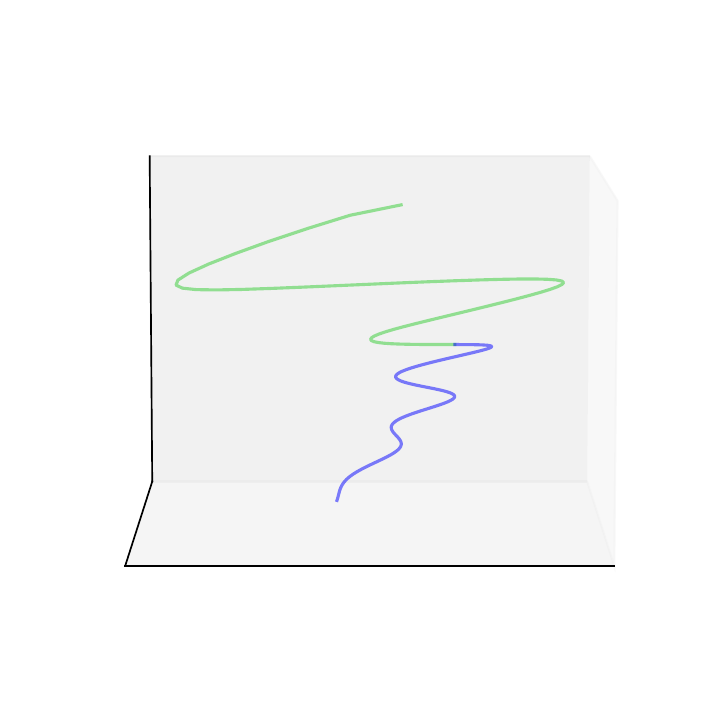}
     \includegraphics[width=0.11\textwidth]{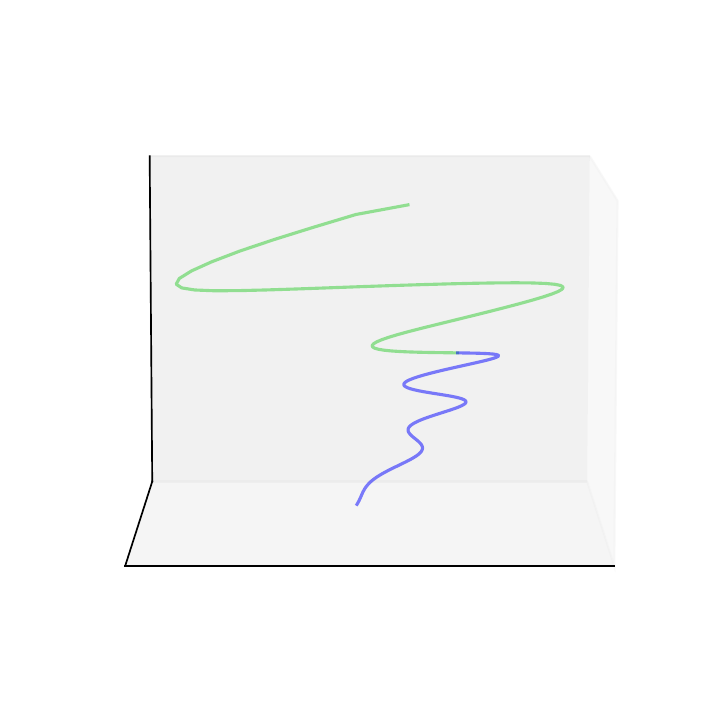}
     \includegraphics[width=0.11\textwidth]{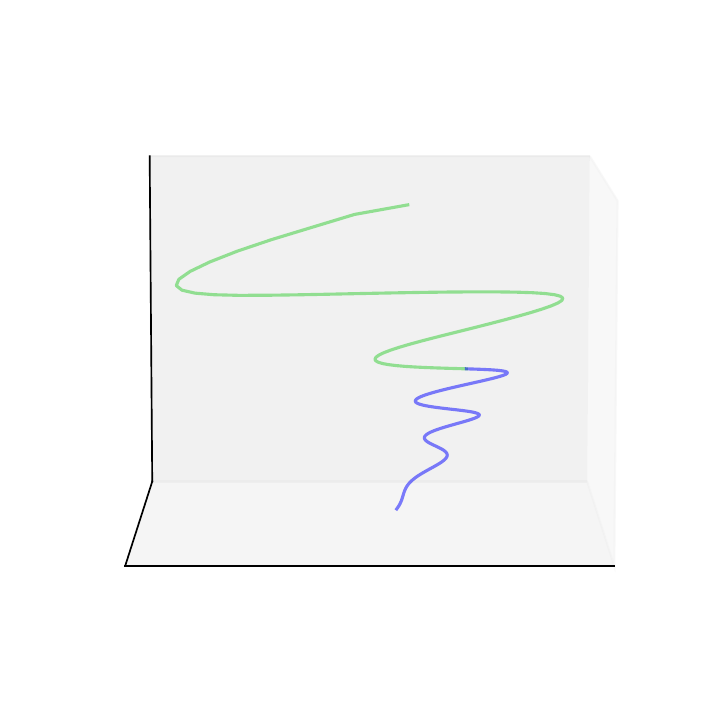}
     \includegraphics[width=0.11\textwidth]{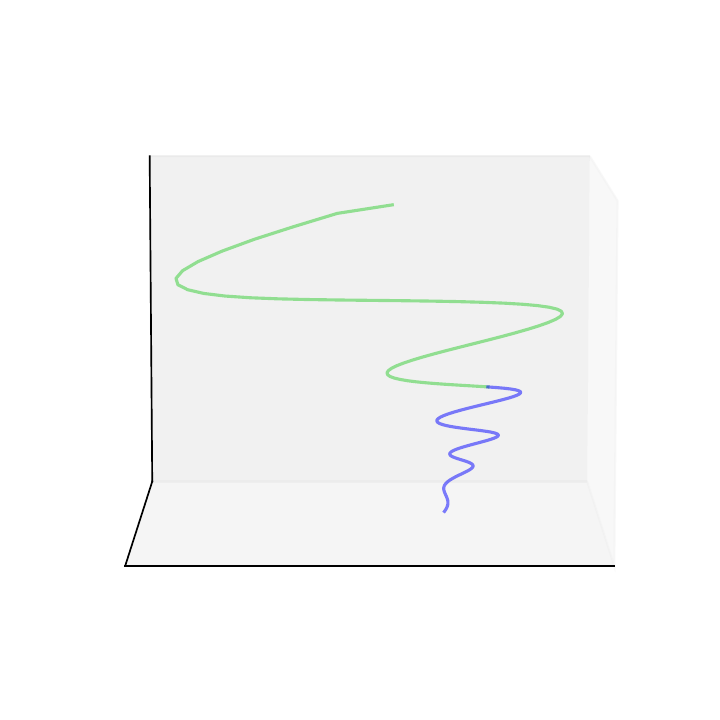}
     \includegraphics[width=0.11\textwidth]{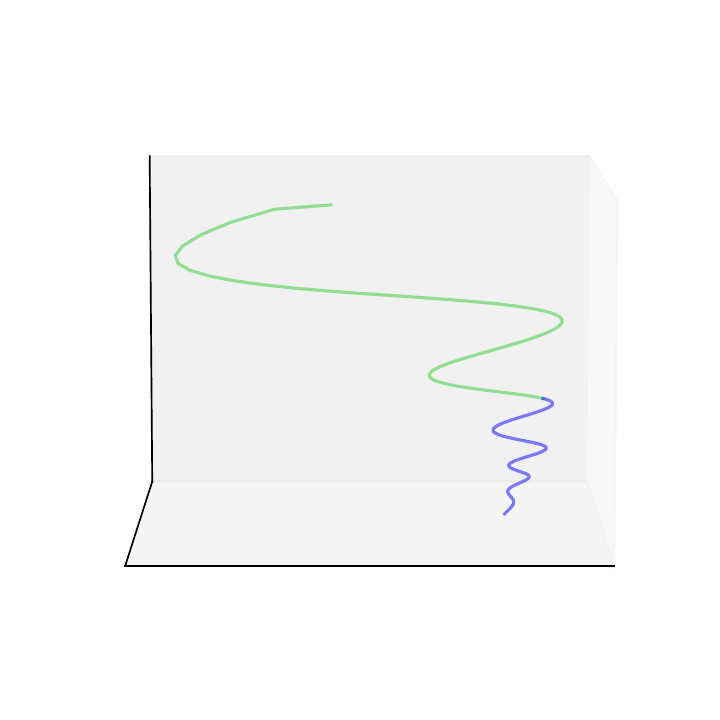}
     \includegraphics[width=0.11\textwidth]{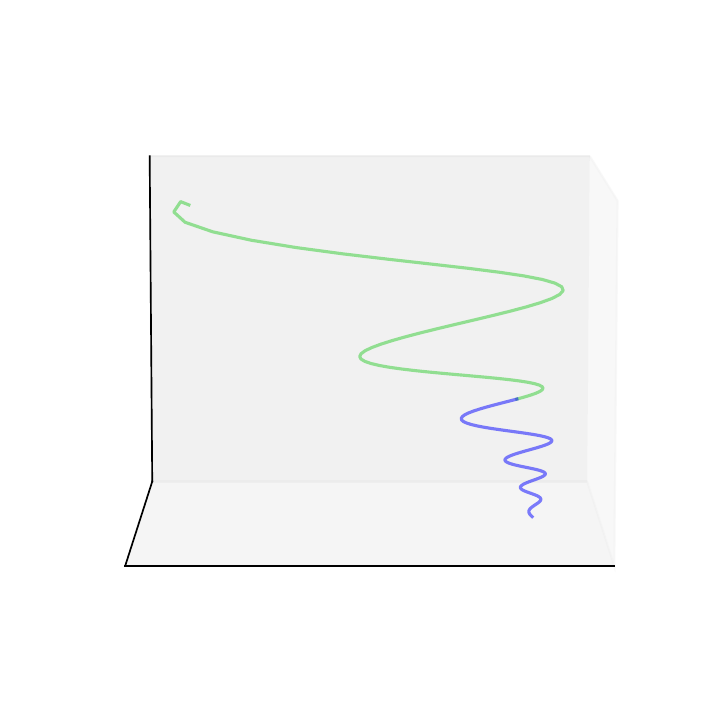}
\vspace{-0.25cm} \\ 
\rotatebox{90}{\hspace{0.5cm} $\B x ^{\mathcal O} (t)$}  &
     \includegraphics[width=0.11\textwidth]{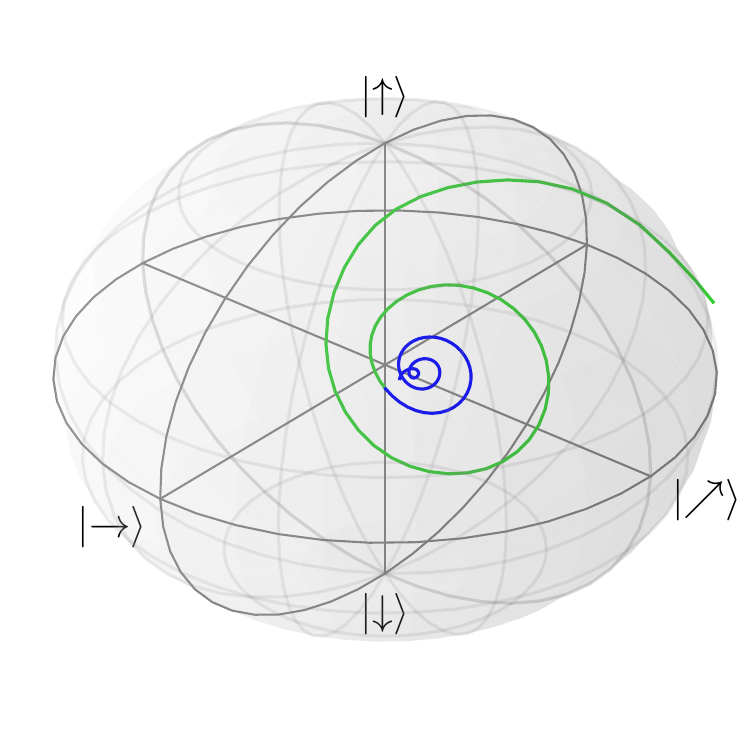}
     \includegraphics[width=0.11\textwidth]{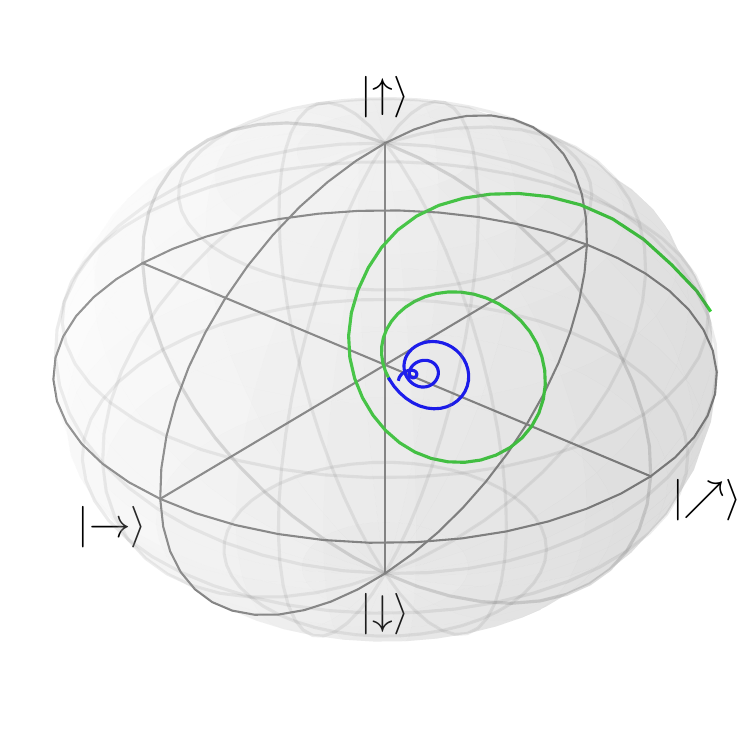}
     \includegraphics[width=0.11\textwidth]{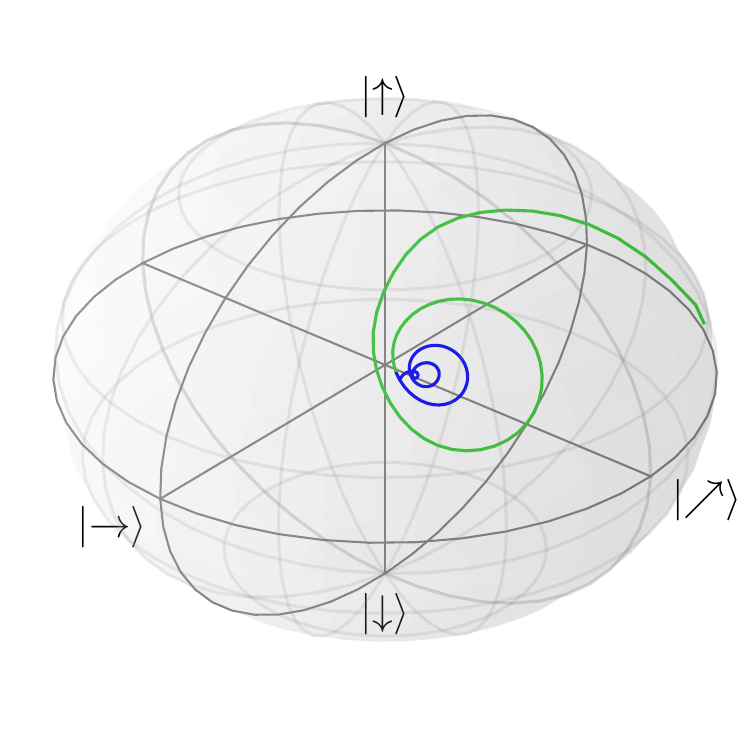}
     \includegraphics[width=0.11\textwidth]{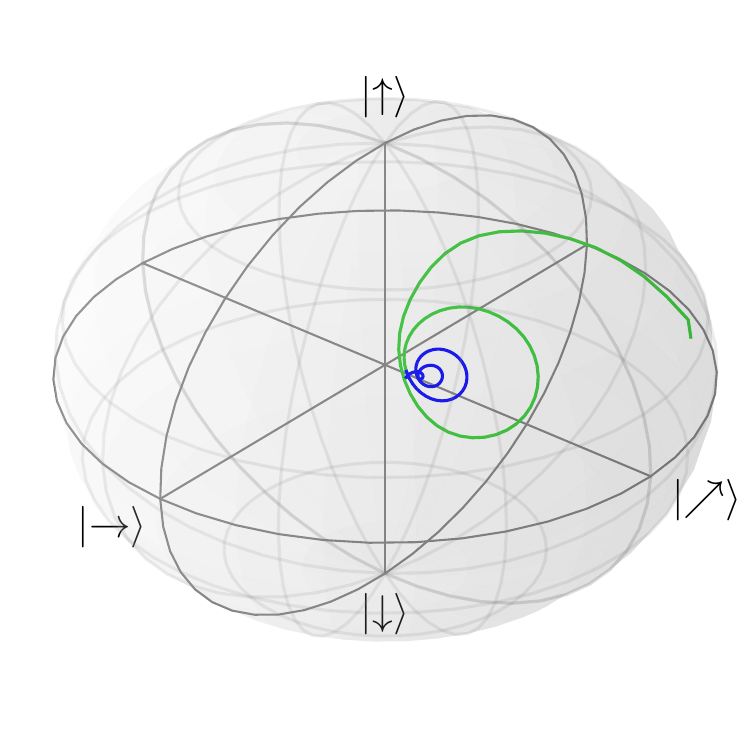}
     \includegraphics[width=0.11\textwidth]{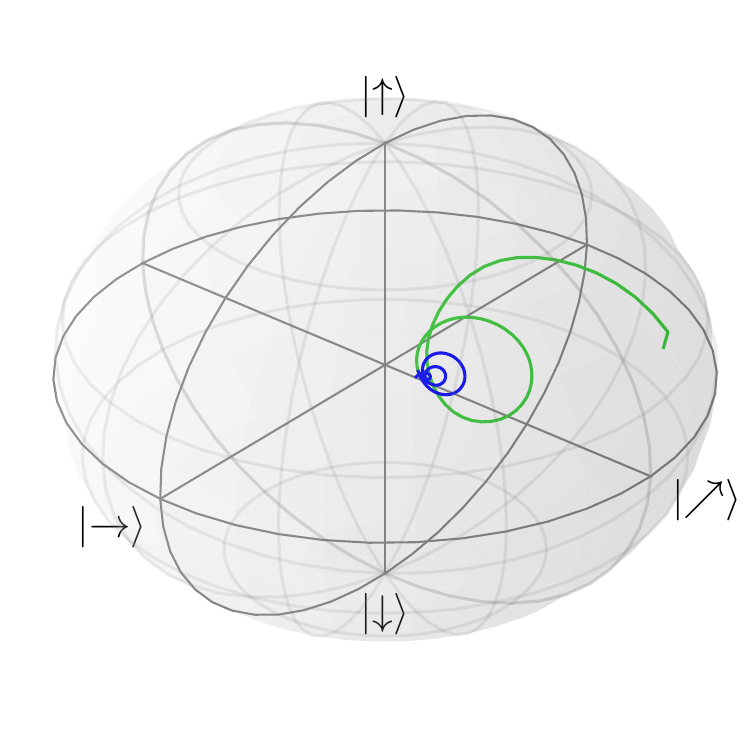}
     \includegraphics[width=0.11\textwidth]{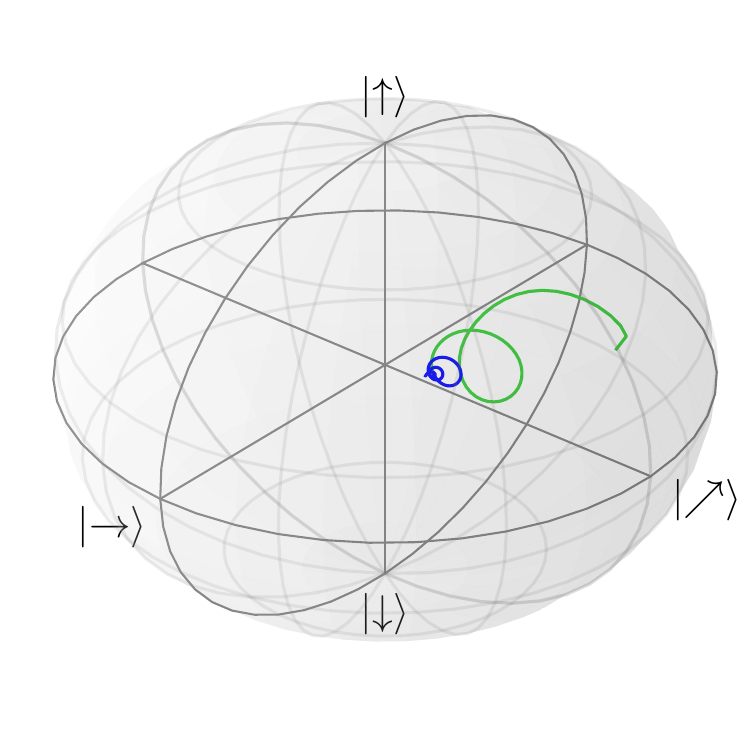}
     \includegraphics[width=0.11\textwidth]{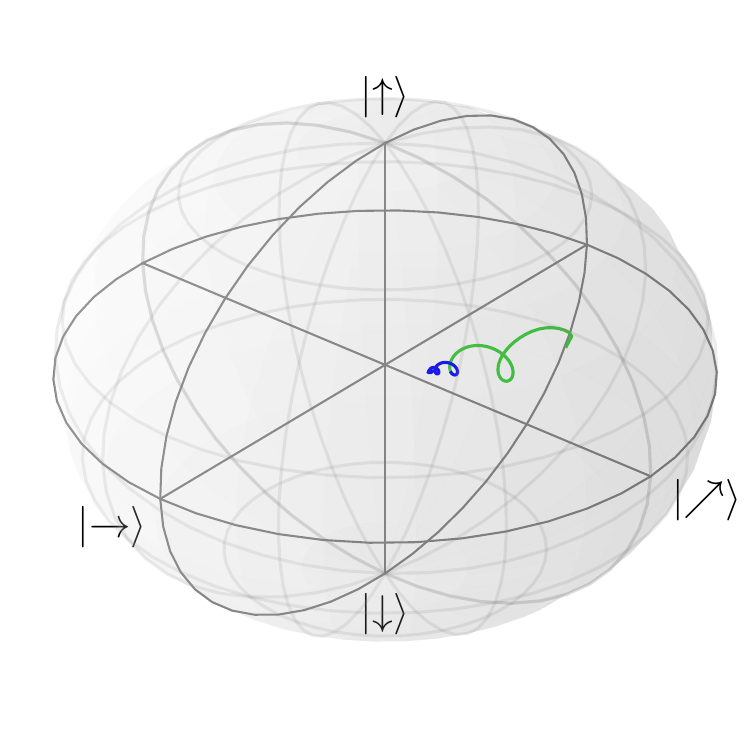}
     \includegraphics[width=0.11\textwidth]{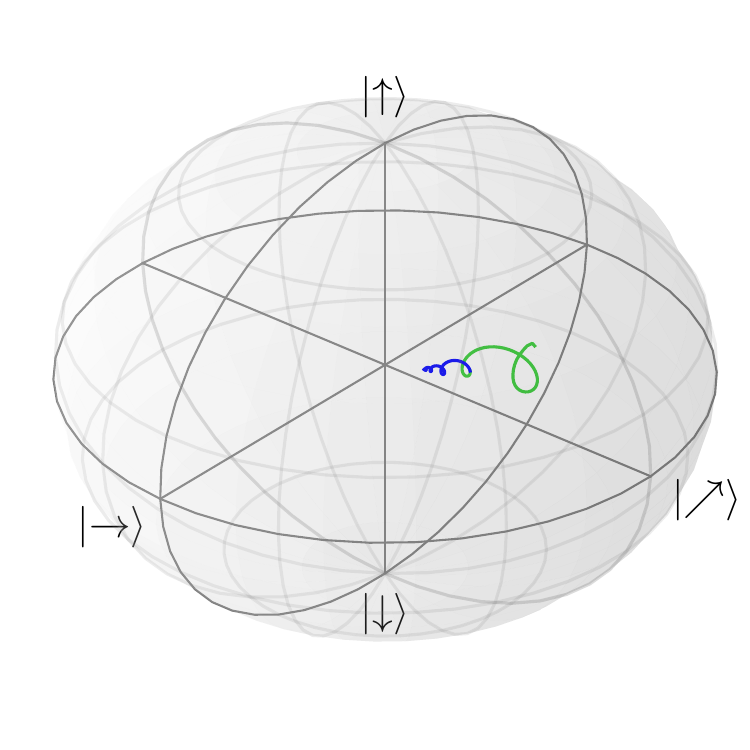}
\vspace{-0.25cm}
\\ 
\rotatebox{90}{\hspace{-.15cm} $||\Psi||_{2}$}  &
     \includegraphics[width=0.11\textwidth]{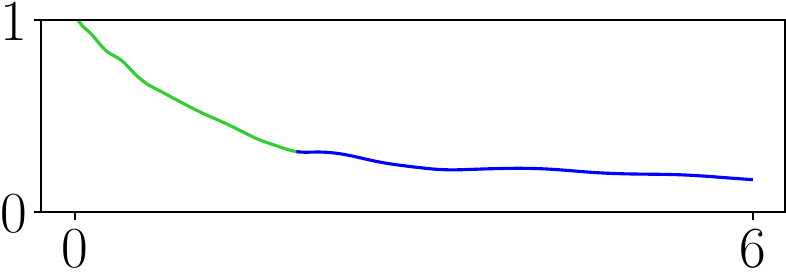}
     \includegraphics[width=0.11\textwidth]{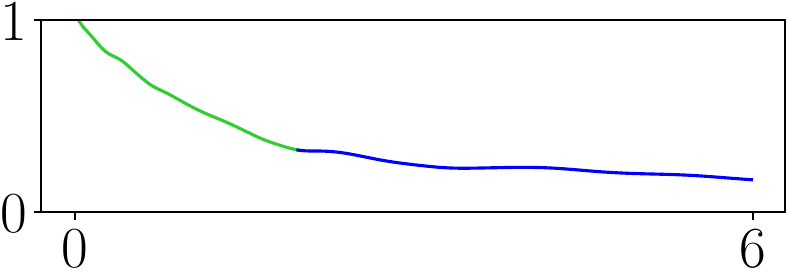}
     \includegraphics[width=0.11\textwidth]{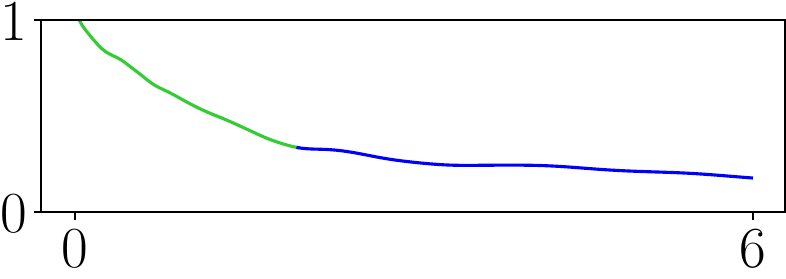}
     \includegraphics[width=0.11\textwidth]{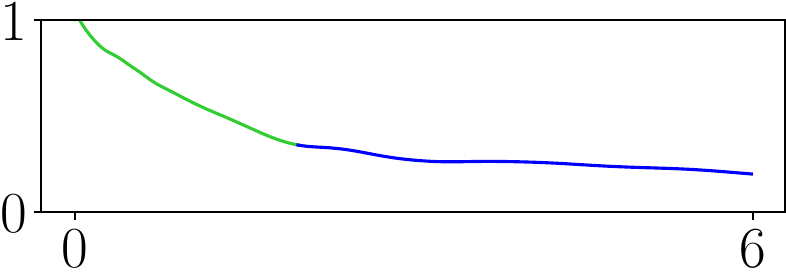}
     \includegraphics[width=0.11\textwidth]{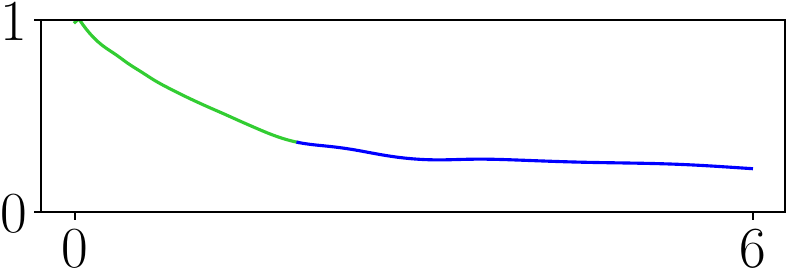}
     \includegraphics[width=0.11\textwidth]{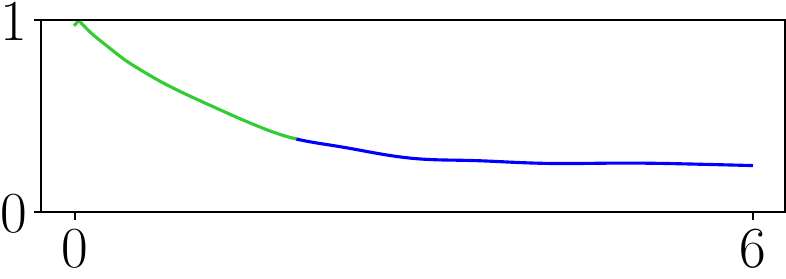}
     \includegraphics[width=0.11\textwidth]{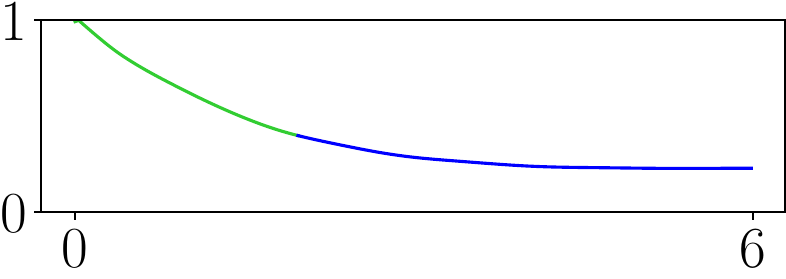}
     \includegraphics[width=0.11\textwidth]{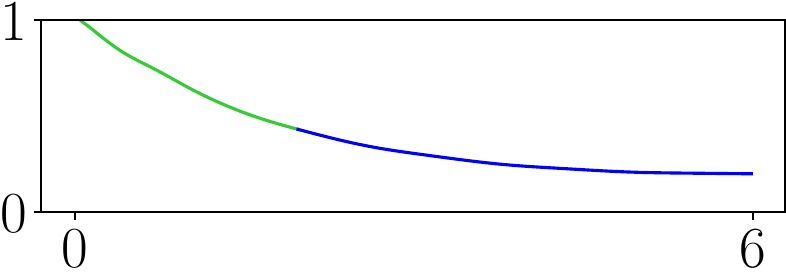}
\\ 
\noalign{\smallskip} \hline \hline \noalign{\smallskip} 
\rotatebox{90}{\hspace{0.25cm} $\hidden ^{\mathcal O}(t)$ }  &
     \includegraphics[width=0.11\textwidth]{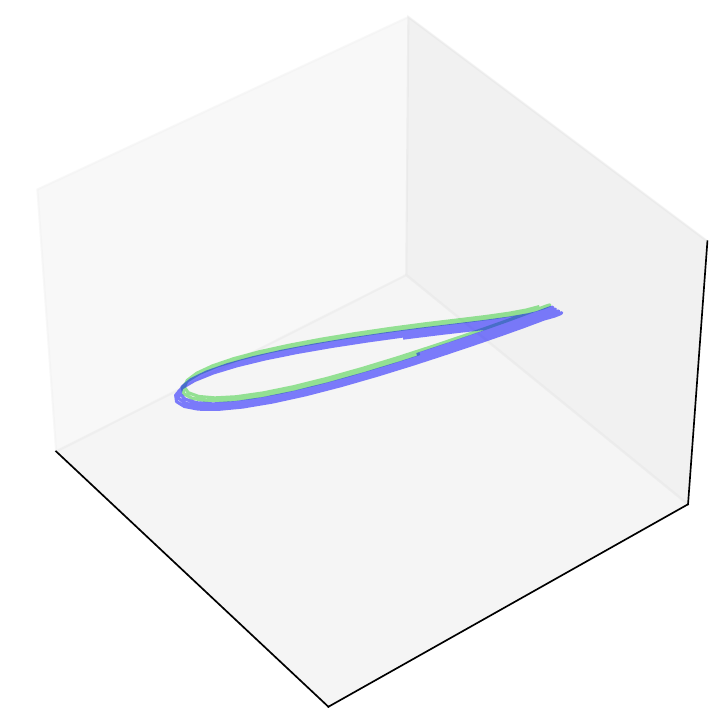}
     \includegraphics[width=0.11\textwidth]{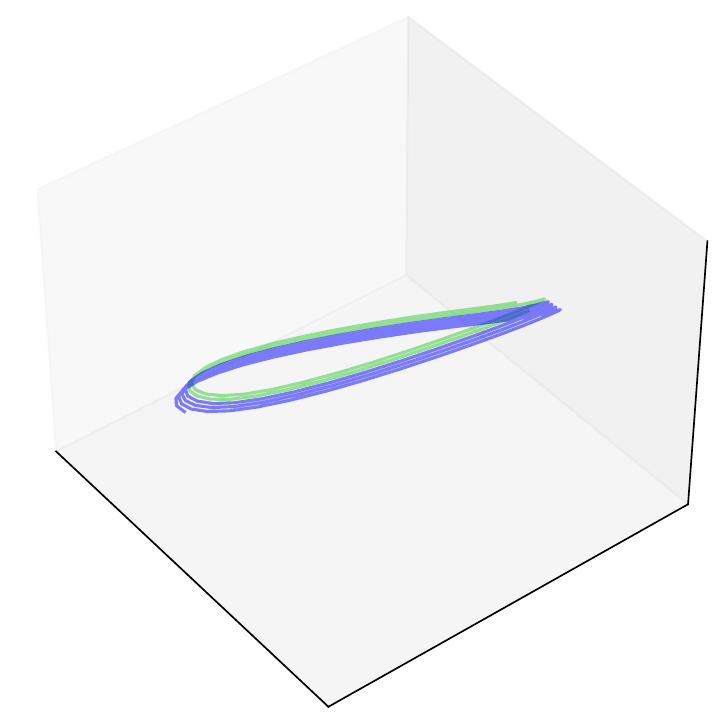}
     \includegraphics[width=0.11\textwidth]{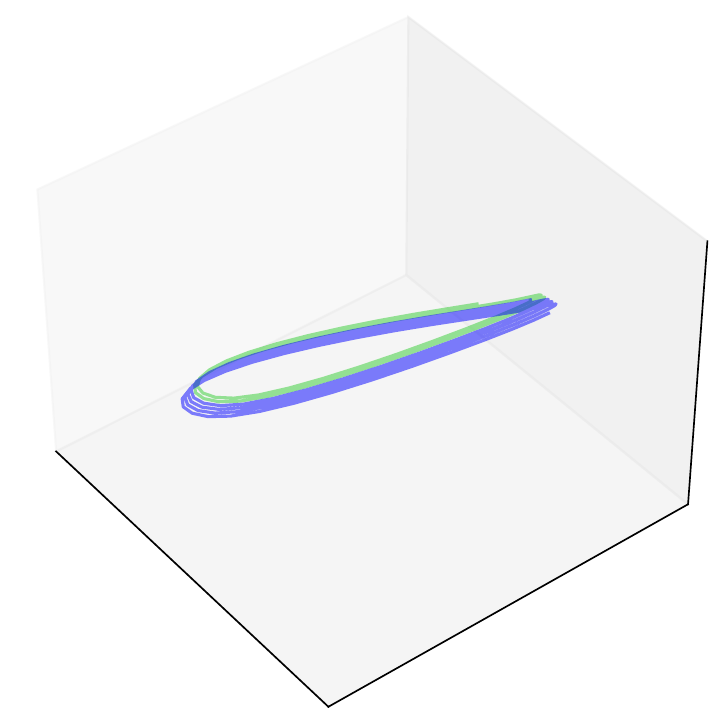}
     \includegraphics[width=0.11\textwidth]{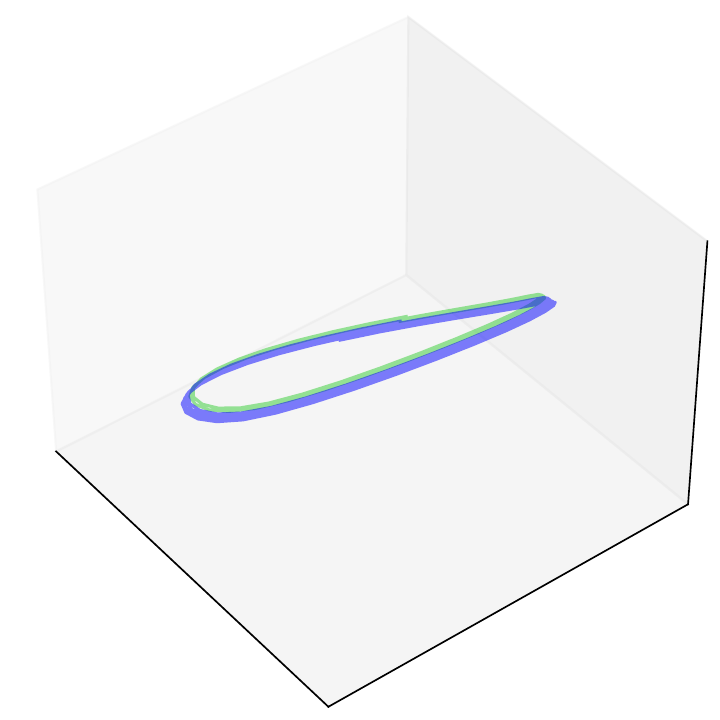}
     \includegraphics[width=0.11\textwidth]{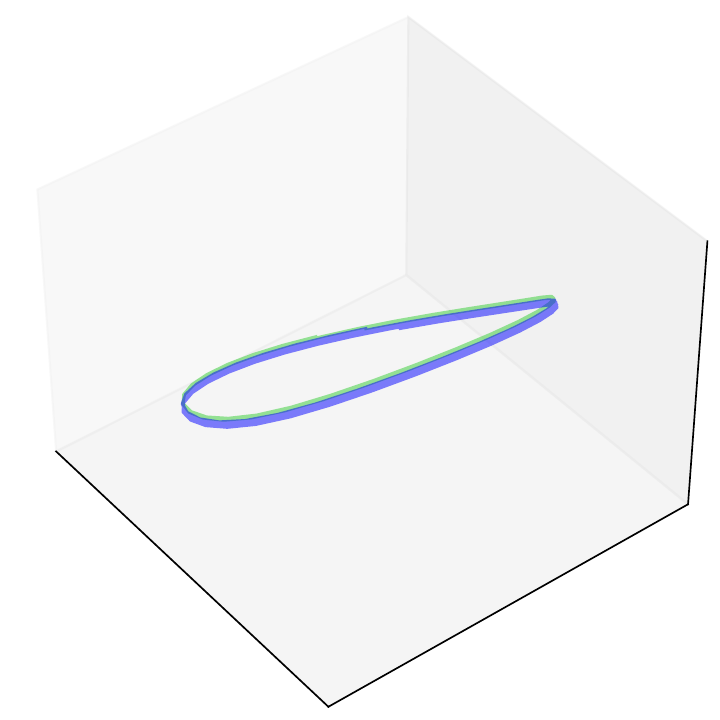}
     \includegraphics[width=0.11\textwidth]{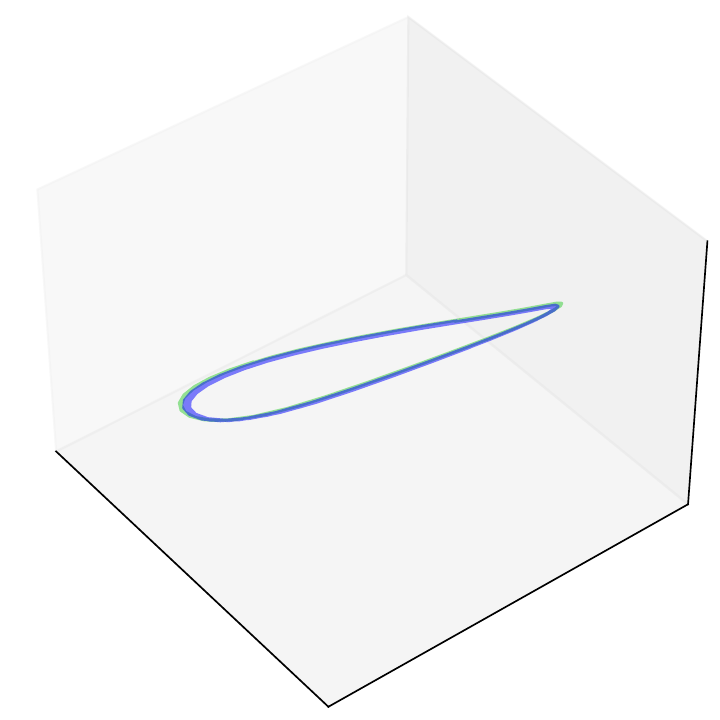}
     \includegraphics[width=0.11\textwidth]{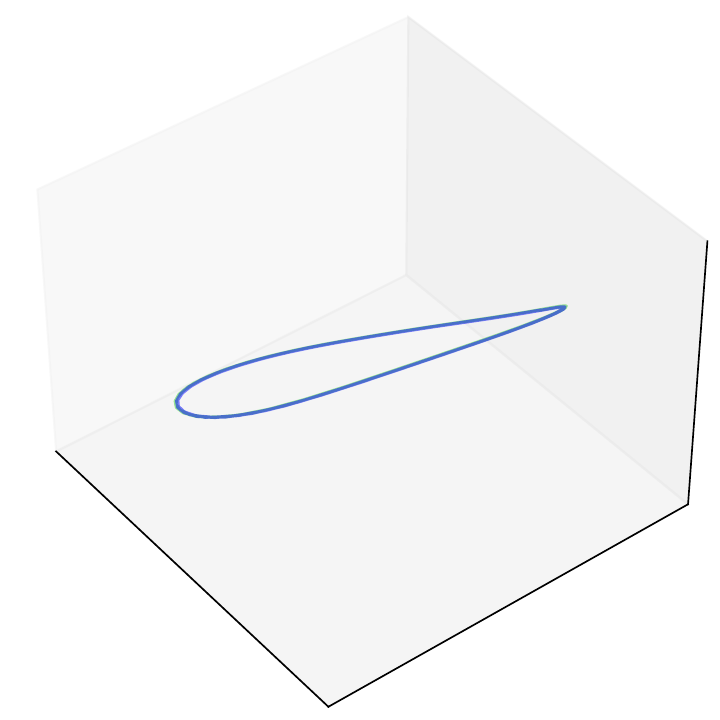}
     \includegraphics[width=0.11\textwidth]{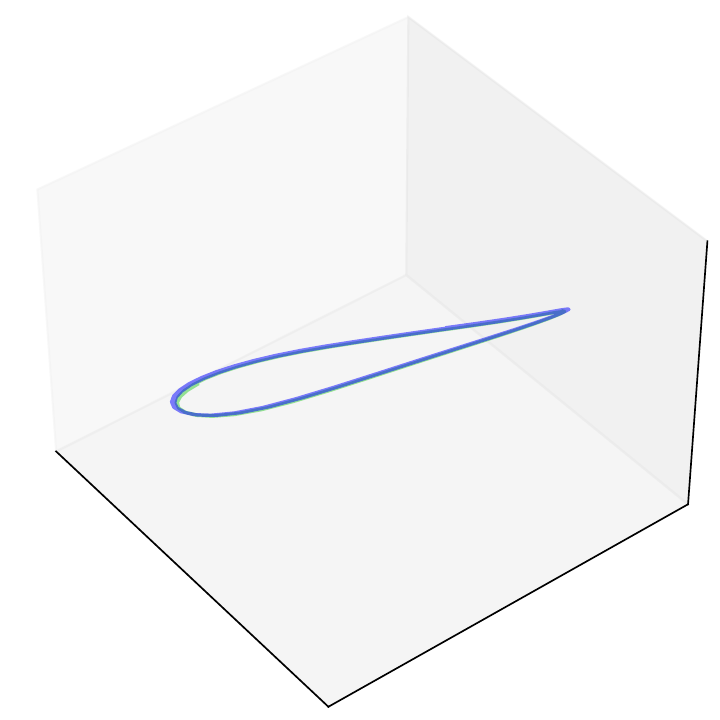}
\vspace{-0.125cm}
\\ 
\rotatebox{90}{\hspace{0.5cm} $\B x ^{\mathcal O} (t)$}  &
     \includegraphics[width=0.11\textwidth]{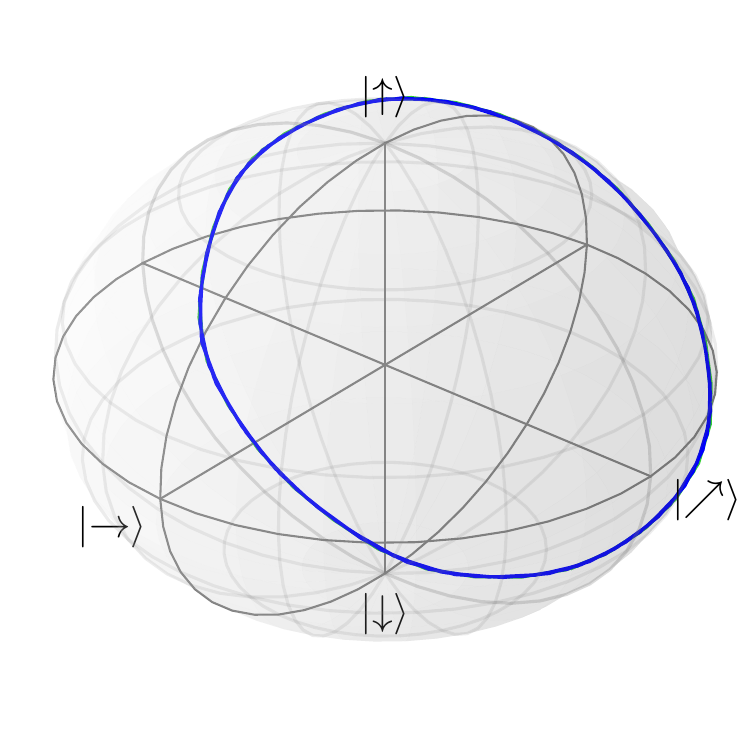}
     \includegraphics[width=0.11\textwidth]{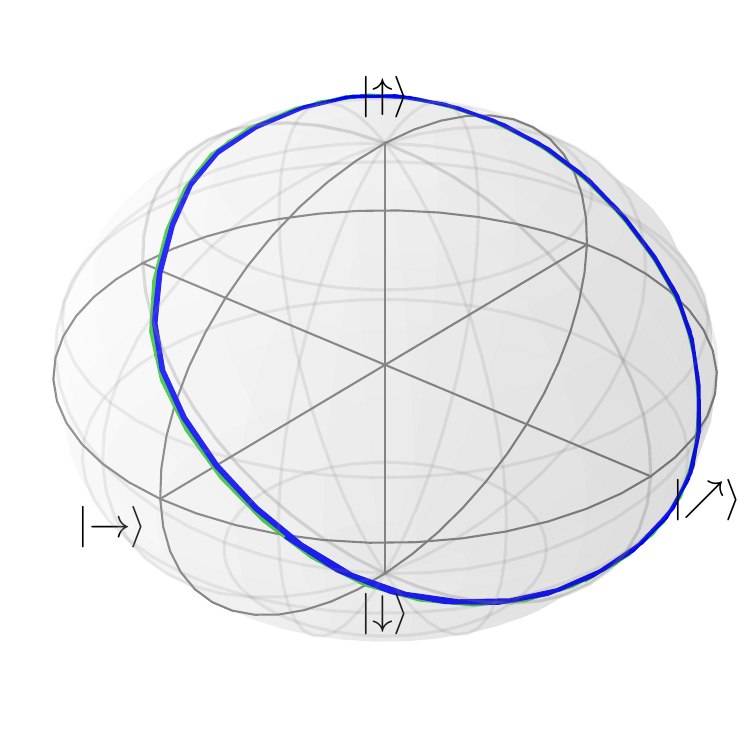}
     \includegraphics[width=0.11\textwidth]{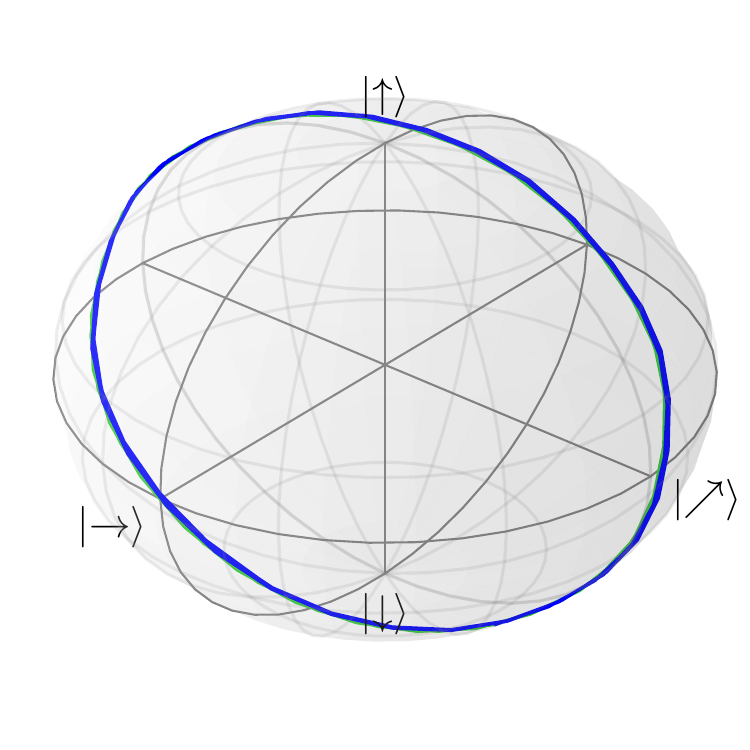}
     \includegraphics[width=0.11\textwidth]{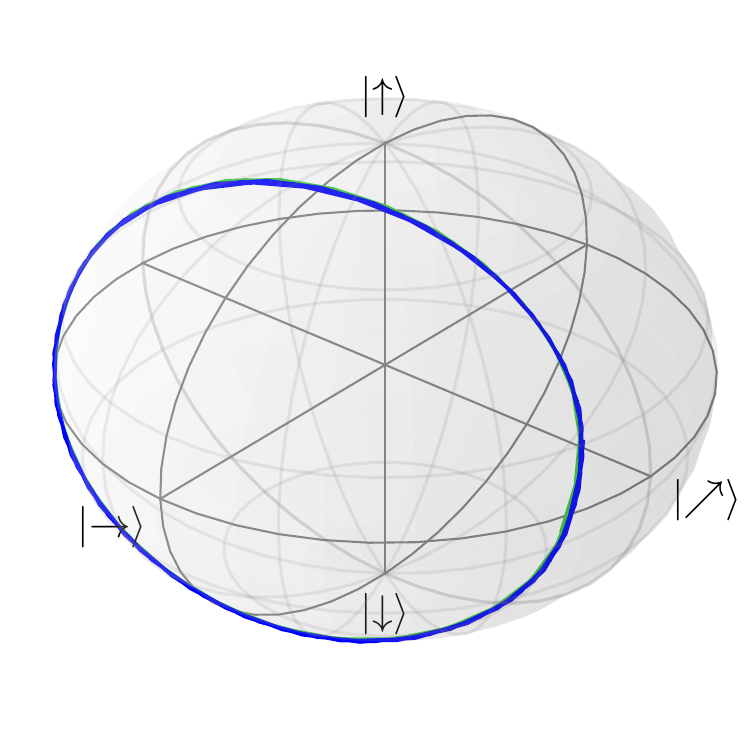}
     \includegraphics[width=0.11\textwidth]{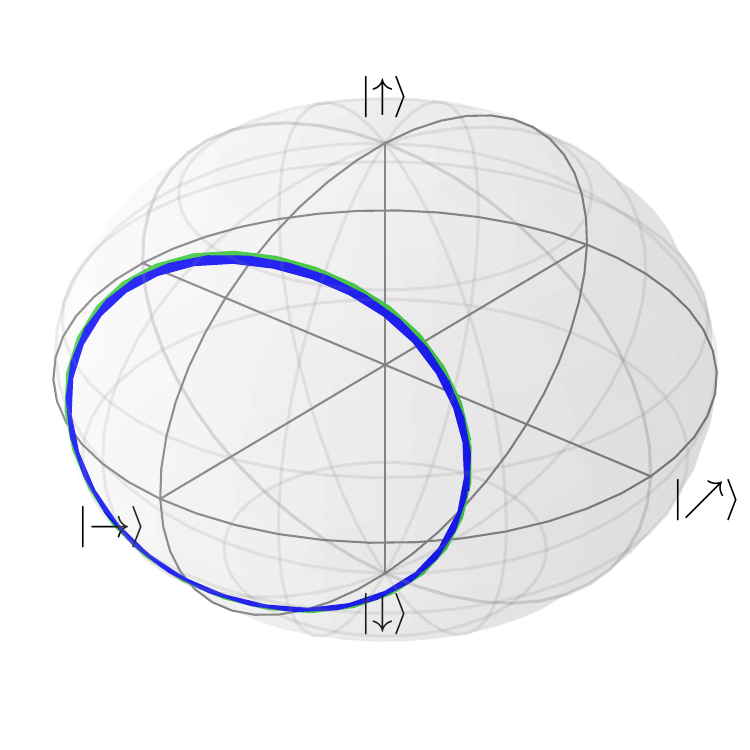}
     \includegraphics[width=0.11\textwidth]{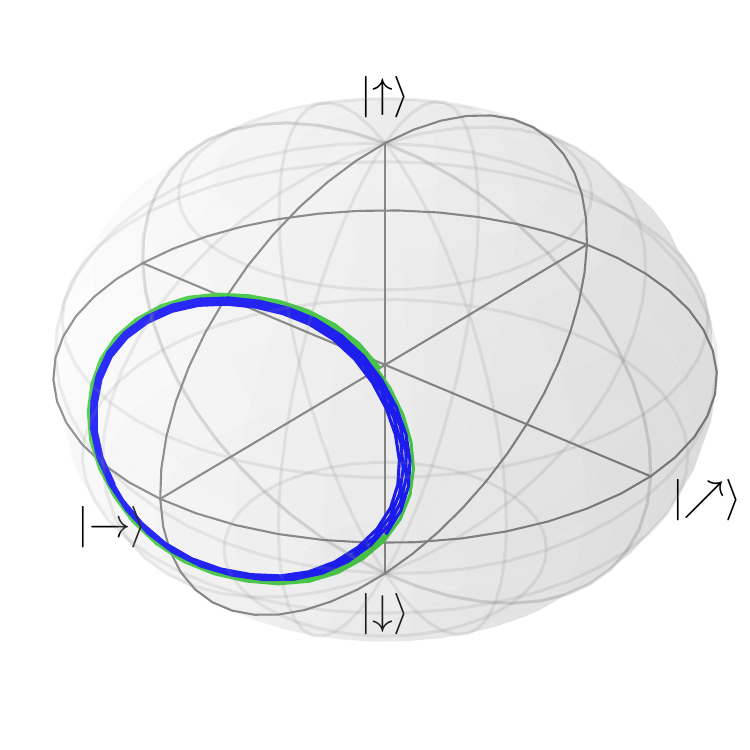}
     \includegraphics[width=0.11\textwidth]{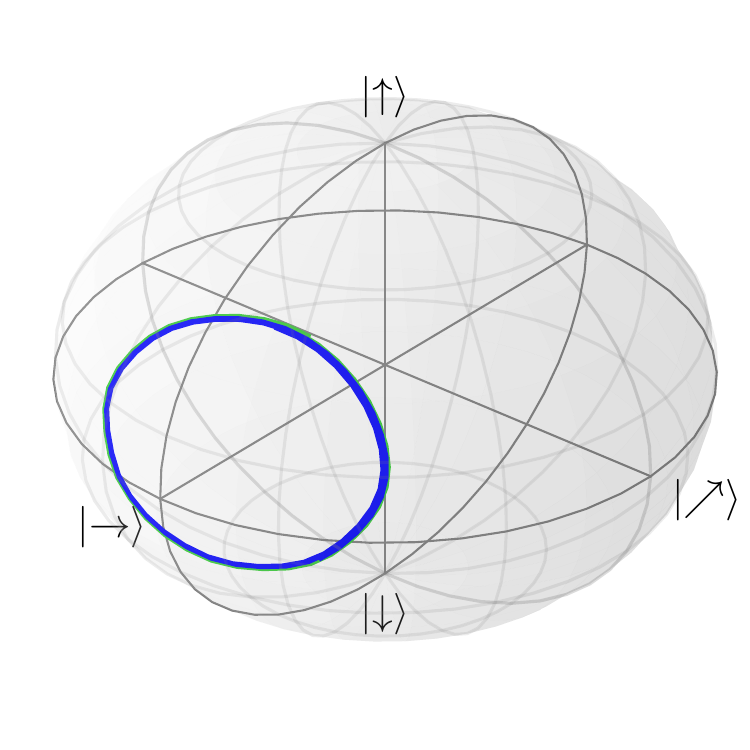}
     \includegraphics[width=0.11\textwidth]{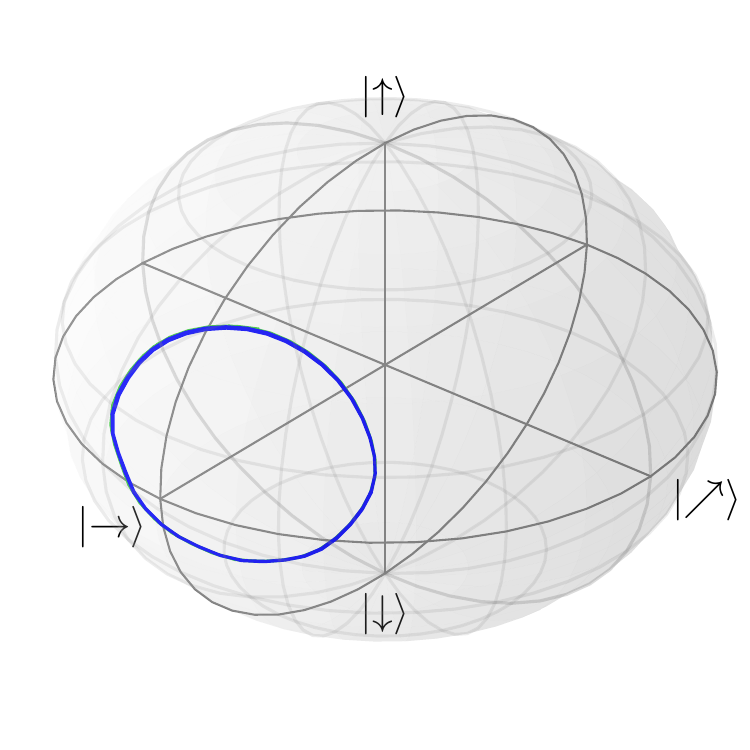}

\vspace{-0.25cm}
\\ 
\rotatebox{90}{\hspace{-.15cm} $||\Psi||_{2}$}  &
     \includegraphics[width=0.11\textwidth]{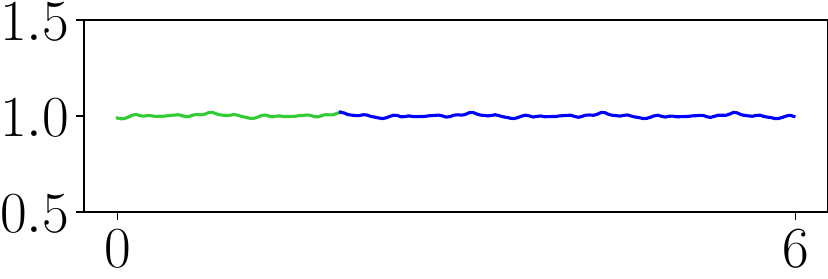}
     \includegraphics[width=0.11\textwidth]{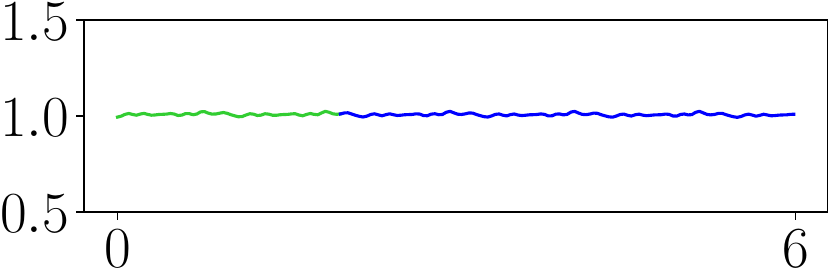}
     \includegraphics[width=0.11\textwidth]{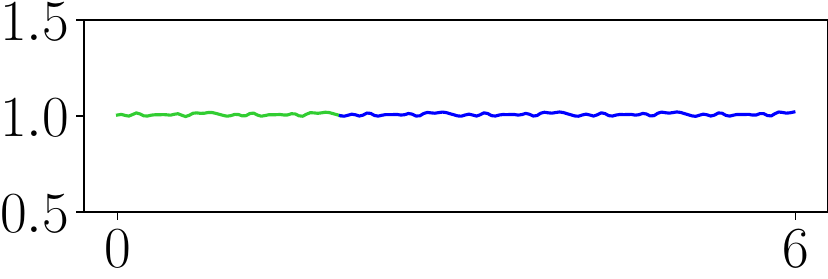}
     \includegraphics[width=0.11\textwidth]{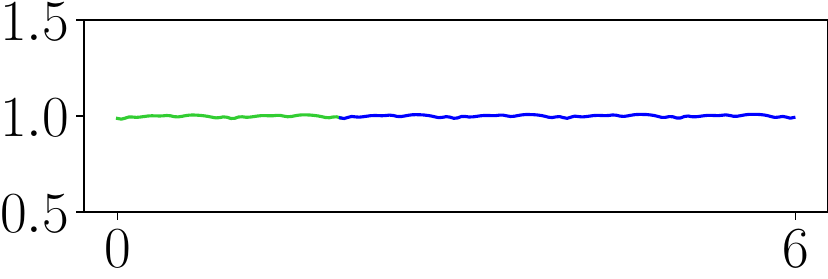}
     \includegraphics[width=0.11\textwidth]{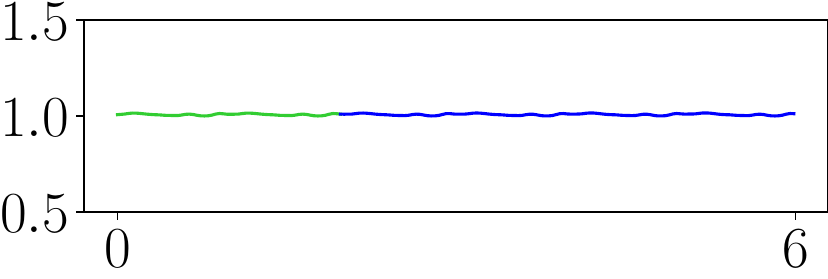}
     \includegraphics[width=0.11\textwidth]{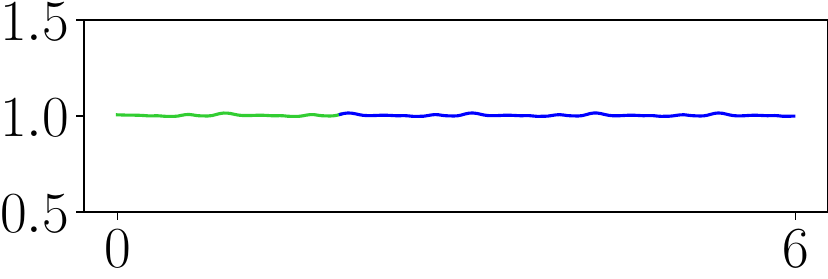}
     \includegraphics[width=0.11\textwidth]{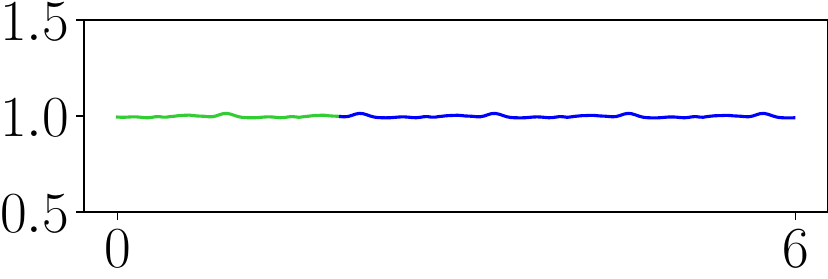}
     \includegraphics[width=0.11\textwidth]{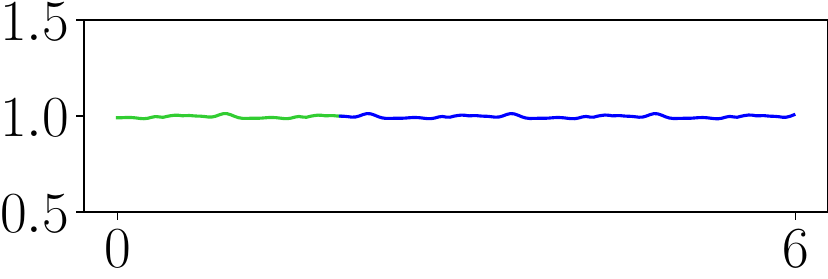}
\\ \noalign{\smallskip} \hline  \noalign{\smallskip}
&   $\hidden _1 ^{\mathcal O}(t_0) \hspace{1.2cm} \hidden _2 ^{\mathcal O}(t_0)  \hspace{1.1cm} \hidden _3 ^{\mathcal O}(t_0)  \hspace{1.1cm} \hidden _4 ^{\mathcal O}(t_0)  \hspace{1.1cm} \hidden _5 ^{\mathcal O}(t_0)  \hspace{1.1cm} \hidden _6 ^{\mathcal O}(t_0)  \hspace{1.1cm} \hidden _7 ^{\mathcal O}(t_0)  \hspace{1.1cm} \hidden _8 ^{\mathcal O}(t_0)   $ \\
\end{tabular}
\caption{\textbf{Latent space interpolations}. A table of two interpolations in the latent space of the QNODE with the two different training dynamics : the top super row is from the open system and bottom super row is from the closed system. In each super row there are three sub rows showing, for each interpolation point, 1) the plotting latent dynamics 2) decoded quantum dynamics 3) time series of the quantum state's norm. 
Each column is from a different interpolation point in the latent space. 
The green solid line is two \textbf{as} of trained dynamics and blue is 4 \textbf{as} of extrapolated dynamics.
}
\label{fig:latent_interpol}
\end{figure*}

\subsection{Data generation}
To generate time series data for a quantum system, we first consider a simple two-level quantum system
$ \hat{H}_1=(\omega\hat{\sigma}_z +\Delta \hat{\sigma}_x)/2$, where $\omega$ is the energy splitting of the two-level system,
$\Delta$ is the detuning, and $\hat{\sigma}$'s are the usual Pauli matrices. 
$\omega,\Delta$ are sampled from a Gaussian distribution (see SM \ref{SM:methods_data}).
For each set of $\{\omega,\Delta\}$, we obtain time series data of the three expectation values that define the Bloch vector :
$\mean{\hat{\sigma}_x (t)}, \mean{\hat{\sigma}_y (t)}, \mean{\hat{\sigma}_z (t)},$ for $t\in[0,t_N]$, 
by numerically solving the corresponding von Neumann equation or Lindblad master equation using QuTiP's \cite{johansson2012qutip} numerical solver, 
given an arbitrary initial quantum state, i.e., $\ket{\psi(0)}=\hat{\mathcal{U}}_R\ket{0}$-- we produce two datasets one each for the open and closed systems.
Here, $\hat{\mathcal{U}}_R$ refers to the $2\times2$ Haar random unitary matrix, and $t_N$ is the total training time.
In this way, one can take the time-series data as a series of projective measurements made across various times $[0,t_N]$ on the statistical ensemble, which is prepared in an initial quantum state $\hat{\rho}_S (0)$. 

Similarly, for two-qubit system, we use the following Hamiltonian $\hat{H}_2 = (\omega_1\hat{\sigma}_z ^1 +\Delta_1 \hat{\sigma}_x ^1)/2 + (\omega_2\hat{\sigma}_z ^2 +\Delta_2 \hat{\sigma}_x ^2)/2 + J \sigma_x ^1 \sigma_x ^2$, where $\omega$'s, $\Delta$'s, $J$'s are all sampled from the same Gaussian distribution as in the single-qubit case. 
The initial states are sampled from the Haar random matrix for each qubit space, and tensor product afterward.

\section{Main results}

We conduct three main experiments using the QNODE trained on both closed and open two-level quantum systems--- 
1) Generating quantum dynamics from random positions in the QNODE's latent space  
2) testing if the QNODE's generated dynamics preserve the Heisenberg uncertainty principle (HUP), and 
3) testing if the QNODE learns an interpretable latent space by performing interpolations and assessing its learned physics. 

In the following, for each trajectory generated by QNODE the green is two arbitrary time units (\textbf{as}, i.e., when $\hbar=\omega=1$) of trained dynamics and blue is 4 \textbf{as} of extrapolated dynamics. Black and red lines are the real quantum dynamics with black being the actual training region or two \textbf{as} and red being four \textbf{as} of dynamics unseen to the QNODE.

\begin{figure*}[t]
\rotatebox{90}{\hspace{0.7cm} \text{QNODE}}
\rotatebox{90}{$\brakets{\sigma_{z2}}$ $\brakets{\sigma_{z1}}$ $\brakets{\sigma_{x2}}$ $\brakets{\sigma_{x1}}$} 
\includegraphics[width=0.23\textwidth]{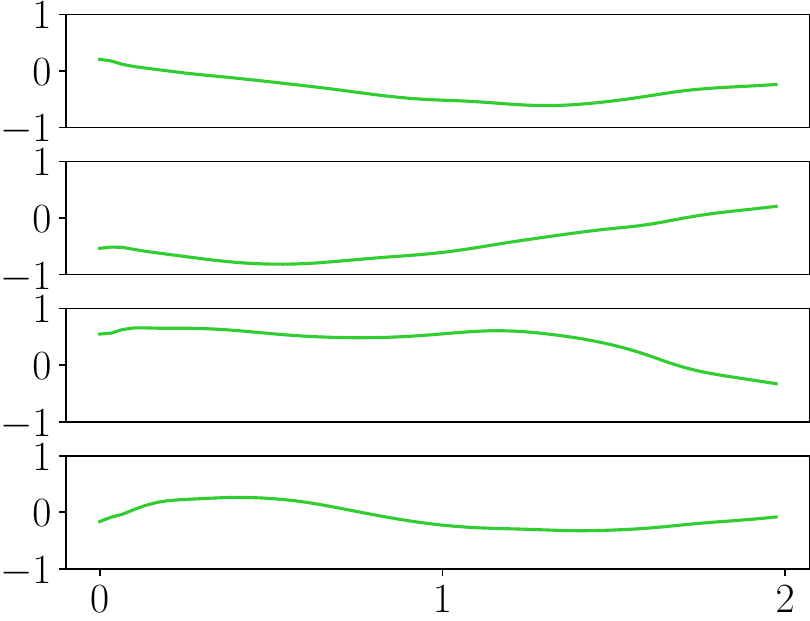}
\includegraphics[width=0.23\textwidth]{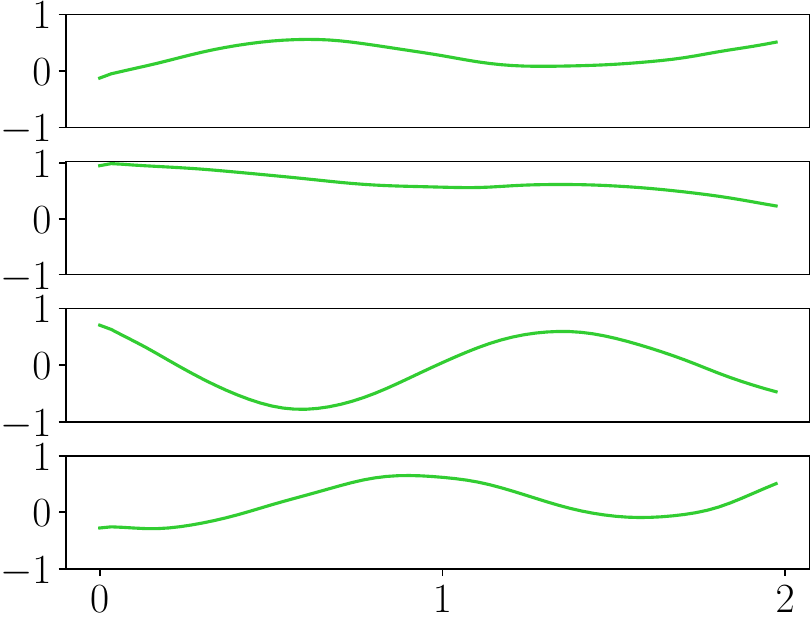} 
\includegraphics[width=0.23\textwidth]{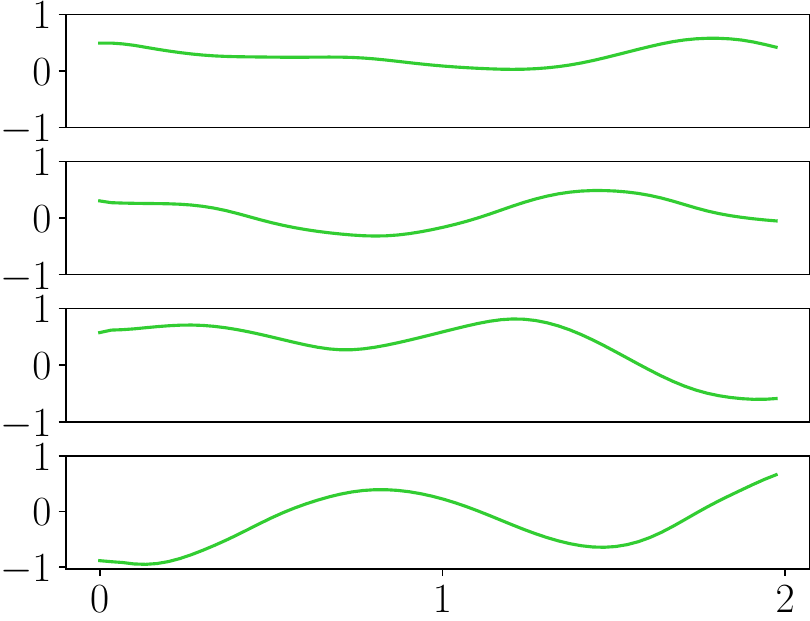} 
\includegraphics[width=0.23\textwidth]{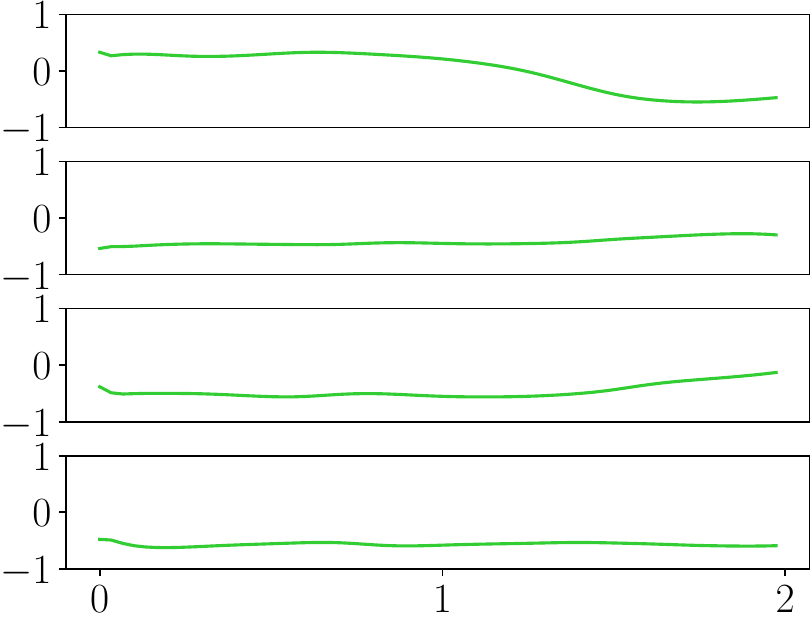} 

\vspace{0.2cm}
\rotatebox{90}{\hspace{1.1cm} \text{DATA}} 
\rotatebox{90}{$\brakets{\sigma_{z2}}$ $\brakets{\sigma_{z1}}$ $\brakets{\sigma_{x2}}$ $\brakets{\sigma_{x1}}$} 
\includegraphics[width=0.23\textwidth]{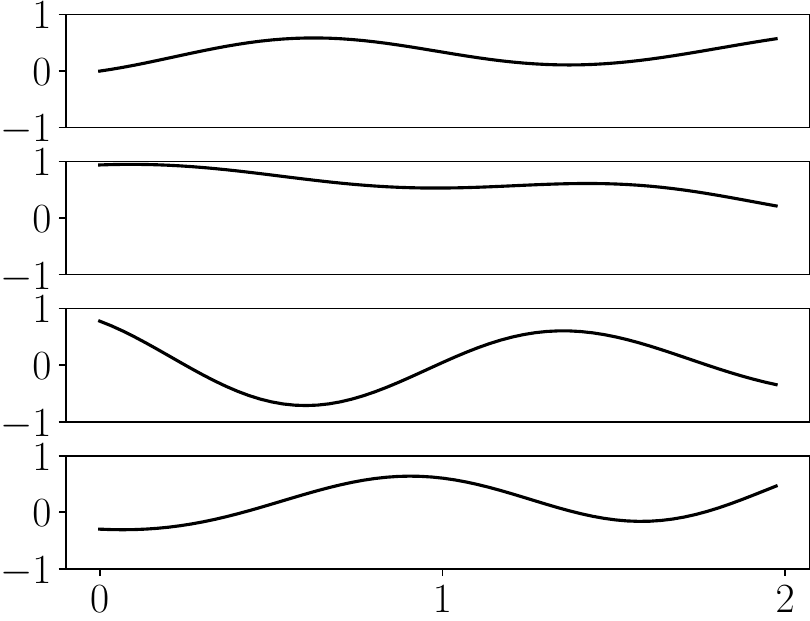}
\includegraphics[width=0.23\textwidth]{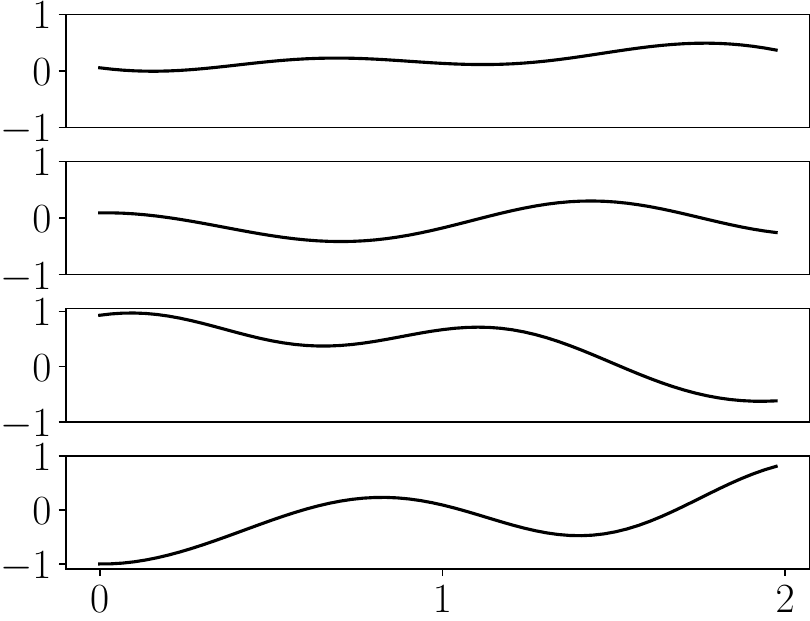}
\includegraphics[width=0.23\textwidth]{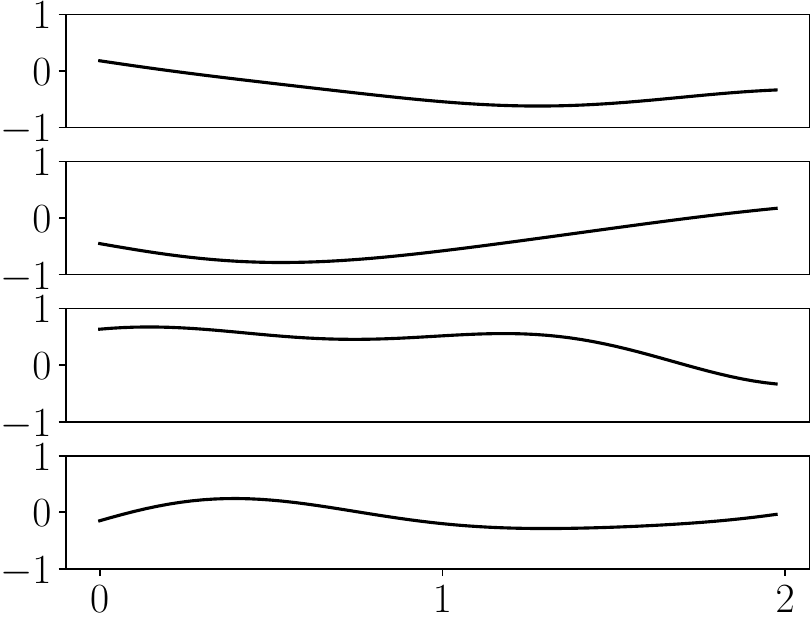}
\includegraphics[width=0.23\textwidth]{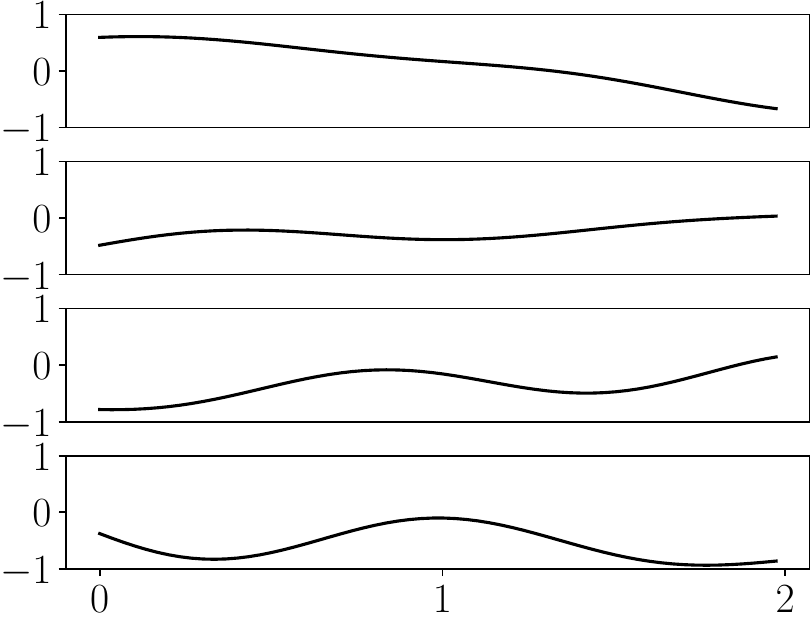}

\caption{\textbf{Two-qubit samples} The green and black trajectories represent expectations values of generated random samples and the training data respectively. Our model creates random samples similar to the training data, showing that it learns valid dynamics. The x-axis traverses from $0$ to $2$ \textbf{as}.}
\label{fig:two_qubit_samples}
\end{figure*}

\subsection{Generated dynamics}
After training the QNODE on trajectories from both closed and open quantum system dynamics for the single-qubit Hamiltonian $\hat{H}_1$, we test if the model will produce dynamics that resemble the training data and satisfy the von Neumann (time-local Lindblad master) equation for the closed (open) quantum system case. 
To generate quantum dynamics from the QNODE we initialize the dynamics randomly in the latent space by sampling $\hidden ^{\mathcal O}(t_0) $ from a standard Gaussian distribution.  
Using that random latent point, we generate quantum dynamics and extrapolate forwards in time for an additional four \textbf{as}. 
We plot the dynamics generated by the QNODE on the Bloch sphere and compare them with its training data in \figref{fig:the_model} (b) for the QNODE trained on the closed system and \figref{fig:gen_hup} (a) for the QNODE trained on the open system. 
Inspecting the dynamics in both, it is apparent that the QNODE can generate similar dynamics to the training data. 
Importantly, the generated dynamics obeys quantum mechanics since in \figref{fig:the_model} (b) $||\Psi(t)||_{2} \approx 1$ for the training and extrapolation time while in \figref{fig:gen_hup} (b) $||\Psi(t)||_{2} $ decays towards $0$ as expected in an open quantum system, where we have bit-flip and dephasing noises, i.e., $\textbf{A}_1=\hat{\sigma}_-$ and $\textbf{A}_2=\hat{\sigma}_z$, with sets of parameters for $\Gamma$, which is sampled from a Gaussian Distribution (see SM \ref{SM:methods_data}).  
Here, $||\mathcal{C}||_2$ is $l_2$-norm of the vector $\mathcal{C}$.

Quantitatively, we find that the QNODE trains and performs very well for both closed and open quantum trajectories. 
The average mean squared error (MSE) between the exact and reconstructed ones are shown in \figref{fig:closed_reconstruct} and \figref{fig:open_reconstruct}.
They are in the order of $10^{-2}$ for the worst cases.

\subsection{Heisenberg uncertainty principle}
Next, we test if the QNODE is capable of learning the Heisenberg uncertainty principle.
For this purpose, we generate $50$ trajectories from the QNODE exactly as before and compute the variance in $\hat{\sigma}_x$ and $\hat{\sigma}_z$ denoted as
$\text{var}(x)$ and $\text{var}(z)$ from 0 to 6 \textbf{as}. 
We plot the variances over time in \figref{fig:gen_hup}(b) and (c) for the closed and open quantum system respectively, with the same coloring as before (green for training and blue for extrapolation). 
We see that over time, including during which the model is extrapolating, the variances are almost entirely bounded by the plane $\text{var}(z)+\text{var}(x)=1$ 
signaling the model does learn to produce dynamics that satisfy the uncertainty principle.

\subsection{Latent space interpolation} 
We then analyze the smoothness of the latent space to see if it learns a representation that is interpretable while satisfying quantum mechanics. 
Specifically, we test if trajectories that are close in the latent space have similar quantum dynamics and also preserve physics. 
This can be done by interpolating in the latent space from one latent point $\hidden _i ^{\mathcal O}(t)$ to another $\hidden _j ^{\mathcal O}(t)$, and decoding the quantum dynamics on the path in the latent space between the two endpoints points 
$\hidden _1 ^{\mathcal O}(t_0)$ and $\hidden _8 ^{\mathcal O}(t_0)$ in \figref{fig:latent_interpol}. 
There, we plot one latent space interpolation from the QNODE trained on the open system (the top three rows) and one from the closed system (the bottom three rows).
We use spherical linear interpolation on the path and decode at 6 equally spaced steps along it, corresponding to $\hidden _2 ^{\mathcal O}(t_0), \dots , \hidden _7 ^{\mathcal O}(t_0) $  in the figure. 
For each interpolation, in the three rows, at the top we first plot the latent dynamics at each interpolation point $\hidden _i ^{\mathcal O}(t)$, in the second row we plot the decoded quantum dynamics ,and in the third row is the quantum state's norm $||\Psi(t)||_{2}$. We use the same coloring as before for training time and extrapolation time. 
To see if the model has learned a notion of similarity between trajectories corresponding to their quantum dynamics, we perform interpolations from very different training quantum dynamics. 
Both interpolations show a smooth transition between the different dynamics that preserves physics, given the behavior of the time series of $||\Psi(t)||_{2}$ in each case.

\section{Discussion}
In this work, we propose the latent neural ODE for quantum dynamics: QNODE which is capable of learning and generating quantum trajectories of closed and open quantum systems without any prior knowledge of quantum physics. 
Based on the projective measurement data along the evolving time series, we find that QNODE can learn, reconstruct and extrapolate the quantum trajectories with high accuracy.
Furthermore, the Heisenberg uncertainty principle of a qubit is recovered from the QNODE's generated dynamics.
The evidence from our numerical experiments demonstrates that the QNODE is capable of generating quantum dynamics that preserve physical laws even when they extrapolate outside of the training time. 

The QNODE is capable of learning systems beyond the simple two-level quantum system presented so far. The difficulty lies in generating enough data from a larger dimensional Hilbert space. 
To demonstrate such point, we have carried out two-qubit numerical experiments with the system Hamiltonian $\hat{H}_2$.
The results are shown in \figref{fig:two_qubit_samples}, where we show that samples generated by the QNODE is valid and physical. 
The quantitative analysis regarding the two-qubit case can be seen in \figref{fig:two_qubit_reconstruct}.
Potentially, multi-qubit data could be gathered from recent superconducting qubits experiments \cite{arute2019quantum,zhong2020quantum,wu2021strong} with randomized quantum gates, to produce some physical observables in time.
We believe that the QNODE is one step closer to machine-assisted discovery of scientific principles such as quantum phenomenon modeled by many interacting classical worlds \cite{hall2014quantum}, or could be used to implement dynamical decoupling schemes \cite{viola1999dynamical} on the fly during quantum dynamical evolutions as a kind of inverse design \cite{vargas2020inverse}, thereby paving the way towards large-scale NISQ devices. 

The main code used to obtain data presented here can be
found in a public repository \cite{choi2021learning}.


\section*{Acknowledgements}
We acknowledge fruitful discussions with Areeya Chantasri, Howard Wiseman ,and Rodrigo Vargas.
T.H.K. and A.A.-G. acknowledge funding from Dr. Anders G. Fr{\o}seth.
A.A.-G. also acknowledges support from the Canada 150 Research Chairs Program, the Canada Industrial Research Chair Program, and Google, Inc. in the form of a Google Focused Award.
Computations and ML training of open, closed ,and two-qubit quantum systems were performed on the Niagara supercomputer at the
SciNet HPC Consortium \cite{loken2010scinet,ponce2019deploying}. SciNet is funded by the Canada Foundation for Innovation; the Government of Ontario; Ontario Research Fund - Research Excellence; and the University of Toronto.

\begin{widetext}
\appendix


\setcounter{equation}{0}
\setcounter{figure}{0}
\setcounter{table}{0}
\makeatletter
\renewcommand{\theequation}{A\arabic{equation}}
\renewcommand{\thefigure}{A\arabic{figure}}

\section{Quantitative Analysis}

\begin{figure*}[h]
\rotatebox{90}{\hspace{0.2cm} $\brakets{\sigma_z}$ \hspace{0.2cm} $\brakets{\sigma_y}$ \hspace{0.2cm} $\brakets{\sigma_x}$} 
\includegraphics[width=0.23\textwidth]{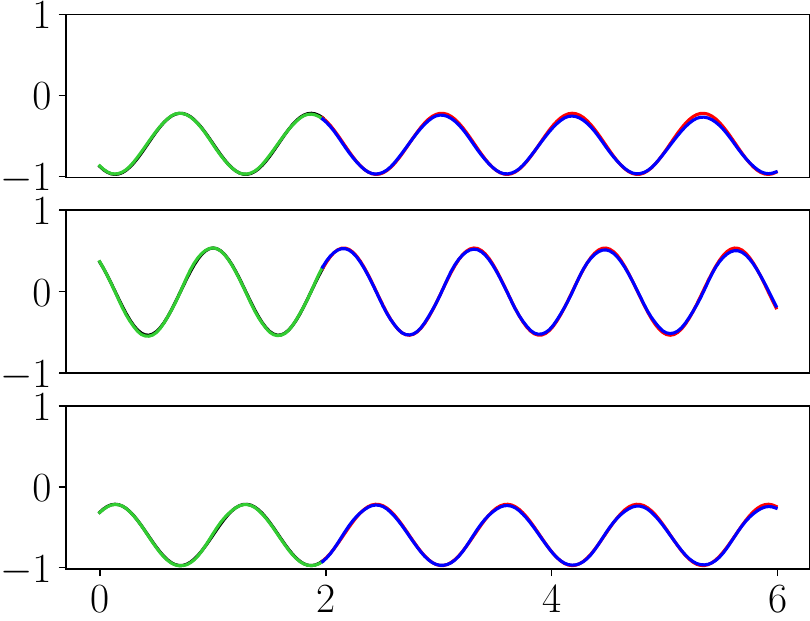}
\includegraphics[width=0.23\textwidth]{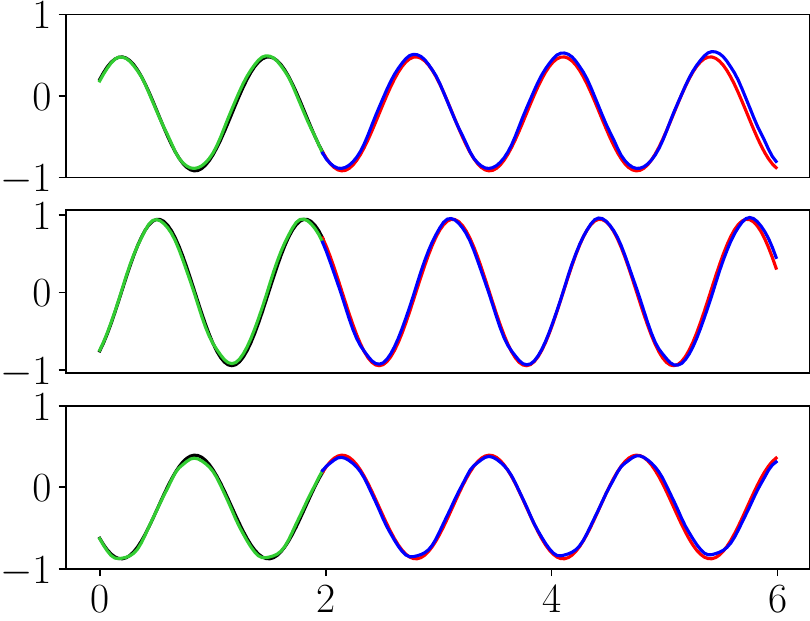}
\includegraphics[width=0.23\textwidth]{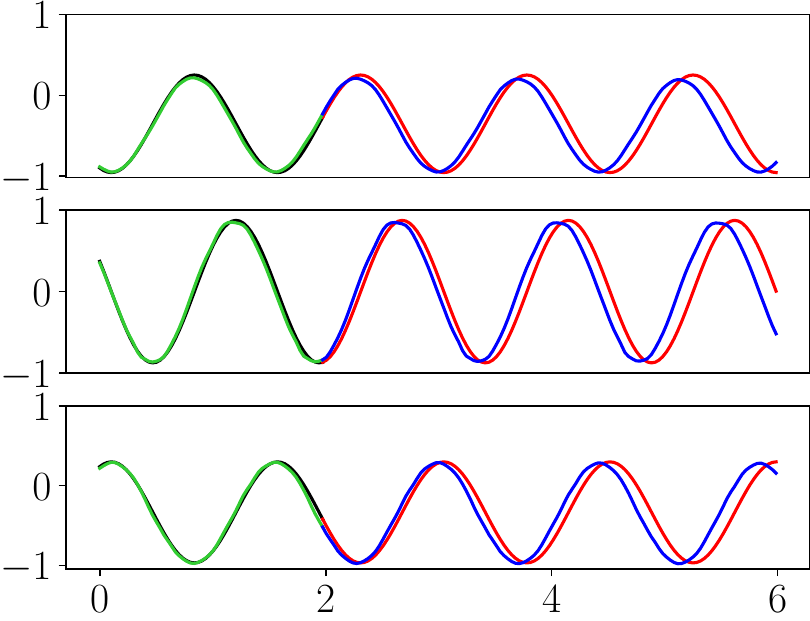} 
\includegraphics[width=0.23\textwidth]{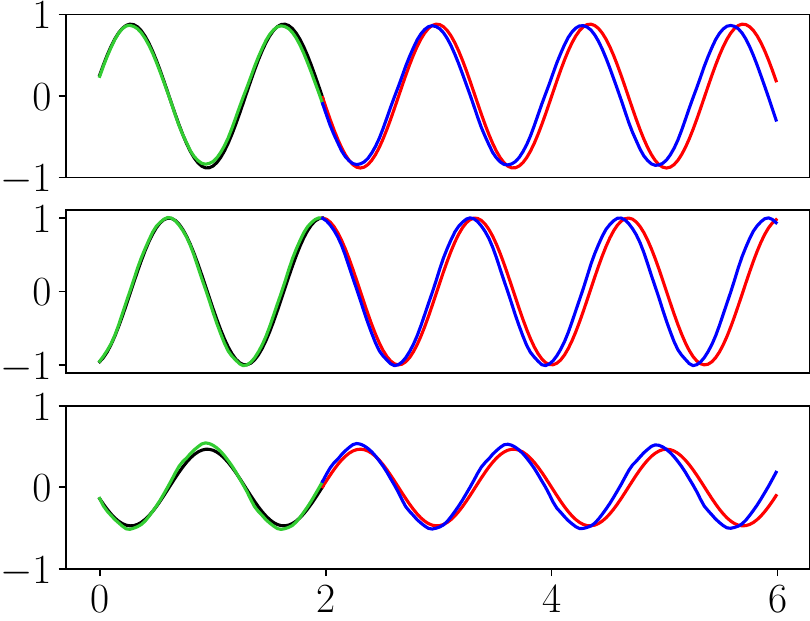}

\hspace{7mm} \color{red} MSE $=6.70\times 10^{-4}$ \hspace{11mm} \color{green} MSE $=8.09\times 10^{-3}$ \hspace{13mm} \color{purple} MSE $=1.51\times 10^{-2}$ \hspace{13mm} \color{magenta} MSE $=1.93\times 10^{-2}$

\color{black}
\vspace{2mm}
(a) Reconstructed trajectories
\vspace{0.3cm}
\hspace{-0.5cm}

\rotatebox{90}{\hspace{0.5cm} $\brakets{\sigma_z}$ \hspace{0.6cm} $\brakets{\sigma_y}$ \hspace{0.6cm} $\brakets{\sigma_x}$} 
\includegraphics[width=0.32\textwidth]{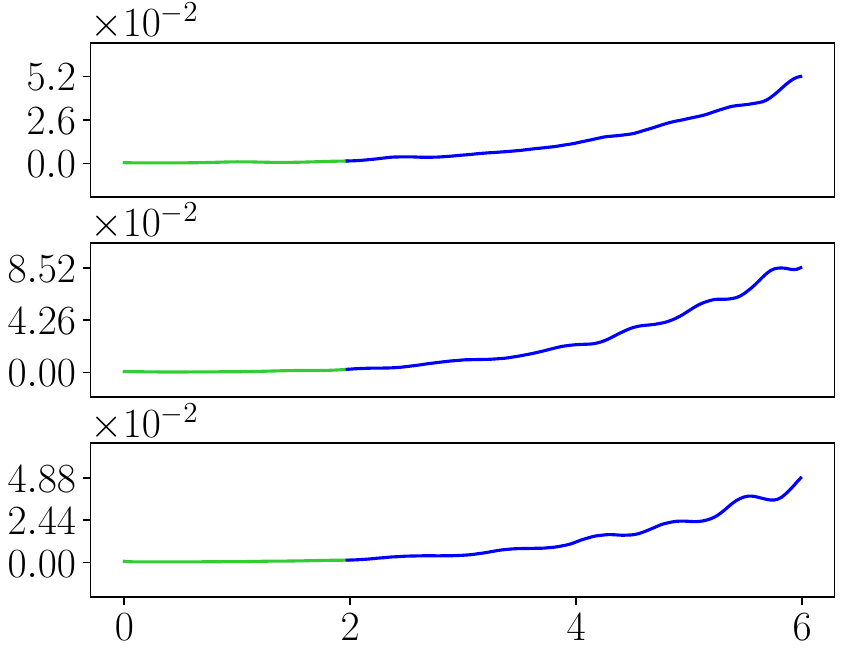}
\hspace{2.1cm}
\includegraphics[width=0.29\textwidth]{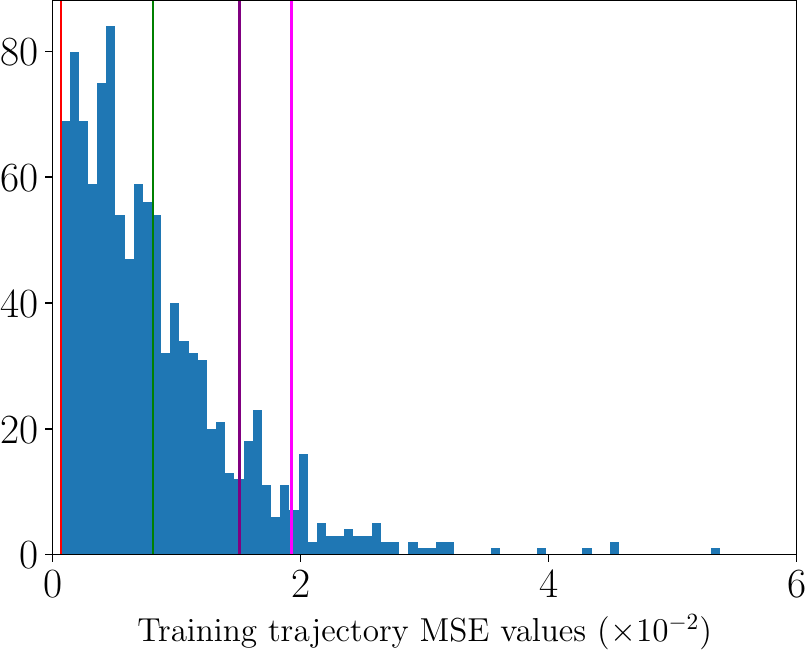}

\hspace{1.6cm} (b) Average MSE over time \hspace{2.4cm}
(c) Histogram of MSE over training data

\caption{\textbf{Closed system reconstructions} (a) Reconstructions with MSEs ranging from best to worst performing (left to right). (b) The average MSE over time. The plot is bounded by the maximum MSE over 6 \textbf{as}. (c) A histogram of all MSE values over training data. The colored lines refer to the MSE of the reconstructions of the same color plotted in (a).}
\label{fig:closed_reconstruct}
\end{figure*}

\begin{figure*}[!h]
\rotatebox{90}{\hspace{0.2cm} $\brakets{\sigma_z}$ \hspace{0.2cm} $\brakets{\sigma_y}$ \hspace{0.2cm} $\brakets{\sigma_x}$} 
\includegraphics[width=0.23\textwidth]{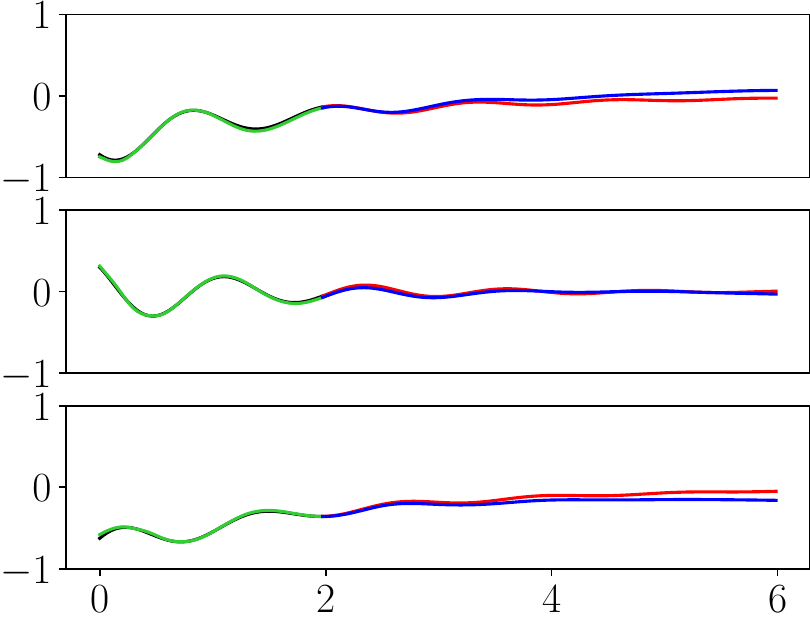}
\includegraphics[width=0.23\textwidth]{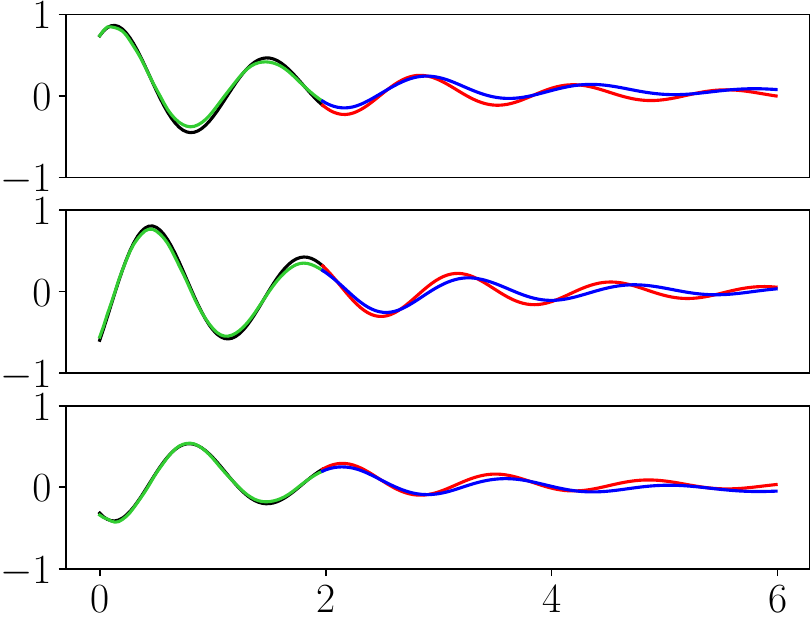}
\includegraphics[width=0.23\textwidth]{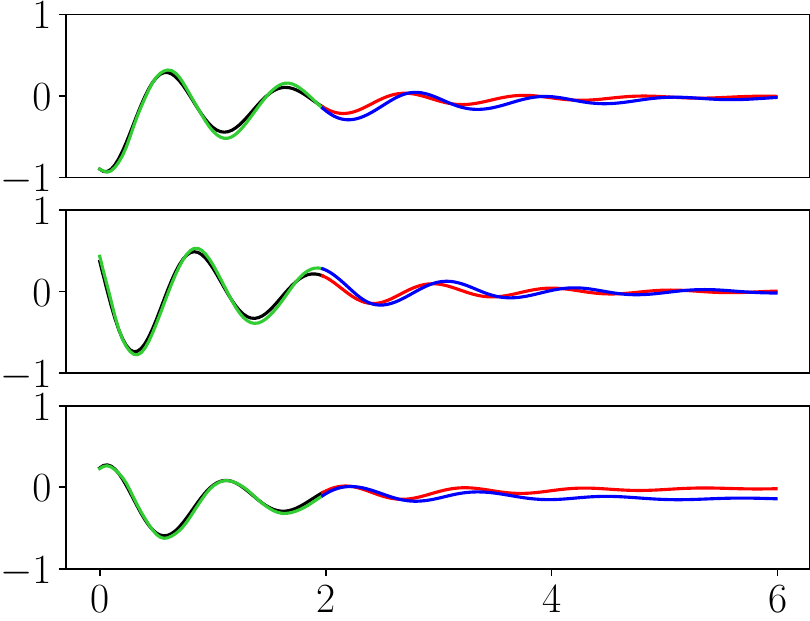} 
\includegraphics[width=0.23\textwidth]{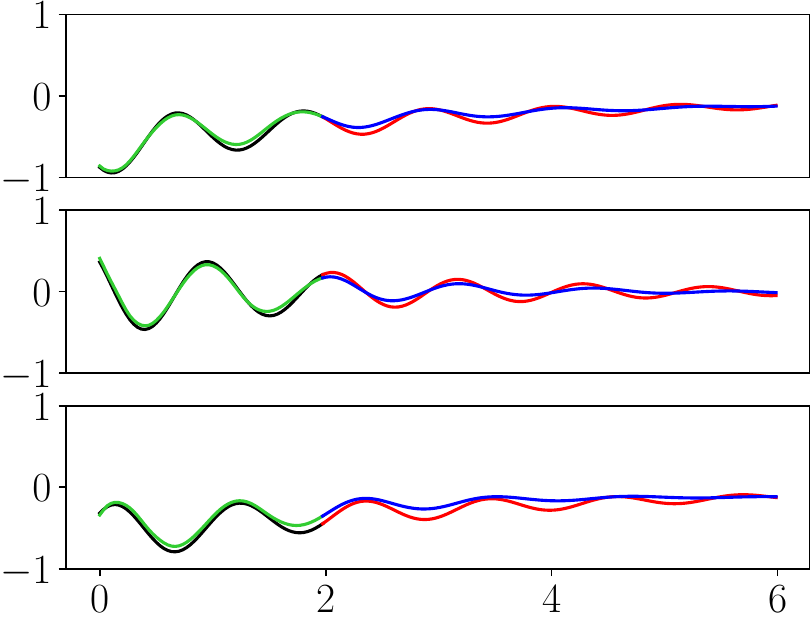}

\hspace{7mm} \color{red} MSE $=1.40\times 10^{-3}$ \hspace{11mm} \color{green} MSE $=1.13\times 10^{-2}$ \hspace{13mm} \color{purple} MSE $=1.50\times 10^{-2}$ \hspace{13mm} \color{magenta} MSE $=1.99\times 10^{-2}$

\color{black}
\vspace{2mm}
(a) Reconstructed trajectories

\vspace{0.3cm}
\hspace{-0.5cm}
\rotatebox{90}{\hspace{0.5cm} $\brakets{\sigma_z}$ \hspace{0.6cm} $\brakets{\sigma_y}$ \hspace{0.6cm} $\brakets{\sigma_x}$} 
\includegraphics[width=0.32\textwidth]{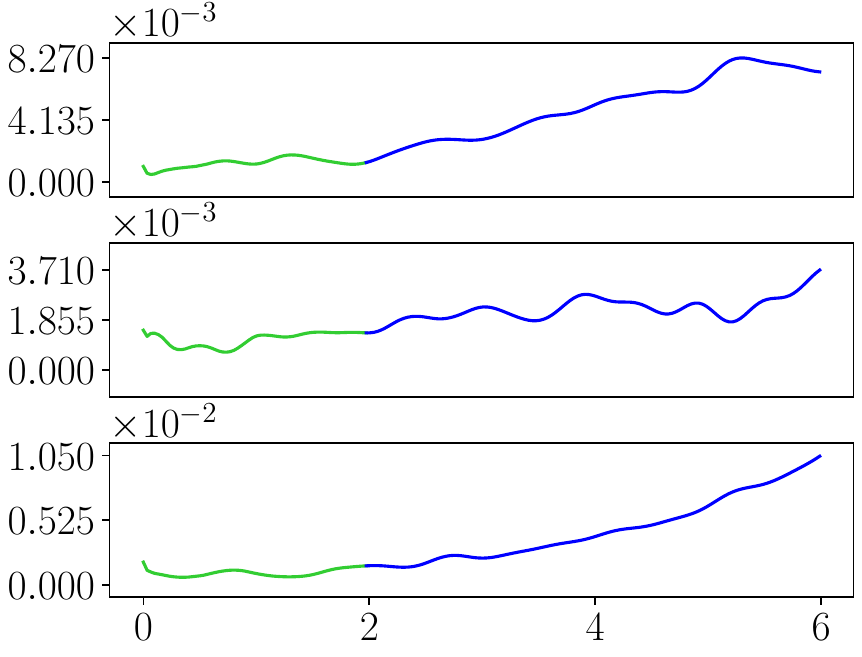}
\hspace{2.1cm}
\includegraphics[width=0.29\textwidth]{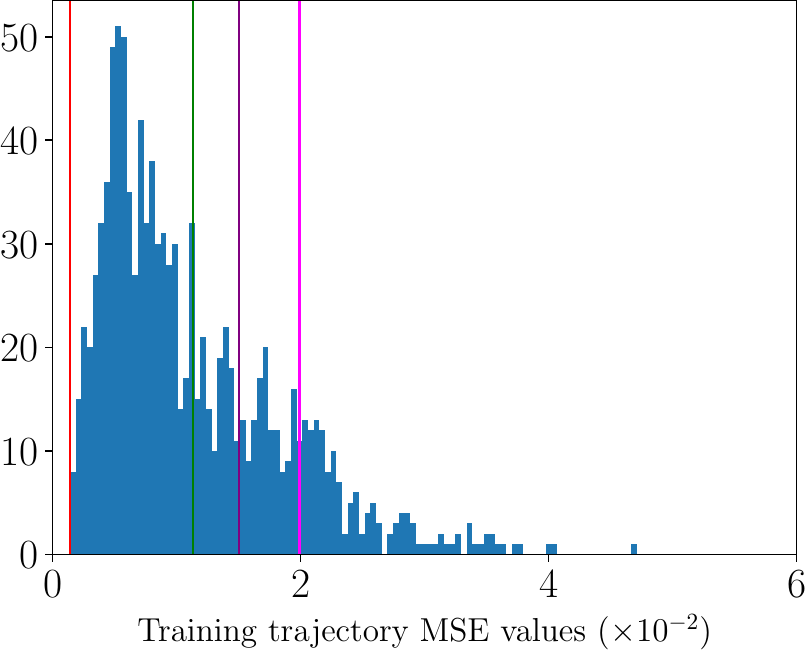}

\hspace{1.6cm} (b) Average MSE over time \hspace{2.4cm}
(c) Histogram of MSE over training data

\caption{\textbf{Open system reconstructions}{(a) Reconstructions with MSEs ranging from best to worst performing (left to right). (b) The average MSE over time. The plot is bounded by the maximum MSE over 6 \textbf{as}. (c) A histogram of all MSE values over training data. The colored lines refer to the MSE of the reconstructions of the same color plotted in (a).}
}
\label{fig:open_reconstruct}
\end{figure*}

\begin{figure*}[!h]
\rotatebox{90}{$\brakets{\sigma_{z2}}$ $\brakets{\sigma_{z1}}$ $\brakets{\sigma_{x2}}$ $\brakets{\sigma_{x1}}$} 
\includegraphics[width=0.23\textwidth]{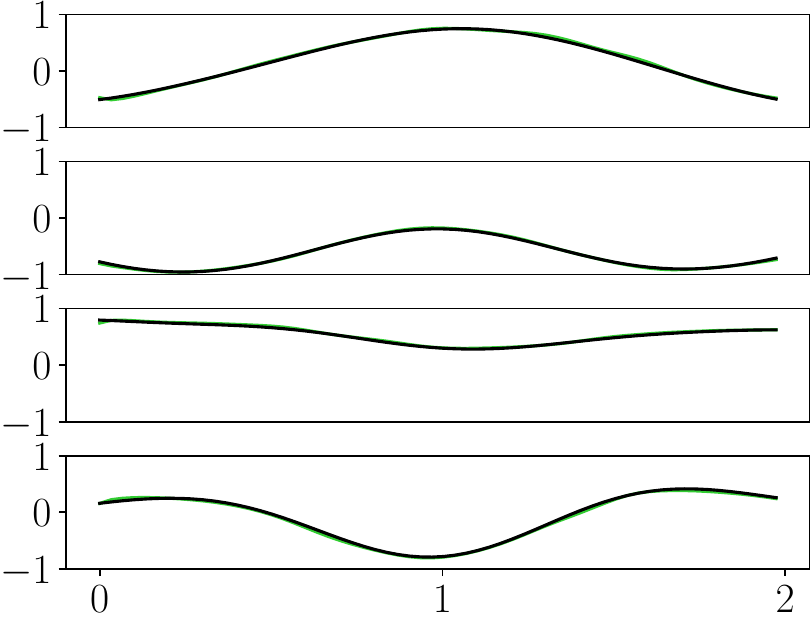}
\includegraphics[width=0.23\textwidth]{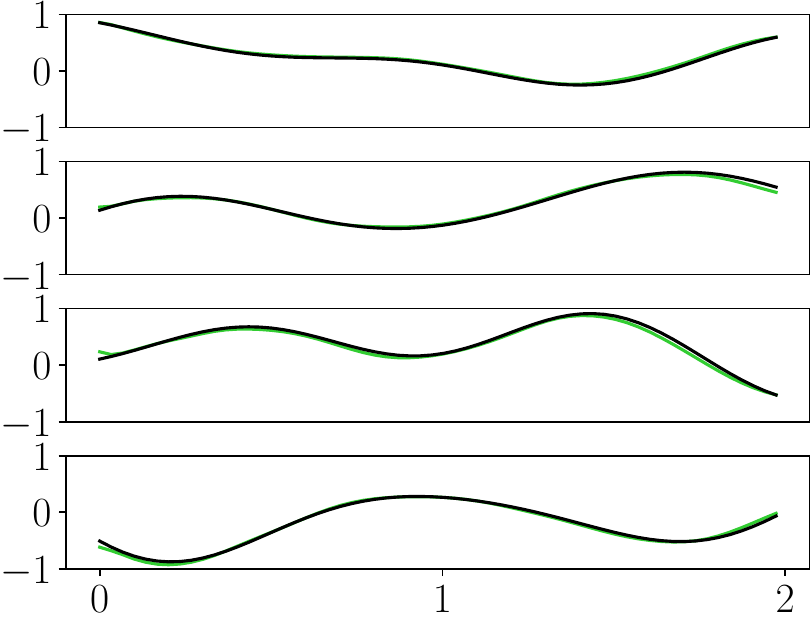}
\includegraphics[width=0.23\textwidth]{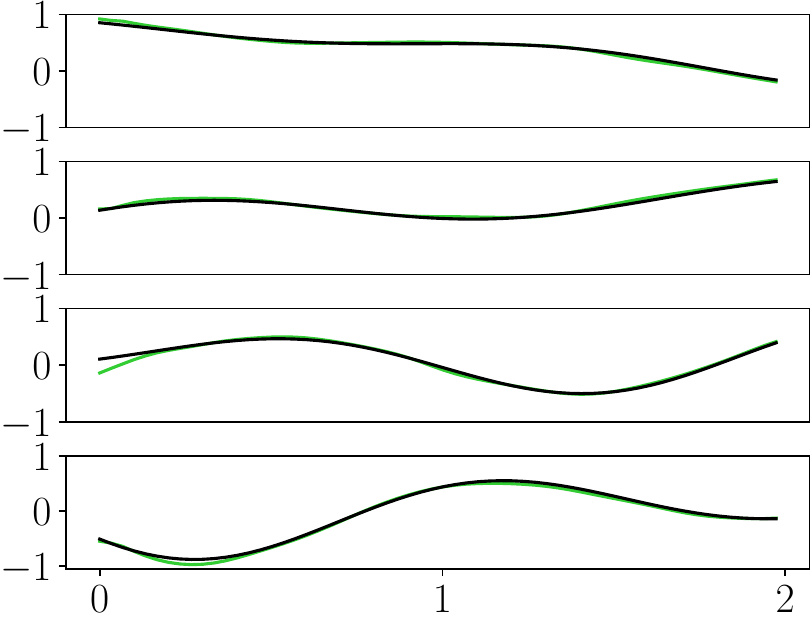} 
\includegraphics[width=0.23\textwidth]{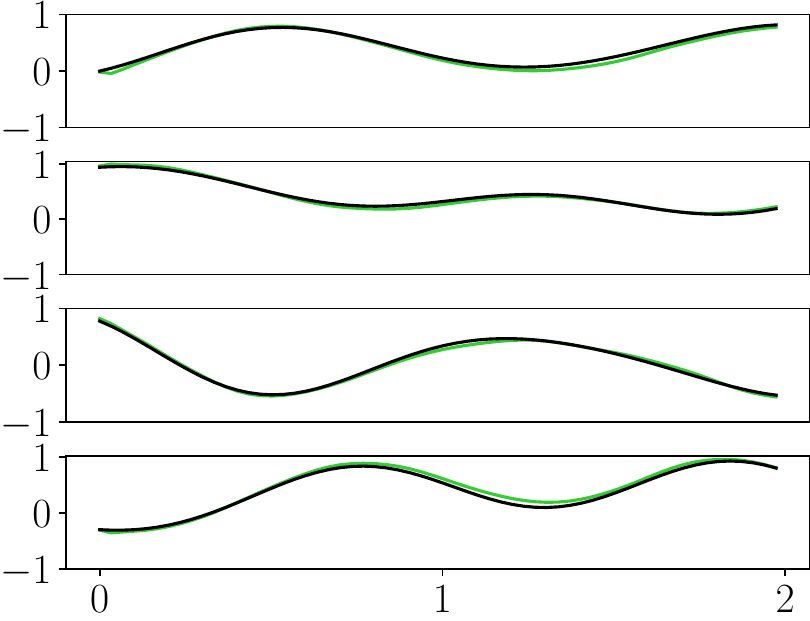}

\hspace{7mm} \color{red} MSE $=3.36\times 10^{-3}$ \hspace{11mm} \color{green} MSE $=1.20\times 10^{-2}$ \hspace{13mm} \color{purple} MSE $=1.49\times 10^{-2}$ \hspace{13mm} \color{magenta} MSE $=2.00\times 10^{-2}$

\color{black}
\vspace{2mm}
(a) Reconstructed trajectories

\vspace{0.3cm}
\hspace{-0.5cm}
\rotatebox{90}{\hspace{0.5cm}$\brakets{\sigma_{z2}}$ \hspace{0.15cm} $\brakets{\sigma_{z1}}$ \hspace{0.15cm} $\brakets{\sigma_{x2}}$ \hspace{0.15cm} $\brakets{\sigma_{x1}}$} 
\includegraphics[width=0.35\textwidth]{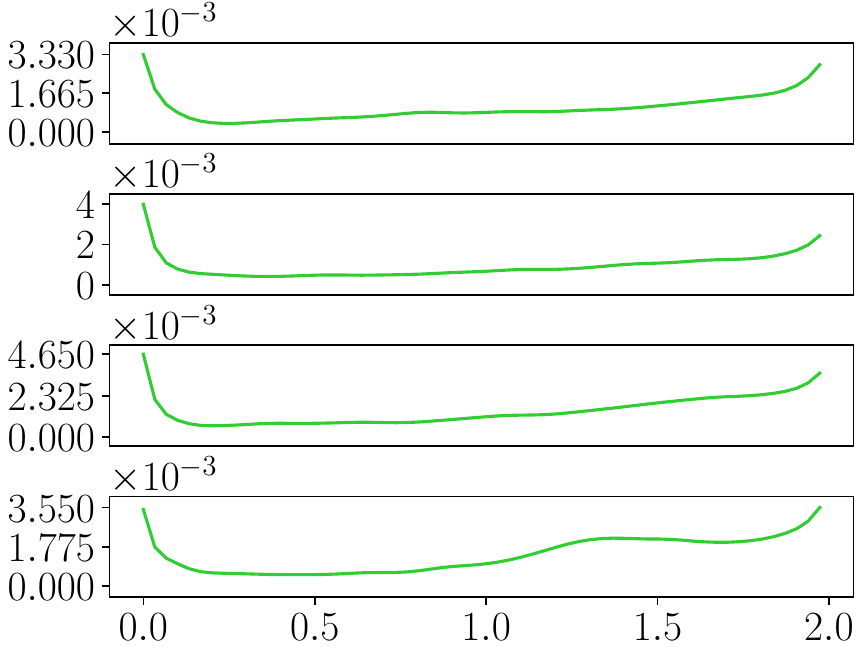}
\hspace{2.1cm}
\includegraphics[width=0.32\textwidth]{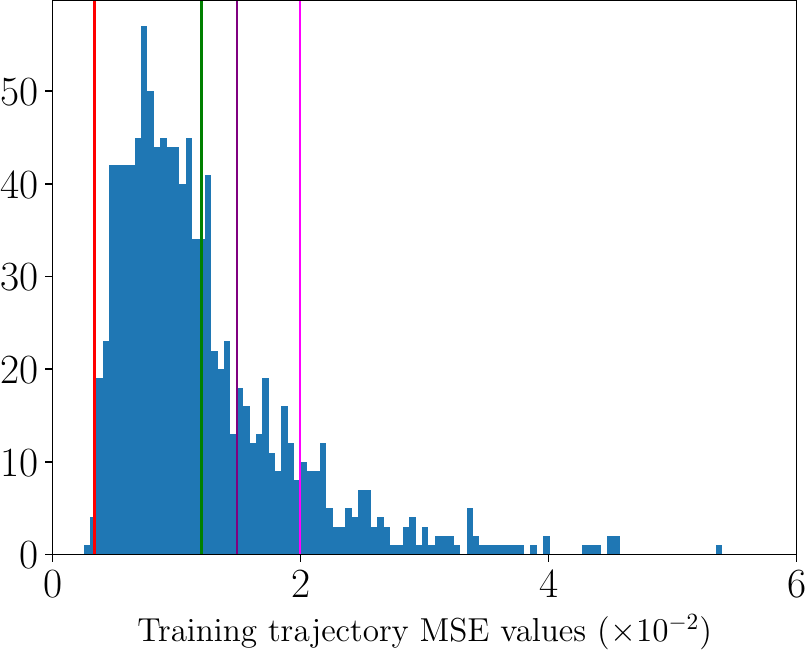}

\hspace{1.1cm} (b) Average MSE over time \hspace{2.7cm}
(c) Histogram of MSE over training data

\caption{\textbf{Two-qubit reconstructions} {(a) Reconstructions with MSEs ranging from best to worst performing (left to right). (b) The average MSE over time. The plot is bounded by the maximum MSE over 6 \textbf{as}. (c) A histogram of all MSE values over training data. The colored lines refer to the MSE of the reconstructions of the same color plotted in (a).}
}
\label{fig:two_qubit_reconstruct}
\end{figure*}

\begin{figure}[!h]
\includegraphics[width=0.4\textwidth]{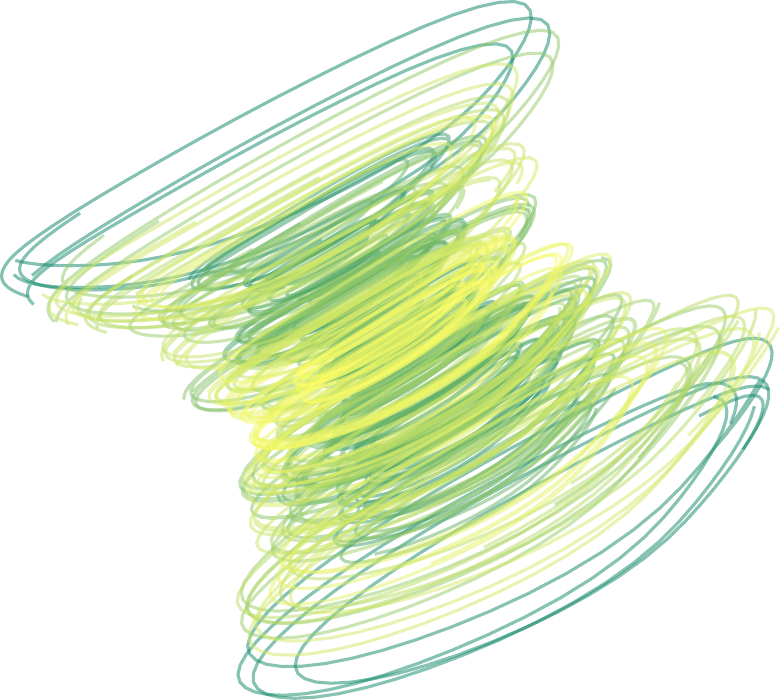}
\includegraphics[width=0.4\textwidth]{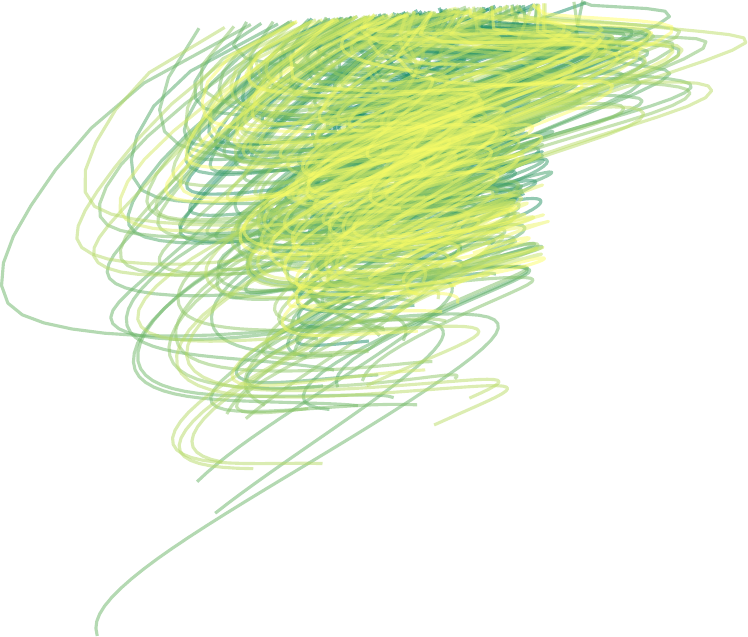}

(a) Closed qubit system \hspace{3.5cm} (b) Open qubit system 

\caption{\textbf{QNODE's latent dynamics of a simple two-level quantum system}.
Latent trajectories learned by the QNODE for all training data in the closed quantum system (a) and open quantum system (b). }
\label{fig:QNODE_latent}
\end{figure}

\section{Methods}

\subsection{Creating the Training Data}\label{SM:methods_data}

As mentioned in the manuscript, both datasets come from a similar type of hamiltonian, more specifically, $\hat{H}=(\omega\hat{\sigma}_z +\Delta \hat{\sigma}_x)/2$, where $\omega$ is energy splitting of the two-level system, $\Delta$ is the detuning, and $\hat{\sigma}$'s are the usual Pauli matrices. 
$\omega$ and $\Delta$ are sampled from a Gaussian Distribution with a range of $1.5$ to $2.5$. Additionally, for the open system, $\Gamma$ is sampled from a Gaussian Distribution with a range of $0.1$ to $0.3$. We create 30 of these hamiltonians and evolve them with 36 sampled initial states from the surface of the bloch sphere for 60 time steps from 0 to 2 seconds. We then take the expectation values of the evolution with respect to $\hat{\sigma}_x$, $\hat{\sigma}_y$, $\hat{\sigma} _z$. This creates our dataset with the size of $(30 * 36, 60, 3)$. 

\subsection{Training the Models}
\label{SM:methods_training}
The results in the paper are the best models based on lowest Evidence Lower Bound Objective (ELBO) and Mean Squared Error (MSE). The table below highlights the hyperparameters e.g. RNN Hidden Units (HU) and metrics for the best closed, open, and two-qubit models trained.

 \begin{figure}[h]
 \begin{tabular}{cccccccc}
         Model & RNN HU & ODE HU & MLP HU & Learning Rate & Epochs & Total MSE & Average MSE\\
         \hline
         Closed & 48 & 48  & 48 & 4e-3  & 7500 & 2.765e-3 & 2.560e-06 \\
         Open & 53 & 53 & 53 & 7e-3 & 7500 &  9.964e-4 & 9.226e-07 \\
         Two Qubit & 170 & 170 & 170 & 2e-3 & 7200 & 2.375e-02 & 2.199e-06 
\end{tabular} 
\end{figure}

\newpage
\section{Microscopic derivation of open quantum system master equation}
\label{append:master_eq}

In this section, to remind our readers who are not familiar with open quantum system treatment, we present a microscopic derivation formalism to arrive at a proper time-local Lindblad master equation for a general open quantum system. 
Suppose we consider a scenario where a quantum system $S$ weakly interacts with a bath environment $B$.  

In the interaction picture, the system evolution can be written as
\begin{equation}
\frac{d\hat{\rho}_S (t)}{dt}= -\int_0 ^t d\tau \hspace{0.2cm} \rm{Tr}_B [\hat{H}_I (t),[\hat{H}_I (\tau),\hat{\rho} (\tau)]],
\end{equation}
where $\hat{\rho}_S$ represents the system density operator, the subscript $B$ represents the bath and $\hat{\rho}$ is the combined system and bath density operator. In application of the Born approximation, we assume the interaction between the system and the bath is so small that the bath degrees of freedom $\hat{\rho}_B$ is negligibly affected by the system-bath interaction. Thus, the total system at time $t$ can be approximated by $\hat{\rho}(t)\approx \hat{\rho}_S (t) \otimes \hat{\rho}_B$. Therefore, we arrive at 
\begin{equation}
\frac{d\hat{\rho}_S (t)}{dt}= -\int_0 ^t d\tau \hspace{0.2cm} \rm{Tr}_B [\hat{H}_I (t),[\hat{H}_I (\tau),\hat{\rho}_S (\tau)\otimes \hat{\rho}_B]].
\end{equation}
Furthermore, we assume environmental excitations decay over time and cannot be resolved. With the Markovian approximation, we arrive at the Redfield equation
\begin{equation}
 \frac{d\hat{\rho}_S (t)}{dt}=-\int_0 ^t d\tau \hspace{0.2cm} \rm{Tr}_B [\hat{H}_I (t),[\hat{H}_I (\tau),\hat{\rho}_S (t)\otimes \hat{\rho}_B]].\label{Redfield_Eq}
\end{equation}
We then substitute $\tau$ by $t-\tau$ and change the upper limit of the integral to $\infty$. This is allowable provided the integrand vanishes sufficiently fast for $\tau \gg \tau_B$ and the timescale over which the state of the system varies is large compared to the timescale over which the bath correlation functions decay. Thus, we arrive at the Markovian quantum master equation
\begin{equation}
 \frac{d\hat{\rho}_S (t)}{dt}=-\int_0 ^\infty d\tau \hspace{0.2cm} \rm{Tr}_B [\hat{H}_I (t),[\hat{H}_I (t-\tau),\hat{\rho}_S (t)\otimes \hat{\rho}_B]],\label{Born_Markov}
\end{equation} 
where the time evolution is given by the present state $\hat{\rho}_S (t)$ and not dependent on the system state in the past. Thus, there is no memory effect. 

The above procedure is termed the Born-Markov approximation. In general, it does not guarantee Eq. \ref{Born_Markov} provides the generator of a dynamical semi-group. Thus, further secular approximation is needed \cite{breuer2002theory}. We proceed by decomposing the interaction Hamiltonian into two parts:
\begin{equation}
	\hat{H}_I = \sum_\alpha \hat{A}_\alpha \otimes \hat{B}_\alpha,
\end{equation}
with $\hat{A}_\alpha ^\dagger =\hat{A}_\alpha$ and $\hat{B}_\alpha ^\dagger =\hat{B}_\alpha$. The secular approximation is achieved if the interaction Hamiltonian is decomposed in terms of the eigenoperators of the system Hamiltonian $\hat{H}_S$. Let us denote the projection onto the eigenspace belonging to the eigenvalue $\epsilon$ in $\hat{H}_S$ as $\hat{\Pi}(\epsilon)$. Then,
\begin{equation}
	\hat{A}_\alpha (\omega) = \sum_{\epsilon' -\epsilon =\omega} \hat{\Pi}(\epsilon)\hat{A}_\alpha \hat{\Pi}(\epsilon').
\end{equation}
The sum is extended over all energy eigenvalues $\epsilon'$ and $\epsilon$ of $H_S$ with a fixed energy difference $\omega$. As a consequence, we have $[\hat{H}_S, \hat{A}_{\alpha}(\omega)]=-\omega \hat{A}_\alpha (\omega), [\hat{H}_S, \hat{A}_{\alpha}^\dagger (\omega)]=+\omega \hat{A}_\alpha ^\dagger (\omega)$. The corresponding interaction picture operators take the form
\begin{eqnarray}
	e^{i\hat{H}_S t}\hat{A}_\alpha (\omega) e^{-i\hat{H}_S t}&=& e^{-i\omega t}\hat{A}_\alpha (\omega),\\
	e^{i\hat{H}_S t}\hat{A}_\alpha (\omega) e^{-i\hat{H}_S t}&=& e^{-i\omega t}\hat{A}_\alpha (\omega),
\end{eqnarray}
with $[\hat{H}_S, \hat{A}^\dagger _\alpha (\omega)\hat{A}_beta (\omega)]=0$ and $\hat{A}^\dagger _\alpha (\omega)=\hat{A}_\alpha (-\omega)$. We note that $\hat{A}$'s satisfy the completeness relationship: $\sum_\omega \hat{A}_\alpha (\omega)=\sum_\omega \hat{A}_\alpha ^\dagger (\omega)=\hat{A}_\alpha$. Eventually, the interaction Hamiltonian in the interaction picture is then
\begin{equation}
	\hat{H}_I (t) = \sum_{\alpha,\omega} e^{-i\omega t} \hat{A}_\alpha (\omega)\otimes \hat{B}_\alpha (t)=\sum_{\alpha,\omega} e^{i\omega t} \hat{A}_\alpha ^\dagger (\omega)\otimes \hat{B}_\alpha ^\dagger (t),
\end{equation}
where $\hat{B}_\alpha (t) = e^{i\hat{H}_B t}\hat{B}_\alpha e^{-i\hat{H}_B t}$. By substituting $\hat{H}_I$ back to Eq. \ref{Born_Markov}, we arrive at 
\begin{eqnarray}
	\frac{d\hat{\rho}_S (t)}{dt}&=& \int_0 ^\infty d\tau \hspace{0.2cm} \rm{Tr}_B \left[ \hat{H}_I (t-\tau)\hat{\rho}_S (t)\hat{\rho}_B \hat{H}_I (t) -\hat{H}_I (t) \hat{H}_I (t-\tau)\hat{\rho}_S (t)\hat{\rho}_B \right] + h.c. \nonumber\\
	&=&\sum_{\omega ,\omega'}\sum_{\alpha,\beta}e^{i(\omega' -\omega)t}\Gamma_{\alpha \beta}(\omega)(\hat{A}_\beta (\omega)\hat{\rho}_S (t)\hat{A}_\alpha ^\dagger (\omega')- \hat{A}_\alpha ^\dagger (\omega')\hat{A}_\beta (\omega)\hat{\rho}_S (t)) + \rm{h.c.},
\end{eqnarray}
with a bath correlation function
\begin{equation}
	\Gamma_{\alpha \beta}(\omega)= \int_0 ^\infty d\tau e^{i\omega \tau} \langle \hat{B}_\alpha ^\dagger (\tau) \hat{B}_\beta (0)\rangle.
\end{equation}
Typical time scale $\tau_S$ for which the system evolves is defined as $|\omega' -\omega|^{-1}$, where $\omega' \neq \omega$. By neglecting the rapidly evolving term $\omega' \neq \omega$ during which $\rho_S$ varies appreciably, we have
\begin{equation}
	\frac{d\hat{\rho}_S (t)}{dt}= \sum_{\omega}\sum_{\alpha,\beta}\Gamma_{\alpha \beta}(\omega)(\hat{A}_\beta (\omega)\hat{\rho}_S (t)\hat{A}_\alpha ^\dagger (\omega)- \hat{A}_\alpha ^\dagger (\omega)\hat{A}_\beta (\omega)\hat{\rho}_S (t))+ \rm{h.c.}.
\end{equation}
Our QNODE model uses quantum data generated from such time-local Lindblad master equation for open quantum system dynamics.

\end{widetext}

\bibliographystyle{apsrev4-1}
\bibliography{main}

\end{document}